\documentclass[aps,prc,twocolumn,showpacs,floatfix,superscriptaddress,nofootnoteinbib,longbibliography]{revtex4-1}

\usepackage{amsmath}
\usepackage{inputenc}
\usepackage{amssymb}
\usepackage{graphicx}
\usepackage[version=4]{mhchem}
\usepackage{braket}
\usepackage{amsfonts}
\usepackage{amssymb}
\usepackage{cancel}
\usepackage{mathtools}
\usepackage{bbold}
\usepackage{xcolor}
\usepackage{epsfig}
\usepackage{bm}
\usepackage{color}
\usepackage{float}
\usepackage{dcolumn}
\usepackage{multirow} 
\usepackage{siunitx}
\usepackage{appendix}
\usepackage{xtab,afterpage}
\usepackage{makecell}
\usepackage{tikz}
\usetikzlibrary{trees}
\usetikzlibrary{decorations.pathmorphing}
\usetikzlibrary{decorations.markings}
\usetikzlibrary{shapes.misc}
\usepackage[unicode=true,
  linktocpage,
  linkbordercolor={0.5 0.5 1},
  citebordercolor={0.5 1 0.5},
  colorlinks=true,
  linkcolor=blue,
  citecolor=blue,
  urlcolor=blue]{hyperref}  
\usepackage{cgp_macros}
\usepackage{lipsum} 
\usepackage{placeins}  

\tikzset{
gluon/.style={decorate, draw=black,
        decoration={coil,amplitude=4pt, segment length=5pt}},
nucleon/.style={draw=black, postaction={decorate},
        decoration={markings,mark=at position .55 with {\arrow[draw=black, scale=2]{<}}}},
amplitude/.style={draw=black, line width = 1mm},
cross/.style={cross out, draw=black, minimum size=2*(#1-\pgflinewidth), inner sep=0pt, outer sep=0pt},
cross/.default={1pt}}

\graphicspath{{.}}

\newcommand{\beq}{\begin{equation}}
\newcommand{\eeq}{\end{equation}}
\newcommand{\beqa}{\begin{eqnarray}}
\newcommand{\eeqa}{\end{eqnarray}}

\newcommand{\NNLOsat}{NNLO$_{\rm sat}$\;}

\newcommand{\NNLOgodlow}{$\Delta$NNLO$_{\rm GO}$(394)\;}

\begin{document}

\title{Electromagnetic observables of open-shell nuclei from coupled-cluster theory}

\thanks{This manuscript has been authored in part by UT-Battelle, LLC, under contract DE-AC05-00OR22725 with the US Department of Energy (DOE). The publisher acknowledges the US government license to provide public access under the DOE Public Access Plan (http://energy.gov/downloads/doe-public-access-plan).}

\author{F.~Bonaiti}
\affiliation{Institut f\"ur Kernphysik and PRISMA$^+$ Cluster of Excellence, Johannes Gutenberg-Universit\"at Mainz, 55128
  Mainz, Germany}

\author{S.~Bacca}
\affiliation{Institut f\"ur Kernphysik and PRISMA$^+$ Cluster of Excellence, Johannes Gutenberg-Universit\"at Mainz, 55128
  Mainz, Germany}
\affiliation{Helmholtz-Institut Mainz, Johannes Gutenberg-Universit\"at Mainz, D-55099 Mainz, Germany}

\author{G.~Hagen}
\affiliation{Physics Division, Oak Ridge National Laboratory,
Oak Ridge, TN 37831, USA} 
\affiliation{Department of Physics and Astronomy, University of Tennessee,
Knoxville, TN 37996, USA} 

\author{G.~R.~Jansen}
\affiliation{National Center for Computational Sciences, Oak Ridge National Laboratory, Oak Ridge, TN 37831, USA}
\affiliation{Physics Division, Oak Ridge National Laboratory,
Oak Ridge, TN 37831, USA} 

\begin{abstract}
\noindent
We develop a new method to describe electromagnetic observables of open-shell nuclei with two nucleons outside a closed shell. This approach combines the equation-of-motion coupled-cluster method for such systems and the Lorentz integral transform technique, expanding the applicability of coupled-cluster theory for electromagnetic observables beyond closed-(sub)shell nuclei. To validate this new approach, we compute the non-energy-weighted dipole sum rule and the dipole polarizability of $^{16,24}$O in both the closed-(sub)shell and the new equation-of-motion coupled-cluster frameworks, finding agreement within error bars. We then analyze the evolution of the dipole polarizability along the oxygen and calcium isotopic chains. Our predictions agree well with available experimental data and other available theoretical calculations for the closed-(sub)shell $^{16,22}$O and the open-shell $^{18}$O. In the calcium isotopes, we observe that our dipole polarizability predictions for open-shell nuclei are lower than those of closed-(sub)shell nuclei. We expect that our predictions for $^{24}$O and $^{54,56}$Ca will motivate future experimental studies at the dripline.
\end{abstract}

\maketitle

\section{Introduction}
Electromagnetic probes are key observables to test our understanding of the internal dynamics of atomic nuclei. The small value of the electromagnetic coupling constant $\alpha$ ensures the applicability of perturbation theory, allowing a straightforward connection between theoretical calculations and experimental measurements. In the past, this close interplay between theory and experiment has led to remarkable discoveries, such as the observation of giant dipole resonances (GDR) and their interpretation in terms of collective excitations \cite{goldhaber1948,steinwedel1950}. While systematic experimental investigations have been performed over the years on stable nuclei, more recently, radioactive ion beam facilities around the world have enabled studies on unstable nuclei, uncovering fascinating phenomena such as the emergence of the pygmy dipole resonance~\cite{kobayashi1989,aumann2013}.

For a long time the computation of electromagnetic response functions was mostly performed using phenomenological models such as the random phase approximation (RPA) and its extensions (see e.g., \cite{paar2007,rocamaza2018,colo2022} and references therein). However, recently progress has been made in the framework of ab initio calculations~\cite{ekstrom2022}, describing the nucleus as a composite system of protons and neutrons and linking their interactions to quantum chromodynamics via chiral effective field theory ($\chi$EFT) \cite{epelbaum2009,machleidt2011,hammer2020}. Electromagnetic response functions have been studied in light nuclei via hyperspherical harmonics (HH)~\cite{bacca_photon1, bacca_photon2, bacca_el, bacca_monopole, Kegel, acharya2023}, no-core shell model (NCSM)~\cite{stetcu2007,quaglioni2007,stumpf2017}, and symmetry-adapted NSCM \cite{baker2020}. Merging coupled-cluster (CC) theory \cite{coester1958,coester1960,dean2004,wloch2005,hagen2008,hagen2010,binder2014,hagen2014} with the Lorentz integral transform (LIT) technique \cite{efros2007} in the LIT-CC method has first extended such computations to closed-(sub)shell medium-mass nuclei~\cite{bacca2013,bacca2014,miorelli2016, miorelli2018,simonis2019,sobczyk2021,sobczyk2023}, stimulating subsequent theoretical efforts, e.g., via the self consistent Green's function (SCGF) method \cite{raimondi2019}. Very recently, calculations of responses in open-shell nuclei have been performed via the  symmetry-adapted NCSM~\cite{burrows2023}, the Hartree-Fock-Bogoliubov quasiparticle RPA  \cite{taudiere2023} and the projected generator coordinate method (PGCM)~\cite{porro2024a,porro2024b}.

While knowledge of the full electromagnetic response is preferable, its energy moments, known as sum rules, can be more easily computed. Since they can be compared to experiments, they still provide substantial insight into the dynamics of a quantum system. Among them, the electric dipole polarizability stands out as a noteworthy example. As it corresponds to an inverse energy-weighted sum rule of the dipole response function, it is enhanced by pygmy dipole resonances in neutron-rich nuclei. At the same time, it connects the realm of atomic nuclei to neutron stars due to its strong correlation with the symmetry energy parameters appearing in the nuclear matter equation of state~\cite{rocamaza2013,rocamaza2015,piekarewicz2021}.

Comparison with experimental determinations of the dipole polarizability in stable \cite{birkhan2017,miorelli2018,fearick2023} and unstable nuclei \cite{miorelli2016,kaufmann2020} have identified the LIT-CC method~\cite{bacca2013} as a successful tool to describe this observable in closed-(sub)shell nuclei. Recently, such calculations have been pushed towards the dripline, with a study of the low-lying strength and polarizability of $^8$He \cite{bonaiti2022,bonaiti2024}. 
Building on these benchmarks with theory, ongoing experimental campaigns aim at measuring the dipole strength of open-shell calcium and nickel isotopes at iThemba Labs, South Africa, and RCNP, Japan \cite{vnc_private}. Moreover, a future upgrade of FRIB, USA to 400 MeV/u will enable Coulomb excitation experiments of very neutron-rich systems, allowing for the determination of their dipole polarizability \cite{frib400}.

Motivated by these experimental efforts, this paper takes the first steps toward the goal of extending the LIT-CC method to open-shell nuclei. In particular, we will show how the LIT-CC approach can be reformulated to tackle nuclei in the vicinity of closed shells, making use of the equation-of-motion coupled-cluster (EOM-CC) framework \cite{stanton1993,bartlett2003,bartlett2007,gour2005,gour2006,jansen2011,jansen2013,shen2013,shen2014,ajala2017}. We will focus on two-particle-attached (2PA) nuclei, which can be obtained by adding two nucleons to a closed-(sub)shell system.

The structure of this paper is as follows. In Section~\ref{formalism}, after summarizing the main aspects of the LIT and EOM-CC methods, we present a derivation of the LIT-CC equations for 2PA systems, dubbed here 2PA-LIT-CC equations, and we show how they can be solved employing the Lanczos algorithm. In Section~\ref{validation}, we present a strategy to estimate many-body truncation errors for 2PA-LIT-CC calculations, and we benchmark the newly developed  approach on dipole sum rules of $^{16,24}$O, which can be described by both the closed-(sub)shell LIT-CC method and the 2PA-LIT-CC formalism. In Section \ref{oxygenisotopes} and \ref{calciumisotopes}, we examine the evolution of the dipole polarizability along the oxygen and calcium isotopic chains, respectively. In Section \ref{conclusions}, we draw conclusions and present an outlook.

\section{Formalism}
\label{formalism}
The starting point of our calculations is the nuclear intrinsic Hamiltonian
\begin{equation}
    H = \!\left(\! 1- \frac{1}{A^{*}} \!\right)\! \sum_{i = 1}^A \frac{p_i^2}{2m} \!+ \!\!\left( \sum_{i<j}^A V_{ij} - \frac{\vec{p}_i\cdot\vec{p}_j}{mA^{*}}\right) \!+\!\! \sum_{i<j<k}^A \!W_{ijk} \,,
\end{equation}
where $m$ is the nucleon mass, $V_{ij}$ is the nucleon-nucleon force, $W_{ijk}$ is the three-nucleon force, $A$ is the number of nucleons in the reference state, and $A^{*}$ the number of nucleons of the target system. In the case of a 2PA nucleus, $A^{*} = A + 2$. This ensures that the correct kinetic energy of the center of mass is employed when computing properties of 2PA nuclei. In all the calculations presented in this work, we choose to work with the \NNLOgodlow interaction \cite{jiang2020}. This next-to-next-to-leading order chiral interaction includes the $\Delta$-isobar as an explicit degree of freedom, and has been successfully used in several applications \cite{hu2022,kondo2023,sun2024}.

Our aim is to study electromagnetic excitations of the nucleus by computing its response function, defined as
\begin{equation}
    R(\omega) = \sum_{\mu} \braket{\Psi_0|\Theta^{\dagger}|\Psi_{\mu}} \braket{\Psi_{\mu}|\Theta|\Psi_0} \delta(E_{\mu} - E_0 - \omega)\,,
    \label{response}
\end{equation}
where $\ket{\Psi_{0}}$ is the ground state of the nucleus, and $\ket{\Psi_{\mu}}$ are the excited states  that can be reached from the ground state via the transition operator $\Theta$. Their energy is denoted by $E_{\mu}$, while $\omega$ is the photon energy.  The sum in Eq.~(\ref{response}) encompasses both bound and continuum excited states of the nucleus.

\subsection{The Lorentz Integral Transform method}
\label{lit-section}
Calculating the continuum excited states of a many-body nucleus represents a formidable challenge, as it requires the knowledge of all the possible unbound configurations arising at any photon energies above the nuclear break-up threshold. The Lorentz Integral Transform (LIT) method \cite{efros1994,efros2007} represents a viable strategy to avoid this issue. The starting point of the LIT technique is an integral transform with Lorentzian kernel of the response function
\begin{equation}
   L(\sigma, \Gamma) = \frac{\Gamma}{\pi} \int d\omega \frac{R(\omega)}{(\omega - \sigma)^2 + \Gamma^2},
   \label{lit}
\end{equation} 
where $\sigma$ is the centroid and $\Gamma$ is the width of the Lorentzian  kernel. Substituting Eq.~(\ref{response}) in Eq.~(\ref{lit}) and using the completeness relation for the Hamiltonian eigenstates 
\begin{equation}
    \sum_{\mu} \ket{\Psi_{\mu}} \bra{\Psi_{\mu}} = 1
\end{equation}
one obtains
\begin{equation}
\begin{split}
    L(\sigma, \Gamma) = &\frac{\Gamma}{\pi} \bra{\Psi_0}\Theta^{\dagger} \frac{1}{H - E_0 - \sigma + i\Gamma} \\ &\times\frac{1}{H - E_0 - \sigma - i\Gamma}\Theta \ket{\Psi_0}. 
\end{split}
\end{equation}
Introducing the variable $z = E_0 + \sigma + i\Gamma$, the integral transform can then be calculated as 
\begin{equation}
    L(z) = \frac{\Gamma}{\pi} \braket{\Tilde{\Psi}|\Tilde{\Psi}},
\end{equation}
where $\ket{\Tilde{\Psi}}$ represents the solution of the equation
\begin{equation}
    (H-z)\ket{\Tilde{\Psi}} = \Theta \ket{\Psi_0}
    \label{boundstatelike}
\end{equation}
for different values of $\sigma$ and $\Gamma$. In this way, the many-body scattering problem involved in Eq.~(\ref{response}) is reduced to the solution of the Schr\"odinger-like equation (\ref{boundstatelike}), for which bound-state techniques can be employed. Once $L(\sigma, \Gamma)$ is known, $R(\omega)$ can be determined via a numerical inversion procedure~\cite{efros2007}.

\subsection{Coupled-cluster theory}
In single reference coupled-cluster theory~\cite{coester1958,coester1960,cizek1966,kuemmel1978,bartlett2007,hagen2014}, the nuclear many-body ground state $\ket{\Psi_0}$  is calculated starting from an exponential ansatz
\begin{equation}
    \ket{\Psi_0} = e^T \ket{\Phi_0},
\label{ansatz}
\end{equation}
where $\ket{\Phi_0}$ is a reference mean-field solution, typically obtained from a Hartree-Fock computation  expanded on the harmonic oscillator (HO) single-particle basis. $T$ is the cluster operator responsible for introducing correlations in the many-body ground state, which can be expressed in terms of a sum of particle-hole (p-h) operators, i.e.,  $T = T_1 + T_2 + T_3 + \dots + T_A$, where $T_1$ represent 1p-1h, $T_2$  2p-2h, and $T_3$  3p-3h excitations, respectively. Using Eq.~(\ref{ansatz}), the Schr\"odinger equation for the ground state of the A-body interacting nuclear system can be written as 
\begin{equation}
    \overline{H}_N \ket{\Phi_0} = E_0 \ket{\Phi_0},
    \label{se-cc}
\end{equation}
where $\overline{H}_N = e^{-T} H_N e^{T}$ is the similarity-transformed Hamiltonian. $\overline{H}_N$ can be computed from the Hamiltonian $H_N$, which is normal-ordered with respect to $\ket{\Phi_0}$. 
This work includes three-nucleon Hamiltonians in the normal-ordered two-body approximation~\cite{hagen2007,roth2012}. To simplify our notation, 
from this point on, we will drop the subscript $N$ in $\overline{H}_N$.

As the similarity-transformed Hamiltonian is not Hermitian, knowledge of both the left and right eigenstates is required to compute expectation values. The right eigenstate of $\overline{H}$ coincides with the reference state, while the left one can be expressed as
$\bra{0} = \bra{\Phi_0} (1+\Lambda)$~\cite{arponen1982,arponen1983}, where $\Lambda = \Lambda_1 + \Lambda_2 + \dots + \Lambda_A$ can be written as a sum of $n$p-$n$h de-excitation operators.

Due to computational limitations, truncating the particle-hole expansion of the cluster operator is necessary. The most frequently employed approximation scheme is coupled-cluster singles and doubles (CCSD), including up to 2p-2h in the cluster expansion, namely $T = T_1 + T_2$. An analogous truncation is applied on the left side with $\Lambda = \Lambda_1 + \Lambda_2$. Increased precision can be obtained by adding leading-order 3p-3h excitations via the CCSDT-1 approximation~\cite{watts1993}. We will follow these procedures in the calculations of closed-shell nuclei in this paper.

\subsection{Equation-of-motion coupled-cluster theory} 
While ground-state properties can be calculated once the amplitudes of the $T$ and $\Lambda$ operators are known, excited states are computed in CC theory within the framework of the EOM-CC method. In the EOM-CC approach, we assume that the target state $\ket{\Psi_{\mu}}$ can be written as
\begin{equation}
    \ket{\Psi_{\mu}} = R_{\mu} e^T \ket{\Phi_0}\,,
\label{ansatz r}
\end{equation}
where $R_{\mu}$ is an EOM excitation operator. Due to the non-Hermiticity of $\overline{H}$, also for excited states we need to differentiate between left and right eigenstates of the Hamiltonian. We will then have a similar ansatz for the left state as
\begin{equation}
   \bra{\Psi_{\mu}} = \bra{\Phi_0} L_{\mu}  e^{-T}, 
\label{ansatz l}
\end{equation}
where $L_{\mu}$ is an EOM de-excitation operator.
Inserting Eq.~(\ref{ansatz r}) in the Schr\"odinger equation, and multipying on the left by $e^{-T}$ we obtain the EOM-CC equation
\begin{equation}
\overline{H}R_{\mu} \ket{\Phi_0} = E_{\mu} R_{\mu} \ket{\Phi_0}. 
\label{right-eom}
\end{equation}
Using Eq.~(\ref{ansatz l}), a similar expression can be derived for the left states as 
\begin{equation}
\bra{\Phi_0} L_{\mu} \overline{H} = \bra{\Phi_0} L_{\mu} E_{\mu}. 
\end{equation}
Subtracting $E_0 R_{\mu} \ket{\Phi_0} $ from~Eq.~(\ref{right-eom}) removes disconnected terms and we obtain 
\begin{equation}
    (\overline{H}R_{\mu})_C \ket{\Phi_0} = \omega_{\mu} R_{\mu} \ket{\Phi_0},
    \label{eom-cc-eqn}
\end{equation}
where $\omega_{\mu} = E_{\mu} - E_0$ is the excitation energy with respect to the ground-state. 


The EOM-CC ansatz is employed for two types of target states. The target state can be an excited state of a closed-shell nucleus. In this case, $R_{\mu}$ ($L_{\mu}$) is a particle-conserving excitation (de-excitation) operator that can be written as~\footnote{Note that in the case of electric dipole transitions, if the nuclear ground state has $J^{\pi} = 0$, $r_0$ and $l_0$ are identically zero.}
\begin{equation}
    R_{\mu} = r_0 + \sum_{ai} r^{a}_{i, \mu} a^{\dagger}_a a_i + \frac{1}{4} \sum_{abij} r^{ab}_{ij, \mu} a^{\dagger}_a a^{\dagger}_b a_j a_i + \dots
    \label{rmu}
\end{equation}
and
\begin{equation}
    L_{\mu} = l_0 + \sum_{ai} l^i_{a, \mu} a^{\dagger}_i a_a + \frac{1}{4} \sum_{abij} l_{ab, \mu}^{ij} a^{\dagger}_i a^{\dagger}_j a_b a_a + \dots \,.
    \label{lmu}
\end{equation}

This work focuses specifically on 2PA systems, characterized by having two extra nucleons on top of a closed-(sub)shell system. This framework treats states of $A+2$ nuclei as excited states of the doubly-magic (or semi-magic) A-body neighbor. Wave functions for 2PA systems can be written as
\begin{equation}
    \ket{\Psi_{\mu}^{\rm(A+2)}} = R^{A+2}_{\mu} \ket{\Psi_0^{(A)}} = R^{A+2}_{\mu} e^T \ket{\Phi_0^{(A)}},
    \label{2pa-ansatz}
\end{equation}
where the EOM excitation operator $R^{A+2}_{\mu}$ involves a net creation of two nucleons as
\begin{equation}
    R^{A+2}_{\mu} = \frac{1}{2} \sum_{ab} r^{ab}_{\mu} a^{\dagger}_a a^{\dagger}_b + \frac{1}{6} \sum_{abci} r^{abc}_{i, \mu} a^{\dagger}_a a^{\dagger}_b a^{\dagger}_c a_i + \dots \,.
    \label{rmu-2pa}
\end{equation}
Similarly, on the bra side we have
\begin{equation}
    L^{A+2}_{\mu} = \frac{1}{2} \sum_{ab} l_{ab, \mu} a_a a_b + \frac{1}{6} \sum_{abci} l_{abc, \mu}^{i} a^{\dagger}_i a_c a_b a_a + \dots \,.
    \label{lmu-2pa}
\end{equation}
The EOM operators $R^{A+2}_{\mu}$ and $L^{A+2}_{\mu}$ define the 2PA-EOM-CC approach~\cite{shen2013,shen2014,jansen2011,jansen2013}. Note that in Eq.~(\ref{2pa-ansatz}) the reference state $\ket{\Phi_0}$ is computed applying a mass shift $A^{*} = A + 2$ in the Hamiltonian. Eq.~(\ref{eom-cc-eqn}) holds also when the EOM-CC approach is used to describe open-shell nuclei. In this case, we have
\begin{equation}
     (\overline{H}R_{\mu}^{A+2})_C \ket{\Phi_0} = \omega_{\mu} R_{\mu}^{A+2} \ket{\Phi_0}\,
    \label{eom-cc-eqn-2pa}
\end{equation}
where $\omega_{\mu} = E_{\mu} - E_0^{*}$, with $E_0^{*}$ being the CC ground-state energy of the closed-shell core calculated with mass number $A^{*} = A + 2$.

\subsection{Coupling the Lorentz Integral Transform technique to the 2PA-EOM-CC approach}
Let us now rewrite the response function of Eq.~(\ref{response}) in the coupled-cluster formalism using the EOM-CC ansatz. In the case of closed-shell nuclei, we get
\begin{equation}
\begin{split}
    R(\omega) = &\sum_{\mu} \braket{\Phi_0|(1+ \Lambda) \overline{\Theta^{\dagger}} R_{\mu}|\Phi_0} \\ &\times \braket{\Phi_0|L_{\mu}\overline{\Theta}|\Phi_0} \delta(E_{\mu} - E_0 - \omega), 
    \label{response-cc}
\end{split}
\end{equation}
where $\overline{\Theta} = e^{-T} \Theta e^T$ is the similarity-transformed transition operator, and $R_{\mu}$ and $L_{\mu}$ are the EOM-CC excitation operators of Eqs.~(\ref{rmu}) and (\ref{lmu}). 

The possibility of employing the EOM-CC ansatz for 2PA nuclei suggests a straightforward way to extend the LIT-CC method to such systems. In this case, the response function is given by 
\begin{equation}
\begin{split}
        R(\omega) = &\sum_{\mu} \braket{\Phi_0|L^{A+2}_0\overline{\Theta^{\dagger}} R^{A+2}_{\mu}|\Phi_0} \\ &\times \braket{\Phi_0|L^{A+2}_{\mu}\overline{\Theta} R^{A+2}_0|\Phi_0} \delta(E_{\mu} - E_{0} - \omega).
        \label{response-2pa}
\end{split}
\end{equation}
In Eq.~(\ref{response-2pa}), we label the EOM-CC excitation (de-excitation) operators associated with the ground and excited states of the 2PA nucleus as $R^{A+2}_0$ ($L^{A+2}_0$) and $R^{A+2}_{\mu}$ ($L^{A+2}_{\mu}$), respectively. Their particle-hole expansion is given in Eqs.~(\ref{rmu-2pa}) and (\ref{lmu-2pa}).

We now compute the LIT starting from Eq.~(\ref{response-2pa}). Substituting Eq. (\ref{response-2pa}) in Eq. (\ref{lit}), and using the completeness relation of the left and right Hamiltonian eigenstates
\begin{equation}
\braket{\Psi_{\mu}^L|\Psi_{\mu^{'}}^R} = \delta_{\mu,\mu^{'}}, \quad \sum_{\mu} \ket{\Psi_{\mu}^R}\bra{\Psi_{\mu}^L} = 1,
    \label{completeness}
\end{equation}
we obtain
\begin{equation}
\begin{split}
    L(\sigma, \Gamma) \!=\! \frac{\Gamma}{\pi}\!\braket{\Phi_0|L^{A+2}_0 \overline{\Theta^{\dagger}} (\overline{H}-z^{*})^{-1} \!(\overline{H}-z)^{-1} \overline{\Theta} R^{A+2}_0|\Phi_0},
    \label{lit2pa}
\end{split}
\end{equation}
where $z = E_0 + \sigma + i\Gamma$. Eq.~(\ref{lit2pa}) can be rewritten as
\begin{equation}
\begin{split}
    L(z) = \frac{\Gamma}{\pi} \braket{\Psi_L(z^*)|\Psi_R(z)},
\label{lit norm}
\end{split}
\end{equation}
where $\Psi_L(z^*)$ and $\Psi_R(z)$ are the solutions of the following Schr\"odinger-like equations
\begin{equation}
    \bra{\Psi_L(z^*)}(\overline{H}-z^*) = \bra{\Phi_0}L^{A+2}_0\overline{\Theta^{\dagger}},
    \label{right2palitcc}
\end{equation}
\begin{equation}
    (\overline{H}-z)\ket{\Psi_R(z)} = \overline{\Theta}R^{A+2}_0\ket{\Phi_0}.
    \label{left2palitcc}
\end{equation}
We refer to Eqs.~(\ref{left2palitcc}) and (\ref{right2palitcc}) as the 2PA-LIT-CC equations. They represent the coupled-cluster equivalent of Eq.~(\ref{boundstatelike}) introduced in Section \ref{lit-section}. We can recover the LIT-CC equations for the closed-shell case setting $L_0 = 1+\Lambda$ and $R_0 = \mathbb{1}$.

The 2PA-LIT-CC equations can be solved by following a strategy analogous to the original 2PA-EOM-CC method targeting states of open-shell 2PA nuclei. We start by expressing $\ket{\Psi_R(z)}$ in this way
\begin{equation}
\begin{split}
    &\ket{\Psi_R(z)} = \mathcal{R}(z)\ket{\Phi_0}\\
    &= \left(\frac{1}{2}\sum_{ab} r^{ab}(z) a^{\dagger}_a a^{\dagger}_b + 
    \frac{1}{6} \sum_{abci} r^{abc}_i(z) a^{\dagger}_a a^{\dagger}_b a^{\dagger}_c a_i + ... \right) \ket{\Phi_0},
    \label{ansatzright}
\end{split}    
\end{equation}
and analogously for $\bra{\Psi_L(z^{*})}$ we write
\begin{equation}
\begin{split}
    &\bra{\Psi_L(z^{*})} = \bra{\Phi_0}\mathcal{L}(z^{*})\\
    &= \bra{\Phi_0} \left(\frac{1}{2} \sum_{ab} l_{ab}(z^{*}) a_a a_b + \frac{1}{6} \sum_{abci} l_{abc}^{i}(z^{*}) a^{\dagger}_i a_c a_b a_a + \dots \right).
    \label{ansatzleft}
\end{split}    
\end{equation}
Substituting $\ket{\Psi_R(z)}$ in Eq.~(\ref{right2palitcc}) and $\bra{\Psi_L(z^{*})}$ in Eq.~(\ref{left2palitcc}) we can rewrite the 2PA-LIT-CC equations in the form
\begin{equation}
    (\overline{H}-z)\mathcal{R}(z)\ket{\Phi_0} = \overline{\Theta}R^{A+2}_0\ket{\Phi_0}
    \label{simple2palitccright}
\end{equation}
and
\begin{equation}
    \bra{\Phi_0}\mathcal{L}(z^{*})(\overline{H}-z^*) = \bra{\Phi_0}L^{A+2}_0\overline{\Theta^{\dagger}},
    \label{simple2palitccleft}
\end{equation}
which resemble the EOM- equations for 2PA nuclei with the addition of source terms on the right-hand side.

Two comments are in order. First, the two main differences between the 2PA-LIT-CC approach and the standard closed-shell technique lie in the form of the $\mathcal{R}(z)$ and $\mathcal{L}(z^{*})$ operators, which in the closed-shell case correspond to particle-number conserving EOM excitation operators, and in the source term, involving a product between $\overline{\Theta}$ and a 2PA operator for 2PA nuclei. Second, changing the shape of the $\mathcal{R}(z)$ operator accordingly, the same ansatz can be applied to solve the LIT-CC equations not only for nuclei with $A^{*} = A+2$ but in general for open-shell systems with $A^{*} = A \pm k$, with $k = 1,2$.

\subsection{The Lanczos method}
\label{lanczos}
To obtain the LIT, we should solve Eqs.~(\ref{simple2palitccright}) and (\ref{simple2palitccleft}) for every value of $z$ and $z^*$. This is unnecessary if we reformulate the problem in matrix form and apply the Lanczos algorithm, following a procedure which is analogous to the closed-shell LIT-CC case. We will briefly sketch it in the following. Starting from Eq.~(\ref{lit2pa}), we can express $L(z)$ in the form
\begin{equation}
\begin{split}
&L(z) = -\frac{1}{2\pi} \times \\ & \times \mathfrak{Im} \left\{\braket{\Phi_0|L^{A+2}_0 \overline{\Theta^{\dagger}} \left(\frac{1}{\overline{H}-z^*} - \frac{1}{\overline{H}-z} \right) \overline{\Theta} R^{A+2}_0|\Phi_0}\right\}\\
&= -\frac{1}{2\pi} \mathfrak{Im}\{\mathbf{S}^L[(\mathbf{M}-z^*)^{-1} - (\mathbf{M}-z)^{-1}] \mathbf{S}^R\},
\end{split}   
\end{equation}
where the matrix elements $M_{\alpha,\alpha'}$ of $\mathbf{M}$ and the components $S_{\alpha}^R$ and $S_{\alpha}^L$ of the row and column vectors $\mathbf{S}^L$ and $\mathbf{S}^R$ are given by
\begin{equation}
\begin{split}
    & M_{\alpha,\alpha'} = \braket{\Phi_{\alpha'}|\overline{H}|\Phi_{\alpha}}, \\
   & S_{\alpha}^R = \braket{\Phi_{\alpha}|\overline{\Theta}R_0|\Phi_0}, \\
    & S_{\alpha}^L = \braket{\Phi_0|L_0\overline{\Theta^{\dagger}}|\Phi_{\alpha}} .
\end{split}
\end{equation}
The indices $\alpha,\alpha'$ run over the set of 2PA particle-hole states as
\begin{equation}
    \ket{\Phi^{ab}} = a^{\dagger}_a a^{\dagger}_b\ket{\Phi_0}, \ket{\Phi^{abc}_i} = a^{\dagger}_a a^{\dagger}_b a^{\dagger}_c a_i \ket{\Phi_0}, \dots \,.
\label{basis}
\end{equation}
We then use the complex-symmetric variant of the Lanczos algorithm \cite{cullum1995} to evaluate $L(z)$. Introducing the left and right pivot vectors as
\begin{equation}
    \mathbf{v}_0 = \mathbf{S}^R/\sqrt{\mathbf{S}^L \mathbf{S}^R}, \quad
    \mathbf{w}_0 = \mathbf{S}^L/\sqrt{\mathbf{S}^L \mathbf{S}^R},
    \label{rlpivot}
\end{equation}
the LIT becomes 
\begin{equation}
    L(z) = -\frac{1}{2\pi} \mathfrak{Im} \{[\mathbf{S}^L\mathbf{S}^R] \mathbf{w}_0 [(\mathbf{M}-z^*)^{-1} - (\mathbf{M}-z)^{-1}] \mathbf{v}_0\}.
\end{equation}
Calculating the LIT is then equivalent to evaluating the matrix element 
\begin{equation}
    x_{00} = \mathbf{w}_0 (\mathbf{M}-z)^{-1} \mathbf{v}_0.
\end{equation}
This can be computed applying Cramer's rule to the solution of the linear system
\begin{equation}
    \sum_{\beta} (\mathbf{M} - z)_{\alpha \beta} x_{\beta 0} = \delta_{\alpha 0},
\end{equation}
arising from the identity 
\begin{equation}
    (\mathbf{M} - z)(\mathbf{M} - z)^{-1} = I
\end{equation}
on the Lanczos basis \{$\mathbf{v}_i$, $\mathbf{w}_i$, $i = 0,...n$\}. In the Lanczos basis, $\mathbf{M}$ assumes a tridiagonal form with coefficients $a_i$ and $b_i$. In this way, $x_{00}$ can be written as a continued fraction depending on the Lanczos coefficients. We get
\begin{equation}
    x_{00}(z) = \frac{1}{a_0-z+\frac{b_1^2}{a_1-z+\frac{b_2^2}{a_2-z+b_3^2\dots}}}.
\end{equation}
As a consequence, we obtain
\begin{equation}
L(z) = -\frac{1}{2\pi} \mathfrak{Im}\{[\mathbf{S}^L\mathbf{S}^R]  [x_{00}(z^*)-x_{00}(z)]\},
\label{lit-lanczos}
\end{equation}
where
\begin{equation}
    \mathbf{S}^L\mathbf{S}^R = \braket{\Phi_0|L_0\overline{\Theta^{\dagger}}\overline{\Theta}R_0|\Phi_0}.
    \label{m0}
\end{equation}
The use of the Lanczos algorithm makes the calculation of the LIT very efficient. In fact, the LIT for different values of $\sigma$ and $\Gamma$ can be computed by performing the tridiagonalization of the $\mathbf{M}$ matrix only once.

The numerical implementation of the 2PA-LIT-CC approach uses the Nuclear Tensor Contraction Library (NTCL) \cite{ntcl}, which provides a hardware-agnostic tensor contraction interface compatible with multiple hardware architectures. The analytic expressions for the Lanczos pivot diagrams are reported in Appendix~\ref{appendix}.

\section{Validating the new method}
\label{validation}
We test the 2PA-LIT-CC approach in nuclei which can be treated as closed-shell and as two-particle-attached with respect to a neighboring closed-(sub)shell system. In these special cases we can perform calculations with both the 2PA-LIT-CC method and the well-established closed-shell LIT-CC approach, allowing us to gauge the accuracy of the 2PA approximation. We focus in particular on two oxygen isotopes: $^{16}$O and $^{24}$O, which can be computed in the 2PA-EOM-CC approach starting from the closed (sub-)shell nuclei $^{14}$O and $^{22}$O, respectively. 

To validate the 2PA-LIT-CC method, we focus on electric dipole transitions. The electric dipole operator is defined as   
\begin{equation}
    \Theta = \sum_{i=1}^A (\mathbf{r}_i - \mathbf{R}_{\rm CM})\left(\frac{1+\tau^z_i}{2}\right),
\end{equation}
where $\mathbf{r}_i$ and $\mathbf{R}_{\rm CM}$ correspond to the coordinates of the $i$-th particle and the center-of-mass, respectively, and $\tau^z_i$ is the third component of the isospin of the $i$-th particle. In this work, we will consider 2PA systems characterized by $J^{\pi} = 0^+$. In this case, the electric dipole operator only allows for transitions from the $0^+$ ground state to the $1^-$ excited states. To calculate the similarity-transformed operator $\overline{\Theta}$, we consider up to 2p-2h excitations. In fact, it has been shown that 3p-3h contributions are negligible \cite{miorelli2018}.

We compute dipole sum rules defined as moments of the dipole response function distribution as
\begin{equation}
    m_n = \int d\omega\; \omega^n R(\omega).
\end{equation}
These quantities can be computed directly from the LIT. As the Lorentzian kernel of the transform reduces to a Dirac $\delta$-function in the limit of vanishing width, the moments can be determined simply by integrating the LIT in the limit of small width as
\begin{equation}
   m_n = \lim_{\Gamma \rightarrow 0} \int d\sigma\; \sigma^n L(\sigma, \Gamma).
   \label{moments}
\end{equation}
In such a limit, the LIT is a discretized response function. While the latter does not correctly account for the continuum, the use of Eq.~(\ref{moments}) has been proved to be equivalent to the integration of the response \cite{miorelli2016}. This allows us to avoid the LIT inversion, which can contribute substantially to the numerical uncertainty budget.

We point out that the integral in Eq.~(\ref{moments}) runs over the excitation energy of the nucleus with respect to its ground state. While in the closed-shell case, the ground-state energy of the nucleus is, by construction, the EOM-CC reference energy, in the 2PA-EOM-CC framework the latter corresponds to the CC ground-state energy $E_0^{*}$ of the closed-shell starting nucleus. This means that the LIT in Eq.~(\ref{moments}) needs to be calculated starting from $\omega_0 = E_0 - E_0^{*}$ to correctly obtain the excitation energy with respect to the ground state of the 2PA nucleus as integration variable.

For  $^{16,24}$O, we will examine the non-energy-weighted dipole sum rule $m_0$ and the electric dipole polarizability $\alpha_D$, which is related to the inverse-energy-weighted sum rule $m_{-1}$ via
\begin{equation}
    \alpha_D = 2\alpha m_{-1} = 2\alpha \int d\omega\;  \frac{R(\omega)}{\omega} = 2\alpha \lim_{\Gamma\rightarrow0} \int d\sigma\;  \frac{L(\sigma, \Gamma)}{\sigma},
    \label{alphaD}
\end{equation}
with $\alpha$ being the fine structure constant.

\subsection{Uncertainty quantification}
\label{uq}
When using identical nuclear Hamiltonians, we expect the LIT-CC and 2PA-LIT-CC results for $^{16,24}$O to be compatible within error bars, which are determined by the many-body truncation and by the residual dependence of the results on the model-space parameters. We leave an analysis of the $\chi$EFT truncation uncertainty to future work. In both approaches, convergence is controlled by the maximum number of HO shells $N_{max}$ included in the calculation and by the underlying HO frequency $\hbar\Omega$. The residual dependence on $\hbar\Omega$ at the maximum model space size available for both LIT-CC approaches ($N_{max} = 12$) is taken as an estimate of this source of uncertainty.

As for the effect of many-body truncation, we first need to consider the different approximation schemes characterizing the closed-shell and 2PA-LIT-CC methods. In the LIT-CC approach for closed-shell nuclei, approximations enter in the $T$ and $\Lambda$ operators involved in the ground-state calculation and in the $R_{\mu}$ and $L_{\mu}$ operators of Eqs. (\ref{rmu}) and (\ref{lmu}) entering the EOM calculation, respectively. A specific truncation has to be chosen for both the ground state and the excited states, either CCSD or CCSDT-1. In this work, we vary the approximation scheme in the ground-state calculation between CCSD and CCSDT-1 while we choose to truncate the EOM calculation at the CCSD level. The inclusion of 3p-3h excitations in the EOM calculation, which is computationally quite demanding, is expected to give a small contribution~\cite{miorelli2018}. From this point on we will specify only the truncation scheme adopted in the ground-state calculation for convenience. To estimate the many-body truncation error we take half of the difference between the CCSDT-1 and CCSD results as done in Refs.~\cite{simonis2019,acharya2023}. 

Let us now consider the 2PA-LIT-CC approach. In this case, we first have to solve for the CC ground state of the closed-shell reference, for which we use the CCSD approximation. A study of the effect of iterative 3p-3h excitations in the CC reference is left to future work. In order to study properties of 2PA nuclei, we have to resort to truncations of the  $R^{A+2}_{\mu}$ and $L^{A+2}_{\mu}$ operators of Eqs.~(\ref{rmu-2pa}) and (\ref{lmu-2pa}) in both the ground-state and excited-state calculations. Keeping only the first term in the sum of Eqs.~(\ref{rmu-2pa}) and (\ref{lmu-2pa}) leads to the 2p-0h approximation while also adding the second term defines the 3p-1h approximation. The latter is expected to be accurate for states with a dominant 2PA structure \cite{jansen2013}. The implementation of the 4p-2h approximation framework \cite{shen2013,shen2014,ajala2017,shen2021} is beyond the scope of this work. 
Table \ref{tab:I} summarizes the approximation schemes used in this work. 
\begin{table}[h]
\caption{\label{tab:I} Notation used to identify the 2PA-LIT-CC approximations for the ground state (left of '/') and the excited states (right of '/'). }
\begin{ruledtabular}
\begin{tabular}{lll}
 Ground state   & Excited states & Truncation scheme \\
 \hline
   2p-0h  & 3p-1h & 2p-0h/3p-1h \\
   3p-1h  & 3p-1h & 3p-1h/3p-1h \\
\end{tabular}
\end{ruledtabular}
\end{table}

Since, in this work, we are interested in  observables such as the dipole polarizability, we retain at least 1p-1h excitations in the excited-state calculation. In analogy to the closed-shell case, we estimate the many-body truncation uncertainty as half of the difference between the two approximation schemes of Tab.~\ref{tab:I}.

\subsection{Non-energy-weighted dipole sum rule}
First, let us concentrate on the results for the non-energy-weighted sum rule $m_0$. This observable can be calculated in two ways: as an integral of $R(\omega)$ according to Eq.~(\ref{moments}) and as a ground-state expectation value according to Eq.~(\ref{m0}). We verified that both approaches lead to the same results.

\begin{figure*}
    \includegraphics[width=0.85\textwidth]{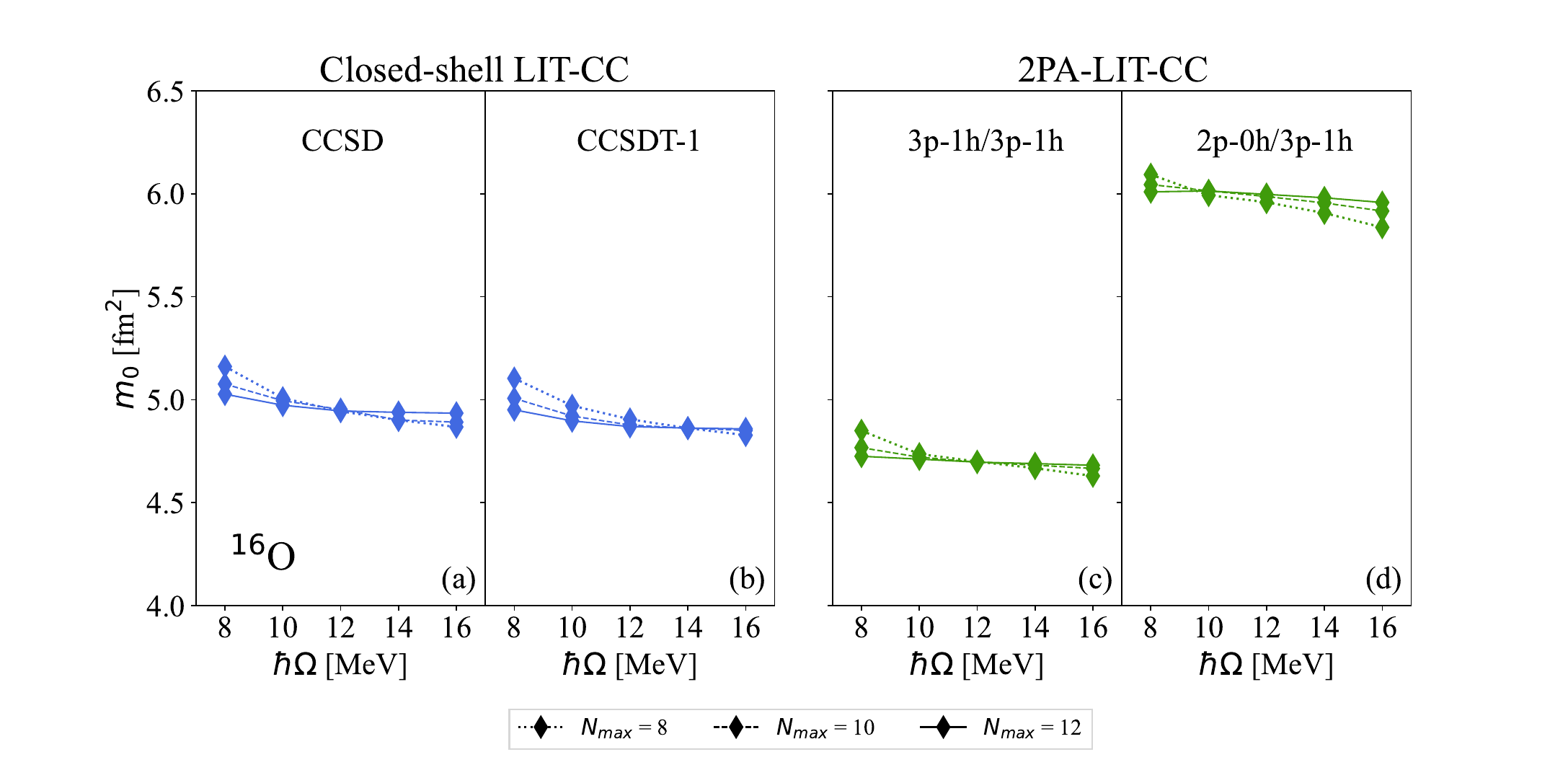}    
    \caption{$^{16}$O results for $m_0$ from the closed-shell CCSD and CCSDT-1 approximations [panels (a) and (b), respectively] and the 2PA 3p-1h/3p-1h and 2p-0h/3p-1h approximations [panels (c) and (d), respectively]. Each panel shows $m_0$ as a function of $\hbar\Omega$ for different model-space sizes. }
    \label{fig:m0_benchmark_16O}
\end{figure*}

\begin{figure*}
    \includegraphics[width=0.85\textwidth]{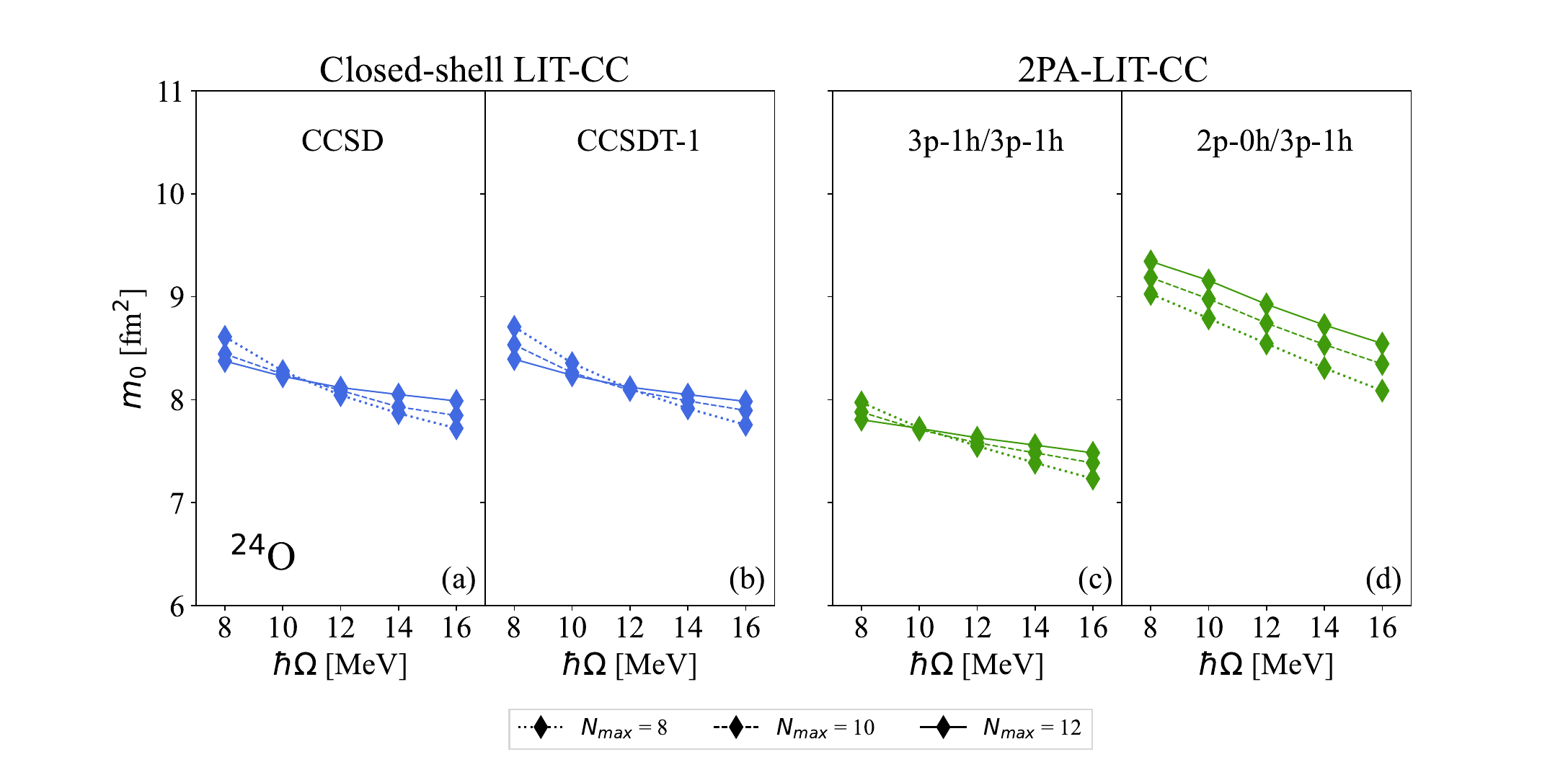}    
    \caption{$^{24}$O results for $m_0$ from the closed-shell CCSD and CCSDT-1 approximations [panels (a) and (b), respectively] and the 2PA 3p-1h/3p-1h and 2p-0h/3p-1h approximations [panels (c) and (d), respectively]. Each panel shows $m_0$ as a function of $\hbar\Omega$ for different model-space sizes. }
    \label{fig:m0_benchmark_24O}
\end{figure*}

In Figs. \ref{fig:m0_benchmark_16O} and \ref{fig:m0_benchmark_24O}, we consider the convergence pattern of $m_0$ as a function of $\hbar\Omega$ for different values of $N_{max}$ in $^{16}$O and $^{24}$O, respectively. For both nuclei, we show CCSD and CCSDT-1 closed-shell results, and 2PA results in the 3p-1h/3p-1h and 2p-0h/3p-1h approximations. In all cases, we vary $N_{max}$ between $8$ and $12$ and span a range of $\hbar\Omega$ values between $8$ and $16$ MeV. We report $m_0$ predictions in the closed-shell and 2PA frameworks for both nuclei in Tab. \ref{tab:II}.


Let us start with $^{16}$O. In all the closed-shell and 2PA approximation schemes considered, $m_0$ is converged at $N_{max} = 12$. Focusing on the closed-shell results, we can identify the optimal frequency leading to a faster convergence with the crossing point of the different $N_{max}$ curves at $\hbar\Omega = 12$ MeV for CCSD and $\hbar\Omega = 14$ MeV for CCSDT-1. Adding triples lowers the value of $m_0$ by $1.5\%$ with respect to the CCSD result. In the 2PA calculations, the optimal frequency is at $\hbar\Omega = 12$ MeV for 3p-1h/3p-1h and $\hbar\Omega = 10$ MeV for 2p-0h/3p-1h. The effect of the many-body truncation is much larger in the 2PA case: the 2p-0h/3p-1h result is around $20\%$ higher with respect to the 3p-1h/3p-1h result. Considering that the difference between the best closed-shell and 2PA approximation schemes (namely, CCSDT-1 and 3p-1h/3p-1h) is around $5\%$, the closed-shell and 2PA results are consistent, as shown in Table~\ref{tab:II}.

\begin{table}[hbt]
\caption{\label{tab:II} Results for the non-energy-weighted sum rule $m_0$ in fm$^2$ for the $^{16,24}$O nuclei in the closed-shell LIT-CC method at the CCSDT-1 level and the 2PA-LIT-CC method in the 3p-1h/3p-1h approximation. The theoretical uncertainty has been obtained as detailed in Section \ref{uq}. }
\begin{ruledtabular}
\begin{tabular}{lll}
 Nucleus   & CCSDT-1 &  3p-1h/3p-1h\\
 \hline 
   $^{16}$O  &  4.86(4)    &  4.7(7) \\
   $^{24}$O  &  8.12(10)   &  7.7(8) \\
\end{tabular}
\end{ruledtabular}
\end{table}

Let us now analyze the $^{24}$O case. The convergence patterns of Fig.~\ref{fig:m0_benchmark_24O} for this nucleus show a more significant residual dependence on the model-space parameters than $^{16}$O. This is a consequence of the more extended size characterizing the neutron-rich $^{24}$O nucleus compared to the doubly-magic $^{16}$O nucleus. Nevertheless, an optimal frequency can be found in most of the approximation schemes considered. When starting from the closed-shell calculations, the optimal frequency varies from $\hbar\Omega = 10$ MeV for CCSD to $\hbar\Omega = 12$ MeV for CCSDT-1. Also, for $^{24}$O, the CCSD value of $m_0$ at the optimal frequency is slightly higher than the CCSDT-1 one, with the difference between the two being less than $2\%$. Turning to the 2PA results, while it is possible to identify an optimal frequency of $\hbar\Omega = 10$ MeV for the 3p-1h/3p-1h approximation, the 2p-0h/3p-1h values are characterized by a slower convergence. In this case, the reference frequency used in the uncertainty estimate is selected by considering the difference between the numerical values of $m_0$ at $N_{max} = 12$ and $N_{max} = 10$. The latter is assumed to be smaller for the $\hbar\Omega$ value closer to the optimal frequency. Since such a difference progressively reduces by going from $\hbar\Omega = 16$ MeV to $\hbar\Omega = 8$ MeV, we choose to estimate the many-body uncertainty by employing $\hbar\Omega = 8$ MeV as reference $\hbar\Omega$ value. We can then say that the value of $m_0$ in the 2p-0h/3p-1h scheme is around $20\%$ larger than the 3p-1h/3p-1h one. Since the CCSDT-1 and 3p-1h/3p-1h values show a difference of $5\%$ as in the $^{16}$O case, the results obtained with the closed-shell and 2PA frameworks are compatible within error bars.

\subsection{Electric dipole polarizability}
\label{alphaDbenchmark}

\begin{figure*}
    \includegraphics[width=0.85\textwidth]{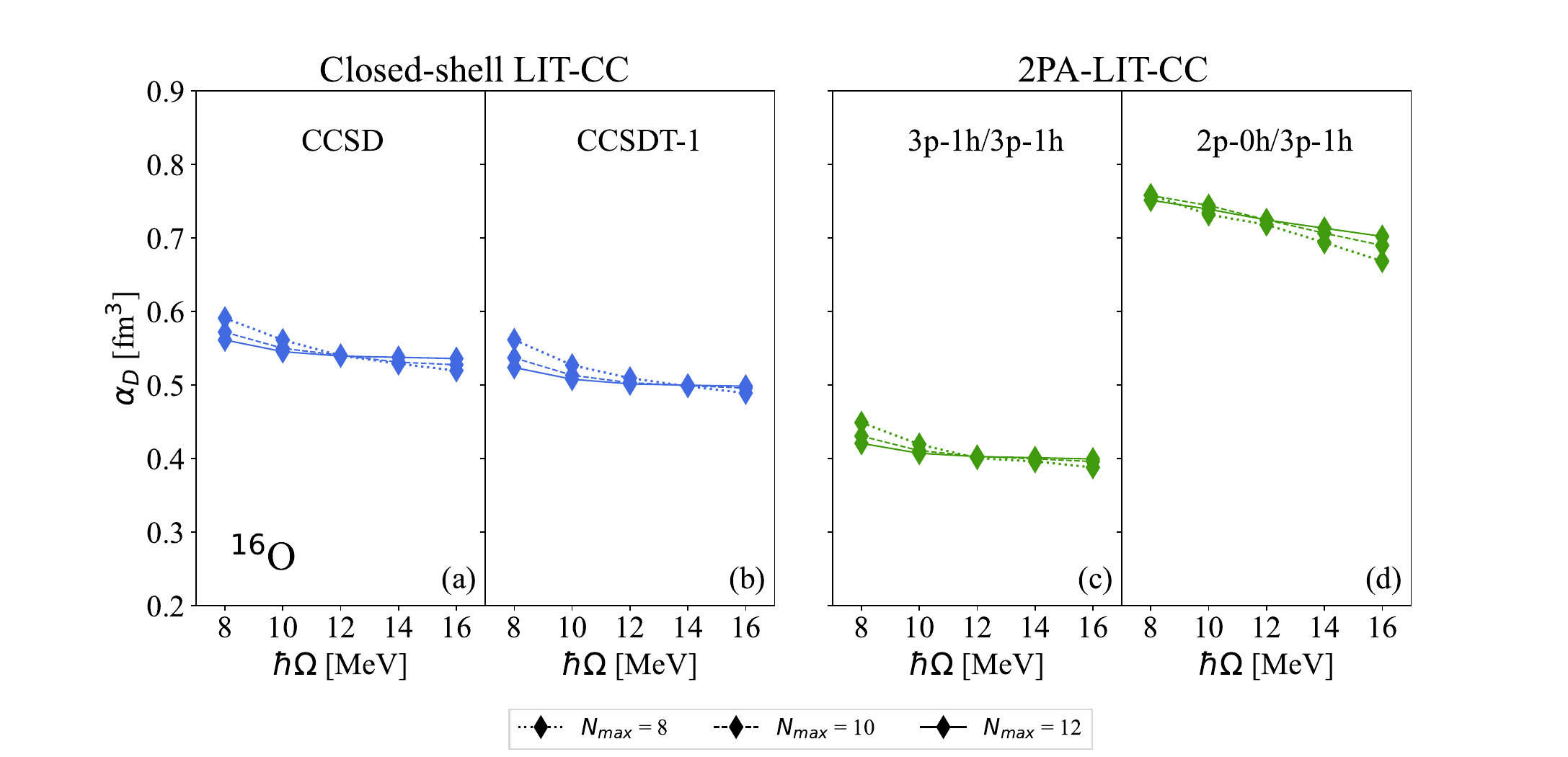}
    \caption{$^{16}$O results for $\alpha_D$ from the closed-shell CCSD and CCSDT-1 approximations [panels (a) and (b), respectively] and the 2PA 3p-1h/3p-1h and 2p-0h/3p-1h approximations [panels (c) and (d), respectively]. Each panel shows $\alpha_D$ as a function of $\hbar\Omega$ for different model-space sizes. }
    \label{fig:alphaD_benchmark_16O}
\end{figure*}
\begin{figure*}
    \includegraphics[width=0.85\textwidth]{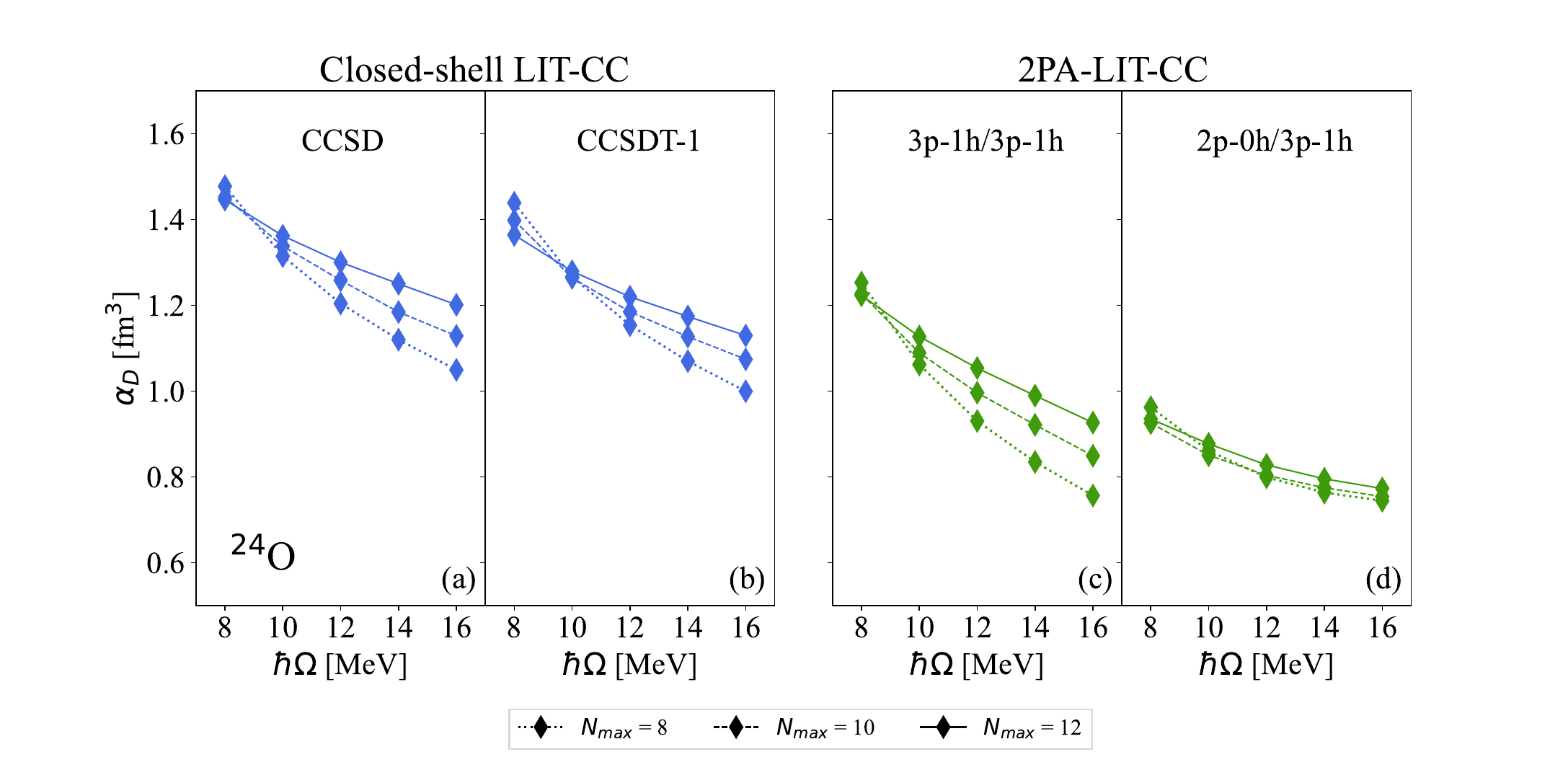}
    \caption{$^{24}$O results for $\alpha_D$ from the closed-shell CCSD and CCSDT-1 approximations [panels (a) and (b), respectively] and the 2PA 3p-1h/3p-1h and 2p-0h/3p-1h approximations [panels (c) and (d), respectively]. Each panel shows $\alpha_D$ as a function of $\hbar\Omega$ for different model-space sizes. }
    \label{fig:alphaD_benchmark_24O}
\end{figure*}

Let us now analyze the results for the electric dipole polarizability of $^{16,24}$O in the LIT-CC and 2PA-LIT-CC frameworks.  This observable sheds light on the effects of using a different coupled-cluster expansion not only in the ground state but also in the dipole-excited states. 
In analogy to the case of $m_0$, Figs.~\ref{fig:alphaD_benchmark_16O} and \ref{fig:alphaD_benchmark_24O} illustrate the dependence on $\hbar\Omega$ of both  at different levels of approximation varying $N_{max}$ between $8$ and $12$. Closed-shell and 2PA predictions for $\alpha_D$ of $^{16,24}$O can be found in Table~\ref{tab:III}.

For all results, we found a good convergence of $\alpha_D$ with respect to the number of Lanczos coefficients included in the LIT continued fractions of Eq.~(\ref{lit-lanczos}). Typically, $20$ Lanczos coefficients are sufficient to reach a stable result. As shown in Fig.~\ref{fig:alphaD_benchmark_16O}, in all approximation schemes $\alpha_D$ for $^{16}$O converges quite quickly at an optimal frequency of $\hbar\Omega = 12$ MeV. Considering the different LIT-CC and 2PA-LIT-CC truncations available, the $\alpha_D$ results share some commonalities with the behavior observed in the case of $m_0$. First, the CCSDT-1 value for $\alpha_D$ is lower than the CCSD one by around $7\%$. The reduction of $\alpha_D$ when 3p-3h excitations are included has already been observed in the case of $^{16}$O employing the \NNLOsat interaction \cite{miorelli2018}, and it is a feature observed in various medium-mass nuclei \cite{simonis2019,kaufmann2020,fearick2023}. Second, as for the 2PA results, the 3p-1h/3p-1h approximation reduces the value of $\alpha_D$ with respect to the 2p-0h/3p-1h scheme by about $40\%$. The resulting large error bar then covers the difference between the CCSDT-1 and 3p-1h/3p-1h optimal frequency values, which amounts to $20\%$, as shown in Table \ref{tab:III}. Moreover, we point out that both the closed-shell and 2PA predictions are in good agreement with the \NNLOsat result of Ref.~\cite{miorelli2018}, corresponding to $\alpha_D = 0.508$ fm$^3$ in the CCSDT-1 scheme.

Let us now focus on the case of $^{24}$O. As shown in Fig.~\ref{fig:alphaD_benchmark_24O}, for all the approximation schemes considered, the optimal frequency moves towards lower values of $\hbar\Omega$, ranging between $8$ and $10$ MeV. In analogy to the $m_0$ results, the larger spatial extension of $^{24}$O leads to a more significant residual dependence on $\hbar\Omega$ compared to the $^{16}$O case. The spread of the different $N_{max}$ curves as a function of $\hbar\Omega$ appears similar among the different approximation schemes, except for the 2p-0h/3p-1h case, showing a more compressed pattern. 
In the closed-shell calculations, 3p-3h excitations reduce $\alpha_D$ by about $7\%$, the same amount observed for $^{16}$O. In the 2PA case, improving the precision of the calculation from 2p-0h/3p-1h to 3p-1h/3p-1h increases $\alpha_D$, bringing it closer to the CCSDT-1 value. The difference between the CCSDT-1 and 3p-1h/3p-1h results amounts to $15\%$. Combining the many-body truncation and convergence contributions, the LIT-CC and 2PA-LIT-CC results are in good agreement, as shown in Table~\ref{tab:III}.

\begin{table}[hbt]
\caption{\label{tab:III} Predictions for the dipole polarizability $\alpha_D$ in fm$^3$ for the $^{16,24}$O nuclei in the closed-shell LIT-CC method at the CCSDT-1 level and in the 2PA-LIT-CC method in the 3p-1h/3p-1h approximation. 
The theoretical uncertainty has been obtained as detailed in Section \ref{uq}.}
\begin{ruledtabular}
\begin{tabular}{lll}
 Nucleus   & CCSDT-1 &  3p-1h/3p-1h\\
 \hline 
   $^{16}$O  &  0.54(4)   &  0.40(16)\\
   $^{24}$O  &  1.32(6)   &  1.22(16)\\
\end{tabular}
\end{ruledtabular}
\end{table}

\begin{figure*}
    \includegraphics[width=0.9\textwidth]{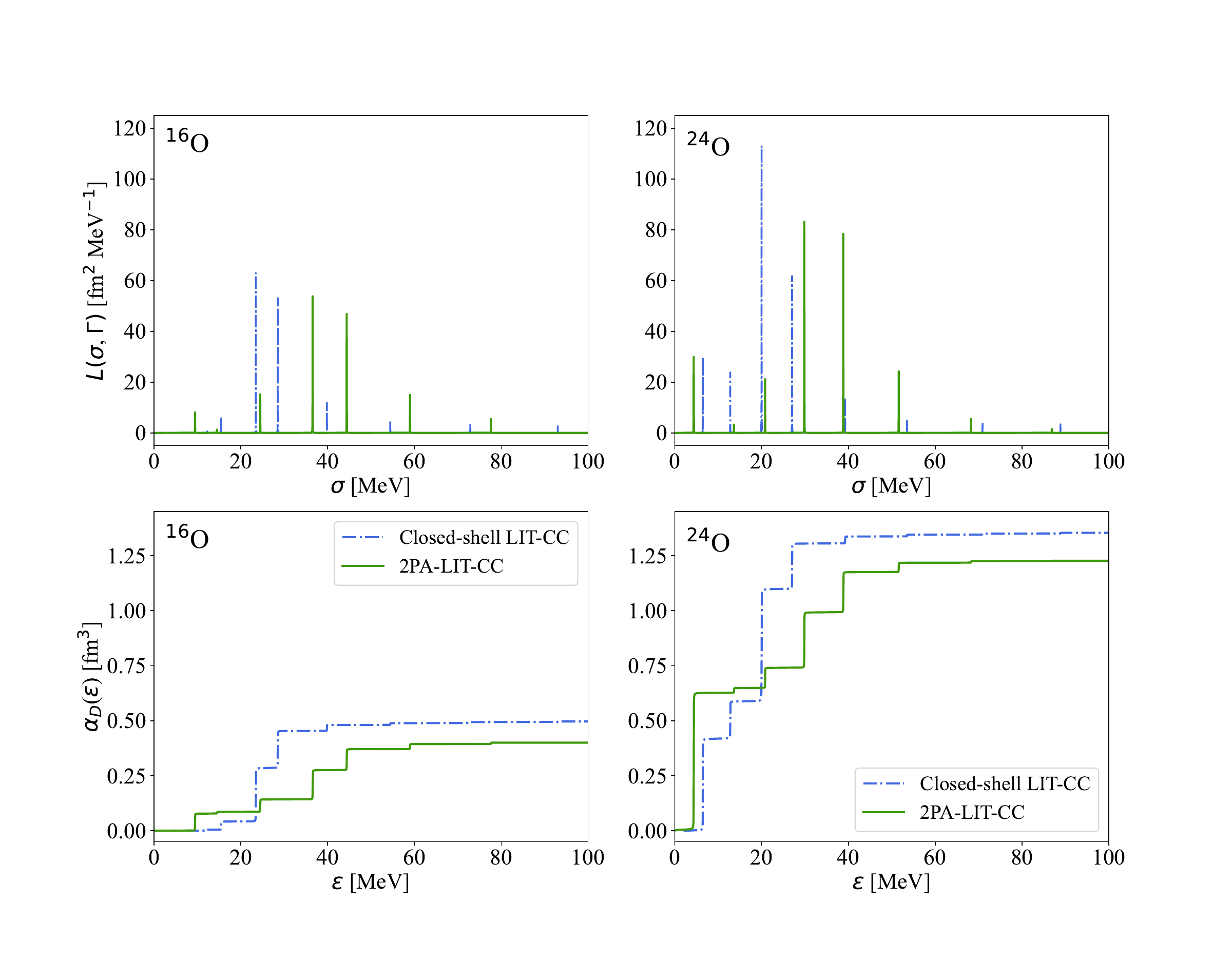}
    \caption{(Upper panels): Comparison of the LIT with $\Gamma = 0.01$ MeV  as a function of excitation energy for  the $^{16}$O (left) and  $^{24}$O (right) nuclei computed with the closed-shell LIT-CC and 2PA-LIT-CC methods.
    (Lower panels): Corresponding $\alpha_D$ running sums as a function of the upper integration limit. See text for details.
    }
    \label{fig:response_1624O}
\end{figure*}

It is interesting to analyze how the difference between the LIT-CC and 2PA-LIT-CC results for the polarizability depends on the upper integration limit of the LIT in Eq. (\ref{alphaD}). To this aim, it is useful to define the $\alpha_D$ running sum, given by 
\begin{equation}
    \alpha_D(\epsilon) = 2\alpha \lim_{\Gamma\rightarrow0} \int_0^{\epsilon} d\sigma\;  \frac{L(\sigma, \Gamma)}{\sigma}.
\end{equation}
This quantity allows one to identify the excitation energy regions where $\alpha_D$ receives the largest contributions. In Fig.~\ref{fig:response_1624O}, the upper left (right) panel shows the LIT for $\Gamma = 0.01$ MeV for $^{16}$O ($^{24}$O) in the two frameworks. These curves have been obtained adopting the CCSDT-1 approximation in the LIT-CC case and the 3p-1h/3p-1h truncation in the 2PA-LIT-CC case. In the lower left (right) panel, the corresponding running sum $\alpha_D(\epsilon)$ is illustrated. Comparing the upper and lower panel for each nucleus, it is possible to see how the discretized strength distribution contributes to the polarizability as a function of excitation energy. The curves shown in Fig.~\ref{fig:response_1624O} correspond to the results at the maximum model space size, $N_{max} = 12$ and at the optimal frequency for the $\alpha_D$ convergence ($\hbar\Omega = 12$ MeV for $^{16}$O and $\hbar\Omega = 8$ MeV for $^{24}$O). 

We observe that below $20$ MeV of excitation energy, the low-lying states characterizing the LIT in the closed-shell and 2PA approaches give similar contributions to the polarizability of $^{16,24}$O. However, at higher excitation energies, corresponding to the region of the GDR, the 2PA result systematically underestimates the closed-shell one in both nuclei, leading to a lower value for $\alpha_D$. The origin of this behavior becomes apparent when looking at the LIT: in both $^{16,24}$O the states corresponding to the GDR are shifted towards higher energies. This effect is more pronounced for $^{16}$O, where the states characterized by the highest strength in the LIT appear close to $40$ MeV, while in $^{24}$O, they are located at around $30$ MeV.

To understand the reason for this behavior, we introduce a tool to gauge the quality of the 3p-1h/3p-1h approximation in describing specific nuclear states. In Refs. \cite{jansen2011,jansen2013}, the wave function's partial 2p-0h and 3p-1h norms were identified as possible markers of missing higher-order correlations in the 2PA expansion. It is worth revisiting this here. Given the 2p-0h and 3p-1h amplitudes of a 2PA state with angular momentum $J$, the partial norms are defined as
\begin{equation}
    \begin{split}
        &n(\mathrm{2p0h}) = \frac{1}{2} \sum_{ab} (2J+1) (r^{ab})^2 ,\\
        &n(\mathrm{3p1h}) = \frac{1}{6} \sum_{abci} \sum_{J_{ab}, J_{abc}} (2 J_{abc} +1) (r^{abc}_i)^2,
    \end{split}
\end{equation}
where $n(\mathrm{2p0h}) + n(\mathrm{3p1h}) = 1$. $n(\mathrm{2p0h})$ and $n(\mathrm{3p1h})$ allow one to quantify the wave function's share in 2p-0h and 3p-1h configurations. As a rule of thumb, a 2p-0h partial norm of around $0.9$ indicates an accurate description of the nuclear state of interest within a 3p-1h truncation of the 2PA expansion. A lower 2p-0h norm could suggest including 4p-2h or higher correlations is necessary.

The calculation of the dipole polarizability is affected by the quality of the ground state and the excited states that we can access via the Lanczos algorithm, as explained in Section \ref{lanczos}. For both $^{16,24}$O we computed the partial norm $n(\mathrm{2p0h})$ for these states. In Table \ref{tab:IV}, we report the values of $n(\mathrm{2p0h})$ for the ground state and the first dipole-excited state appearing in the LIT. Since $\alpha_D$ is an inverse energy-weighted sum rule, the latter significantly impacts this observable. For instance, in the 2PA-LIT-CC case, such state covers around $1/4$ of the total value of the polarizability for $^{16}$O and almost $1/2$ for $^{24}$O.

\begin{table}[hbt]
\caption{\label{tab:IV} Partial norms $n(\mathrm{2p0h})$ of the ground state and first dipole-excited state in $^{16,24}$O. }
\begin{ruledtabular}
\begin{tabular}{lll}
 Nucleus   & Ground state &  First $1^{-}$ state\\
 \hline 
   $^{16}$O  &  0.84   &  0.73 \\
   $^{24}$O  &  0.90  &   0.88 \\
\end{tabular}
\end{ruledtabular}
\end{table}

The results of Tab. \ref{tab:IV} suggest that while for $^{24}$O both the ground state and the first $1^{-}$ state, characterized by $n(\mathrm{2p0h})\approx 0.9$, appear to have a dominant 2PA structure, for $^{16}$O the 3p-1h/3p-1h approximation is less accurate. Moreover, it is worth pointing out that considering the dipole-excited states at energies above $20$ MeV and, in particular, the GDR region, $n(\mathrm{2p0h})$ quickly decreases for both nuclei, falling well below 50\%. This likely indicates the need to include higher-order correlations for a more precise description of the GDR.

\section{The oxygen isotopic chain}
\label{oxygenisotopes}
Equipped with the LIT-CC and 2PA-LIT-CC approaches, we can explore how the dipole polarizability evolves along the oxygen isotopic chain. In addition to $^{16}$O and $^{24}$O, which can be addressed with both methods, we can consider $^{14}$O and $^{22}$O in the closed-shell framework and $^{18}$O as a 2PA nucleus with respect to $^{16}$O. We adopt the CCSDT-1 approximation for the closed-shell calculations and the 3p-1h/3p-1h approximation for the 2PA ones.

\label{results_oxygen}
\begin{figure*}[hbt]
    \includegraphics[width=0.6\textwidth]{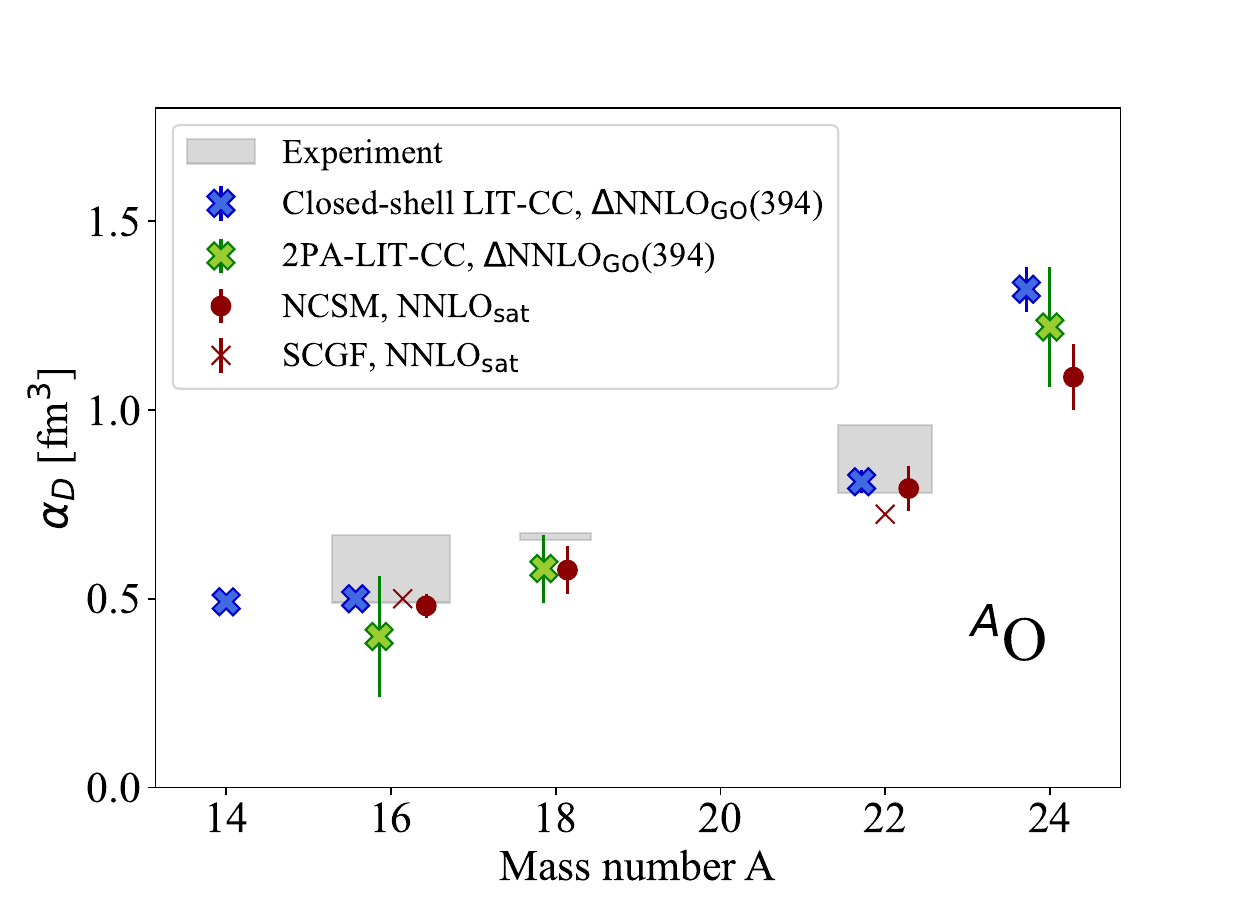} 
    \caption{Electric dipole polarizability in the oxygen isotopic chain obtained either with the closed-shell LIT-CC method in the CCSDT-1 approximation or with the 2PA-LIT-CC method in the 3p-1h/3p-1h approximation, in comparison with other available theoretical calculations~\cite{stumpf2017,stumpf2018,raimondi2019} and experimental data\cite{ahrens1975,ishkanov2002,leistenschneider2001}. See text for details.
    }
    
    \label{fig:alphaD_oxygenchain}
\end{figure*}

Figure \ref{fig:alphaD_oxygenchain} shows our results compared to available experimental data and theoretical predictions. 
It is noteworthy that the polarizability increases significantly  as the neutron to proton ratio increases.  Starting from $^{14}$O and going towards $^{24}$O, the calculated polarizability becomes more than twice as high.  
The uncertainty bars on our results are determined according to the recipe of Section \ref{uq}. For the closed-shell LIT-CC results, the many-body and convergence contribution to the theoretical error are comparable, leading to a total uncertainty that varies from $3\%$ to $5\%$ going from $^{14}$O to the more neutron-rich $^{24}$O. For the open shell 2PA-LIT-CC results, the many-body uncertainty is dominant. While for $^{18}$O and $^{24}$O it amounts to $15\%$ and $13\%$ of the central value, respectively, for $^{16}$O it grows to $40\%$. The larger uncertainty in the case of $^{16}$O could reflect the level of accuracy of the 3p-1h/3p-1h approximation for this nucleus, as explained in Section \ref{alphaDbenchmark}. Figure~\ref{fig:alphaD_oxygenchain} makes  the consistency between the LIT-CC and 2PA-LIT-CC results visually apparent for our benchmark  $^{16}$O and $^{24}$O nuclei. 

We find good agreement between our calculations and the available experimental data. The experimental values for $\alpha_D$ are obtained by energy-integrating the photoabsorption data of Ref.~\cite{ahrens1975} for $^{16}$O, Ref.~\cite{ishkanov2002} for $^{18}$O and Ref.~\cite{leistenschneider2001} for $^{22}$O. While for $^{16}$O data are available up to an excitation energy of 100 MeV, which is sufficient for the integral of Eq.~(\ref{alphaD}) to converge, the maximum excitation energy reached by data is 40 MeV for $^{18}$O and only 18 MeV for $^{22}$O. To obtain a final estimate of $\alpha_D$ for these two nuclei, we calculate $\alpha_D(\omega > 40 \;\mathrm{MeV})$ for $^{18}$O  and $\alpha_D(\omega > 18 \;\mathrm{MeV})$ for $^{22}$O and add these values to the corresponding experimental result. 

In the case of $^{18,22}$O we can also compare our predictions for $\alpha_D$ to experiment focusing on the region where data are available. We report in Tab.~\ref{tab:V} a comparison between experimental values and our theoretical results, integrated between $0$ and $40$ MeV for $^{18}$O and $0$ and $18$ MeV for $^{22}$O.

\begin{table}[hbt]
\caption{\label{tab:V} Predictions for the dipole polarizability $\alpha_D$ in fm$^3$ for $^{18,22}$O compared to experiment in the excitation energy range where data are available.The theoretical uncertainty has been obtained as detailed in Section \ref{uq}.}
\begin{ruledtabular}
\begin{tabular}{llll}
 Nucleus   & Excitation energy range & Theory & Experiment\\
 \hline 
   $^{18}$O  & 0 - 40 MeV &  0.48(10)   &  0.57\\
   $^{22}$O  & 0 - 18 MeV &  0.20(3)   &  0.24(6)\\
\end{tabular}
\end{ruledtabular}
\end{table}

We point out that the photoabsorption data of Ref.~\cite{ishkanov2002} for $^{18}$O are reported without uncertainties. For both $^{18,22}$O, theoretical predictions are compatible with experiments within error bars. 

In Fig.~\ref{fig:alphaD_oxygenchain}, we also report $\alpha_D$ predictions for $^{16, 18, 22,  24}$O in the importance-truncated NCSM approach~\cite{stumpf2017,stumpf2018} and for $^{16, 22}$O in the SCGF approach \cite{raimondi2019}. In both NCSM and SCGF  calculations, the \NNLOsat interaction was employed. NCSM predictions for $^{16, 18, 24}$O agree well with our results. Uncertainties in the importance-truncated NCSM results reflect the residual model-space dependence of $\alpha_D$. In particular, for $^{18}$O,  the excellent accordance of our results with those from NCSM provides an additional benchmark, despite the different interactions used, for the newly developed 2PA-LIT-CC method. The SCGF prediction for $^{16}$O is consistent with our results and experimental data, while their value is slightly lower than the CCSDT-1 one for $^{22}$O.

Being at the dripline, $^{24}$O represents an interesting physics case. With a polarizability of $1.087(87)$ fm$^3$, NCSM calculations give a smaller value compared to the CCSDT-1 result of $1.32(6)$ fm$^3$, while being consistent with the 2PA value of $1.22(16)$ fm$^3$. To our knowledge, no experimental data are available for $^{24}$O. However, the large spread of the predictions obtained with different ab initio methods and nuclear interaction models motivates an experimental investigation of the dipole strength of this nucleus.

\section{The calcium isotopic chain}
\label{calciumisotopes}
\begin{figure*}[hbt]
    \includegraphics[width=0.6\textwidth]{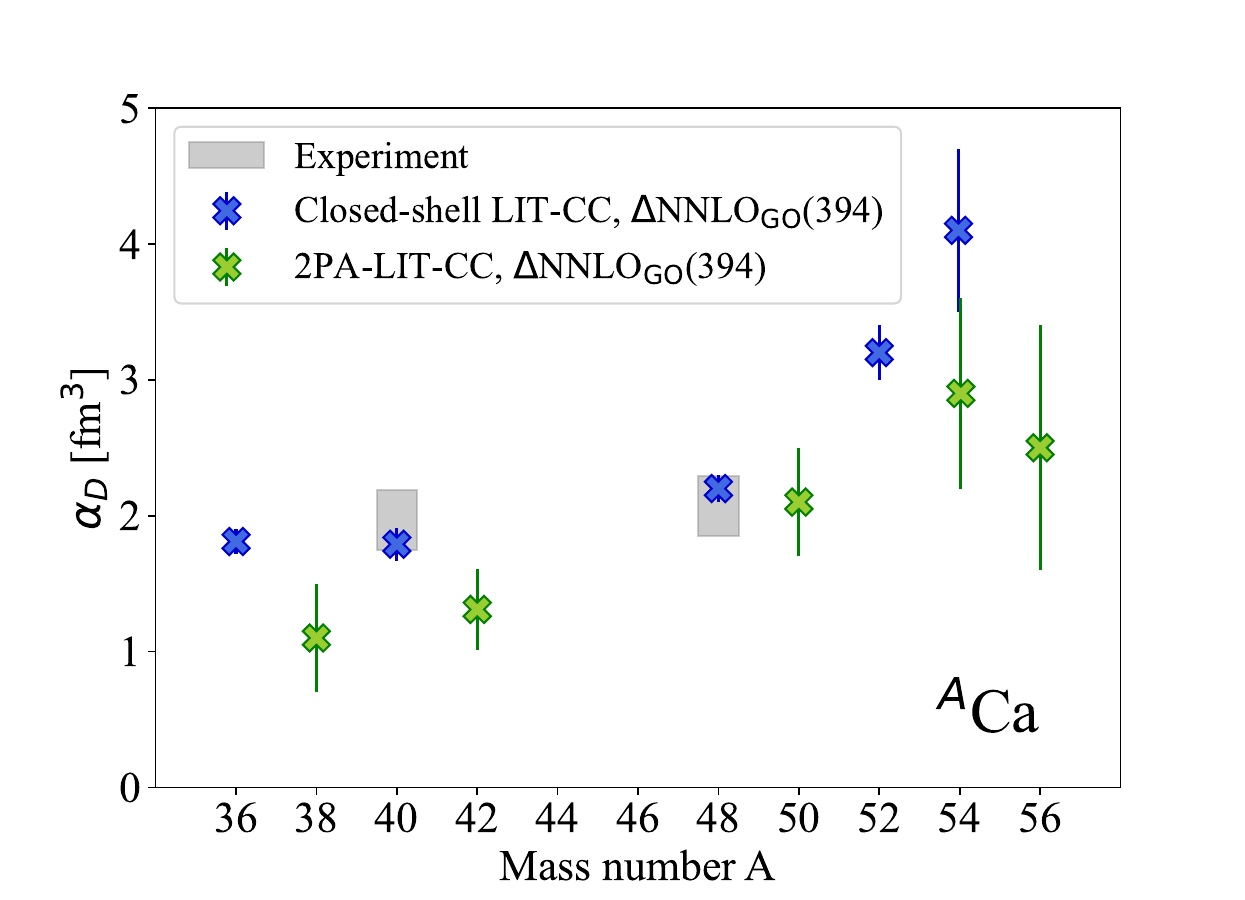}
    \caption{Electric dipole polarizability in the calcium isotopic chain obtained either with the closed-shell LIT-CC method in the CCSDT-1 approximation or with the 2PA-LIT-CC method in the 3p-1h/3p-1h approximation, in comparison with available  experimental data~\cite{fearick2023,birkhan2017}.}    
    \label{fig:alphaD_calciumchain}
\end{figure*}

Moving to heavier systems, we can address the evolution of the dipole polarizability along the calcium isotopic chain. In this case the polarizability of $^{36,40,48,52}$Ca can be computed with the usual closed-shell LIT-CC, while we can provide predictions of $\alpha_D$ for $^{38,42,50,56}$Ca with the 2PA-LIT-CC approach. Both LIT-CC frameworks can be used to calculate $\alpha_D$ for $^{54}$Ca, which serves as a benchmark nucleus in this isotopic chain. As in the previous Section, we adopt the CCSDT-1 approximation for the closed-shell calculations, while we truncate the 2PA expansion at the 3p-1h/3p-1h level. The results are shown in Figure \ref{fig:alphaD_calciumchain} in comparison to available experimental data.
\label{results_calcium}

We observe that predictions for the polarizability of 2PA nuclei appear smaller than the ones obtained for their closed-shell neighbors. The theoretical uncertainty of the closed-shell results, mainly determined by the many-body truncation, varies between $5\%$ and $7\%$ of the corresponding central value. It increases to $15\%$ only for the neutron-rich $^{54}$Ca, where we notice a larger residual dependence on the model-space parameters. Also, in the case of the 2PA calculations, the effect of truncating the many-body expansion dominates the theoretical error. Depending on the nucleus, the uncertainty ranges between $24\%$ and $36\%$ of the central value. This result is also affected by the slower convergence characterizing $\alpha_D$ for $^{50,54,56}$Ca in the 2p-0h/3p-1h scheme, needed in order to evaluate the many-body truncation error. A further refinement of our 2PA results could come by using a CCSDT-1 reference in the 2PA ground state calculation. We leave this analysis to future work.

It is useful to focus on the case of $^{54}$Ca, where we can compare the two CC frameworks. While consistent within estimated uncertainties, the 2PA calculation leads to a smaller central value of $\alpha_D$ compared to the closed-shell result. As in the case of $^{16,24}$O, we observed that the reason for this behavior lies in the shape of the discretized response function, where the GDR appears at higher excitation energies in the 2PA approach than in the closed-shell one. To understand this systematic effect, we have analyzed the partial norm $n(\mathrm{2p0h})$ for the ground state and the dipole-excited states of $^{54}$Ca, finding a situation similar to the case of $^{24}$O. While the ground state and first $1^{-}$ state  have a good 2PA-dominated structure, with $n(\mathrm{2p0h}) \approx 0.9$ in both cases, $n(\mathrm{2p0h})$ becomes rapidly much smaller than $n(\mathrm{3p1h})$ for the dipole-excited states in the GDR region, suggesting the need of higher-order correlations in this energy range. 

Starting from $^{36}$Ca up to $^{54}$Ca, our predictions for closed-shell and 2PA nuclei show an increase of the dipole polarizability with the number of neutrons, as observed for oxygen isotopes. When going from $^{54}$Ca to $^{56}$Ca the central value of $\alpha_D$ for $^{56}$Ca  is slightly lower than that of $^{54}$Ca in the 2PA framework, even though uncertainties increase in this region.

Experimental data in this isotopic chain are available only for $^{40,48}$Ca \cite{birkhan2017,fearick2023}. In these works, the family of interactions from Ref.~\cite{hebeler2011} and \NNLOsat were employed, but we also observe a good agreement between the theoretical prediction obtained using \NNLOgodlow and experiment. Inelastic proton scattering data for $^{42}$Ca from iThemba Labs, currently under analysis \cite{vnc_private}, could soon provide an experimental benchmark for our 2PA prediction. 

\section{Conclusions}
\label{conclusions}
\noindent
Based on the unified  treatment of nuclear structure and electromagnetic reactions, we extended the reach of  coupled-cluster theory
to open-shell nuclei characterized by two valence nucleons outside a closed shell. The newly developed approach, dubbed 2PA-LIT-CC, 
is validated by considering the non-energy-weighted dipole sum rule $m_0$ and electric dipole polarizability $\alpha_D$ of $^{16}$O and $^{24}$O, where both the closed-shell LIT-CC and 2PA-LIT-CC methods can be applied. The two frameworks yield compatible results within uncertainties, giving us confidence in the implementation of the 2PA-LIT-CC method. At the same time, this comparison suggested the need for higher-order excitations in the 2PA expansion to capture more accurately the contribution of the giant dipole resonance region to $\alpha_D$.

Performing calculations with the closed-shell LIT-CC and 2PA-LIT-CC approaches, we considered the evolution of the dipole polarizability along the oxygen and calcium isotopic chains. Focusing on oxygen isotopes, the comparison between coupled-cluster, NCSM, and SCGF predictions with available experimental data shows a good understanding of the physics of $\alpha_D$ for the closed-shell $^{16,22}$O and the open-shell $^{18}$O. Coupled-cluster and NCSM calculations of the polarizability of $^{24}$O, employing different interactions, span a large set of values, from $1$ to $1.4$ fm$^3$, calling for an experimental benchmark on this nucleus. As for calcium isotopes, although our predictions for open-shell nuclei are lower than the closed-shell ones, the tendency of the dipole polarizability to grow with the number of neutrons emerges, except for the case of $^{56}$Ca. This makes it an exciting region to investigate in future experiments.

To improve the accuracy of our many-body framework, we plan to study the effect of employing a CCSDT-1 reference in 2PA-LIT-CC calculations. Future work will also explore different strategies to tackle electromagnetic observables in open-shell nuclei, such as coupling the LIT technique with coupled-cluster computations based on reference states that break rotational invariance~\cite{novario2020,hagen2022,sun2024}. In this case, the most expensive part of the calculation is determined by the necessary symmetry restoration procedure. However, very recent PGCM results for electromagnetic responses suggest that the effect of the angular momentum projection is small in nuclei that are not strongly deformed~\cite{porro}.

\begin{acknowledgments}
\noindent
We gratefully acknowledge insightful discussions with Thomas Papenbrock. F.B. would also like to thank the ORNL Nuclear Theory group for their hospitality and support during different phases of this work, and Christian Forssén and Weiguang Jiang for useful discussions on uncertainty quantification. This work was supported by the Deutsche Forschungsgemeinschaft  (DFG, German Research Foundation) through Project-ID 279384907 - SFB 1245 and through the Cluster of Excellence "Precision Physics, Fundamental Interactions, and Structure of Matter" (PRISMA$^+$ EXC 2118/1, Project ID 39083149), by the Office of Nuclear Physics, U.S. Department of Energy, under SciDAC-5 (NUCLEI collaboration) and contract DE-FG02-97ER41014, and contract No. DE-AC05-00OR22725 with UT-Battelle, LLC (Oak Ridge National Laboratory). Computer time was provided by the Innovative and Novel Computational Impact on Theory and Experiment (INCITE) program and by the supercomputer Mogon at Johannes Gutenberg Universit\"at Mainz. This research used resources of the Oak Ridge Leadership Computing Facility located at ORNL, which is supported by the Office of Science of the Department of Energy under Contract No.~DE-AC05-00OR22725.
\end{acknowledgments}
\newpage
\appendix
\section{Diagrammatic contributions to the 2PA-LIT-CC Lanczos pivots}
\label{appendix}
As explained in Section \ref{lanczos}, the left and right 2PA-LIT-CC equations (\ref{left2palitcc}) and (\ref{right2palitcc}) are solved in an efficient way employing the Lanczos algorithm. The implementation of this strategy requires: 
\begin{enumerate}
    \item[(i)] The left and right Lanczos pivots $\textbf{S}^L$ and $\textbf{S}^R$ of Eq.~(\ref{rlpivot}).
    \item[(ii)] The computational machinery  to perform the matrix-vector product $(\overline{H} R_{\mu}^{A+2})_C$ and its left counterpart, which lies at the core of the Lanczos algorithm. 
\end{enumerate}
In this paper, we work with the 2PA-EOM-CC method in its spherical formulation, introduced in detail in Ref.~\cite{jansen2013}. The diagrammatic contributions to $(\overline{H} R_{\mu}^{A+2})_C$, together with the corresponding angular-momentum-coupled expressions, can be found in Ref.~\cite{jansen2013}. We will focus here on the first point, regarding the computation of the angular-momentum-coupled diagrams contributing to the right and left Lanczos pivots. 
When performing the angular momentum coupling, the similarity-transformed transition operator  is treated as a spherical tensor of rank $K$ and projection $M_K$. As for the ground state 2PA operators $R_0^{A+2}$ and $L_0^{A+2}$, it is key to notice that the ground state of most 2PA open-shell nuclei has $J^{\pi} = 0^{+}$. Therefore, we choose to consider $R_0^{A+2}$ and $L_0^{A+2}$ as scalar tensors. Under this assumption, the tensor products $\overline{\Theta}^K \otimes R_0^{A+2}$ and $L_0^{A+2} \otimes \overline{\Theta^{\dagger}}^K$, whose matrix elements correspond to the components of the right and left Lanczos pivots, become tensors of rank $K$.   

We  summarize below the notation used in the following. The amplitudes of a 2PA-EOM-CC excitation operator are indicated as
\begin{equation}
\begin{split}
    &r^{ab} = \braket{\Phi^{ab}|R|\Phi_0} = \braket{ab|R|0},\\
    &r^{abc}_i = \braket{\Phi^{abc}_i|R|\Phi_0} = \braket{abc|R|i}.
\end{split}
\end{equation}
An analogous notation is used for the amplitudes of the left de-excitation operator. In the derivation of the coupled expressions, we chose the following convention for the Wigner-Eckart theorem 
\begin{equation}
\begin{split}
  &\braket{ab|R^J_M|0} = C_{JM00}^{J_{ab} M_{ab}} \braket{ab||R^J||0} , \\
  &\braket{0|L^J_M|ab} = C_{JM00}^{J_{ab} M_{ab}} \braket{0||L^J||ab},\\
\end{split}
\end{equation}
applied to right $R$ and left $L$ spherical tensors of rank $J$ and projection $M$.  We include up to 3p-1h excitations in the calculation of the pivots. The uncoupled and coupled expressions for the diagrams contributing to the right and left Lanczos pivots are reported in Tables~\ref{table-right-diagrams} and \ref{table-left-diagrams}, respectively.

\begin{table*}
\caption{\label{table-right-diagrams} Coupled-cluster diagrams for the right Lanczos pivot of the 2PA LIT-CC method with both ordinary and reduced expressions. The curly line in the diagrams represents the action of the similarity-transformed operator $\overline{\Theta}$. The angular-momentum-coupled expressions for the permutation operators are derived in Ref.~\cite{jansen2013}. }
\begin{xtabular*}{\textwidth}{l@{\extracolsep{\fill}}lll}
\hline
\hline
Diagram & Uncoupled expression & Coupled expression\\
\hline
\\\begin{tikzpicture}
\draw[nucleon] (0,2) -- (0,0);
\draw[amplitude] (0,0) -- (2,0);
\draw[draw=black, postaction={decorate},
        decoration={markings,mark=at position .55 with {\arrow[draw=black, scale=2]{>}}}] (2,1) -- (2,2);
\draw[black, postaction={decorate},
        decoration={markings,mark=at position .55 with {\arrow[draw=black, scale=2]{>}}}] (2,0) -- (2,1);
\draw [gluon] (2,1) -- (3,1);
\end{tikzpicture}
& $\makecell{P(ab) \sum_e \braket{b|\overline{\Theta}^K_{M_K}|e}\braket{ae|R^0_0|0}}$ & $\makecell{P(ab)  \sum_e \frac{\hat{j}_b}{\hat{j}_a \hat{K}} \braket{b||\overline{\Theta}^K|| e} \\ \times \braket{ae; J_{ae}=0||R^0 ||0} \delta_{j_a, j_e}}$\\    \\\begin{tikzpicture}
\draw[nucleon] (0,2) -- (0,0);
\draw[amplitude] (0,0) -- (2,0);
\draw[draw=black, postaction={decorate},
        decoration={markings,mark=at position .55 with {\arrow[draw=black, scale=2]{>}}}] (1,0) -- (1,2);
\draw[gluon] (2,1) -- (3,1);  
\draw[draw=black, postaction={decorate},
        decoration={markings,mark=at position .55 with {\arrow[draw=black, scale=2]{<}}}] (2,0) .. controls (2.25, 0.5) .. (2,1);
\draw[draw=black, postaction={decorate},
        decoration={markings,mark=at position .55 with {\arrow[draw=black, scale=2]{>}}}] (2,0) .. controls (1.75, 0.5) .. (2,1);
\end{tikzpicture} & $\makecell{\sum_{en}  \braket{n|\overline{\Theta}^K_{M_K}|e} \braket{abe|R^0_0|n}}$ & $\makecell{\sum_{en} \frac{\hat{j_n^2}}{\hat{K^2}} \braket{n||\overline{\Theta}^K||e}\\\times \braket{abe;J_{ab}=K; J_{abe}=j_n||R^0||n}}$ \\
\\\begin{tikzpicture}
\draw[nucleon] (0,2) -- (0,0);
\draw[amplitude] (0,0) -- (2,0);
\draw[draw=black, postaction={decorate},
        decoration={markings,mark=at position .55 with {\arrow[draw=black, scale=2]{>}}}] (1,0) -- (1,2);
\draw[draw=black, postaction={decorate},
        decoration={markings,mark=at position .55 with {\arrow[draw=black, scale=2]{>}}}] (2,0) -- (1.5,2);
\draw[gluon] (2.25,1) -- (3.25,1);
\draw[draw=black, postaction={decorate},
        decoration={markings,mark=at position .55 with {\arrow[draw=black, scale=2]{<}}}] (2,0) -- (2.25,1);
\draw[draw=black, postaction={decorate},
        decoration={markings,mark=at position .55 with {\arrow[draw=black, scale=2]{<}}}] (2.25,1) -- (2.5,2);
\end{tikzpicture} & $\makecell{-\sum_n \braket{n|\overline{\Theta}^K_{M_K}|i}\braket{abc|R^0_0|n}}$ & $\makecell{-\sum_n\braket{n||\overline{\Theta}^K|| i}\\\times \braket{abc;J_{ab};J_{abc}= j_n||R^0 ||n} \delta_{j_n, J_{abc}}}$ \\
\\\begin{tikzpicture}
\draw[nucleon] (0,2) -- (0,0);
\draw[amplitude] (0,0) -- (2,0);
\draw[draw=black, postaction={decorate},
        decoration={markings,mark=at position .55 with {\arrow[draw=black, scale=2]{>}}}] (2,0) -- (2,2);
\draw[draw=black, postaction={decorate},
        decoration={markings,mark=at position .55 with {\arrow[draw=black, scale=2]{>}}}] (3.5,0) -- (2.5,2);
\draw[draw=black, postaction={decorate},
        decoration={markings,mark=at position .55 with {\arrow[draw=black, scale=2]{<}}}] (3.5,0) -- (4.5,2);
\draw[gluon] (3.5,0) -- (4.5,0);
\end{tikzpicture}    
     & $\makecell{P(ab,c) \braket{c|\overline{\Theta}^K_{M_K}|i} \braket{ab|R^0_0|0}}$    & $\makecell{P(ab,c) \braket{c||\overline{\Theta}^K||i}\\ \times \braket{ab; J_{ab} = 0||R^0 ||0} \delta_{j_c, J_{abc}}}$ \\
\\\begin{tikzpicture}
\draw[nucleon] (0,2) -- (0,0);
\draw[amplitude] (0,0) -- (2,0);
\draw[draw=black, postaction={decorate},
        decoration={markings,mark=at position .55 with {\arrow[draw=black, scale=2]{>}}}] (1,0) -- (1,2);
\draw[draw=black, postaction={decorate},
        decoration={markings,mark=at position .55 with {\arrow[draw=black, scale=2]{<}}}] (2,0) -- (1.5,2);
\draw[gluon] (2.25,1) -- (3.25,1);
\draw[draw=black, postaction={decorate},
        decoration={markings,mark=at position .55 with {\arrow[draw=black, scale=2]{>}}}] (2,0) -- (2.25,1);
\draw[draw=black, postaction={decorate},
        decoration={markings,mark=at position .55 with {\arrow[draw=black, scale=2]{>}}}] (2.25,1) -- (2.5,2);
\end{tikzpicture}     & $\makecell{P(ab,c) \sum_e \braket{c|\overline{\Theta}^K_{M_K}|e} \braket{abe|R^0_0|i}}$ & $\makecell{P(ab,c) \sum_e (-1)^{j_c + J_{ab} + K + j_i} \hat{j_i}\hat{j_c} \\ \times
  \left \{ \begin{array}{ll}
  j_e \;\;\; J_{ab} \;\; j_i\\
  J_{abc} \;\; K \;\;  j_c 
  \end{array} \right \} \\ \times \braket{c||\overline{\Theta}^K ||e} \braket{abe; J_{ab}; J_{abe} = j_i||R^0||i}}$\\
\\ \begin{tikzpicture}
\draw[nucleon] (0,2) -- (0,0);
\draw[amplitude] (0,0) -- (2,0);
\draw[draw=black, postaction={decorate},
        decoration={markings,mark=at position .55 with {\arrow[draw=black, scale=2]{>}}}] (2,1) -- (2,2);
\draw[black, postaction={decorate},
        decoration={markings,mark=at position .55 with {\arrow[draw=black, scale=2]{>}}}] (2,0) -- (2,1);
\draw[gluon] (2,1) -- (3,1);
\draw[draw=black, postaction={decorate},
        decoration={markings,mark=at position .55 with {\arrow[draw=black, scale=2]{>}}}] (3,1) -- (2.5,2);
\draw[draw=black, postaction={decorate},
        decoration={markings,mark=at position .55 with {\arrow[draw=black, scale=2]{<}}}] (3,1) -- (3.5,2);
\end{tikzpicture}     &  $\makecell{P(a, bc) \sum_e \braket{bc|\overline{\Theta}^K_{M_K}|ei}\braket{ae|R^0_0|0}}$& $\makecell{P(ab,c) \sum_e \sum_{J_{ei}}  (-1)^{1+K+J_{ei}+J_{abc} + j_c} \\ \times\frac{\hat{J}_{ei}\hat{J}_{ab}}{\hat{j}_c}\left \{ \begin{array}{ll}
  j_i \;\;\; K \;\; J_{abc}\\
  J_{ab} \;\; j_c \;\;  J_{ei} 
  \end{array} 
\right \} \\ \times \braket{ab;J_{ab}||\overline{\Theta}^K||ei;J_{ei}} 
\\ \times \braket{ec;J_{ec}=0||R^0||0} \delta_{j_c,j_e}}$\\ \\
\hline
\end{xtabular*}
\end{table*}

\begin{table*}
\caption{\label{table-left-diagrams} Coupled-cluster diagrams for the left Lanczos pivot of the 2PA LIT-CC method with both ordinary and reduced expressions. The curly line in the diagrams represents the action of the similarity-transformed operator $\overline{\Theta^{\dagger}}$. The angular-momentum-coupled expressions for the permutation operators are derived in Ref. \cite{jansen2013}. }
\begin{xtabular*}{\textwidth}{l@{\extracolsep{\fill}}lll}
\hline
\hline
Diagram & Uncoupled expression & Coupled expression\\
\hline
\\ \begin{tikzpicture}
\draw[draw=black, postaction={decorate},
        decoration={markings,mark=at position .55 with {\arrow[draw=black, scale=2]{<}}}] (0,2) -- (0,0);
\draw[amplitude] (0,2) -- (2,2);
\draw[draw=black, postaction={decorate},
        decoration={markings,mark=at position .55 with {\arrow[draw=black, scale=2]{>}}}] (2,1) -- (2,2);
\draw[black, postaction={decorate},
        decoration={markings,mark=at position .55 with {\arrow[draw=black, scale=2]{>}}}] (2,0) -- (2,1);
\draw [gluon] (2,1) -- (3,1);
\end{tikzpicture} & $\makecell{P(ab) \sum_e \braket{e|\overline{\Theta^{\dagger}}^K_{M_K}|b} \braket{0|L^0_0|ae}}$ & $\makecell{P(ab) \sum_e \frac{\hat{j_b}}{\hat{j_a}\hat{K}}  \braket{e||\overline{\Theta^{\dagger}}^K|| b} \\ \times \braket{0||L^0 ||ae; J_{ae} = 0} \delta_{j_a,j_e}}$\\
\begin{tikzpicture}
\draw[draw=black, postaction={decorate},
        decoration={markings,mark=at position .55 with {\arrow[draw=black, scale=2]{<}}}] (0,2) -- (0,0);
\draw[amplitude] (0,2) -- (2,2);
\draw[draw=black, postaction={decorate},
        decoration={markings,mark=at position .55 with {\arrow[draw=black, scale=2]{>}}}] (1,0) -- (1,2);
\draw[gluon] (2,1) -- (3,1);  
\draw[draw=black, postaction={decorate},
        decoration={markings,mark=at position .55 with {\arrow[draw=black, scale=2]{<}}}] (2,1) .. controls (2.25, 1.5) .. (2,2);
\draw[draw=black, postaction={decorate},
        decoration={markings,mark=at position .55 with {\arrow[draw=black, scale=2]{>}}}] (2,1) .. controls (1.75, 1.5) .. (2,2);
\end{tikzpicture} & $\makecell{\sum_{en} \braket{e|\overline{\Theta^{\dagger}}^K_{M_K}|n} \braket{n|L^0_0|abe}}$ & $\makecell{\sum_{en} \frac{\hat{j_n^2}}{\hat{K^2}} \braket{e||\overline{\Theta^{\dagger}}^K||n}\\ \times \braket{n||L^0||abe;J_{ab}=K; J_{abe}=j_n}}$\\
\begin{tikzpicture}
\draw[draw=black, postaction={decorate},
        decoration={markings,mark=at position .55 with {\arrow[draw=black, scale=2]{<}}}] (0,2) -- (0,0);
\draw[amplitude] (0,2) -- (2,2);
\draw[draw=black, postaction={decorate},
        decoration={markings,mark=at position .55 with {\arrow[draw=black, scale=2]{>}}}] (1,0) -- (1,2);
\draw[draw=black, postaction={decorate},
        decoration={markings,mark=at position .55 with {\arrow[draw=black, scale=2]{<}}}] (2,2) -- (1.5,0);
\draw[gluon] (2.25,1) -- (3.25,1);
\draw[draw=black, postaction={decorate},
        decoration={markings,mark=at position .55 with {\arrow[draw=black, scale=2]{>}}}] (2,2) -- (2.25,1);
\draw[draw=black, postaction={decorate},
        decoration={markings,mark=at position .55 with {\arrow[draw=black, scale=2]{>}}}] (2.25,1) -- (2.5,0);
\end{tikzpicture} & $\makecell{-\sum_n \braket{n|L^0_0|abc} \braket{i|\overline{\Theta^{\dagger}}^K_{M_K}|n}}$ & 
$\makecell{-\sum_n \braket{n||L^0||abc; J_{ab};J_{abc} = j_n} \\ \times \braket{i||\overline{\Theta^{\dagger}}^K||n} \delta_{j_n, J_{abc}}}$ \\
\begin{tikzpicture}
\draw[nucleon] (0,2) -- (0,0);
\draw[amplitude] (0,2) -- (2,2);
\draw[draw=black, postaction={decorate},
        decoration={markings,mark=at position .55 with {\arrow[draw=black, scale=2]{>}}}] (2,0) -- (2,2);
\draw[draw=black, postaction={decorate},
        decoration={markings,mark=at position .55 with {\arrow[draw=black, scale=2]{>}}}] (3.5,2) -- (2.5,0);
\draw[draw=black, postaction={decorate},
        decoration={markings,mark=at position .55 with {\arrow[draw=black, scale=2]{<}}}] (3.5,2) -- (4.5,0);
\draw[gluon] (3.5,2) -- (4.5,2);
\end{tikzpicture} & $\makecell{P(ab,c) \braket{0|L^0_0|ab} \braket{i|\overline{\Theta^{\dagger}}^K_{M_K}|c}}$ & $\makecell{P(ab,c) \braket{0||L^0||ab; J_{ab} = 0}\\ \times \braket{i||\overline{\Theta^{\dagger}}^K||c} \delta_{j_c, J_{abc}}}$ \\
\begin{tikzpicture}
\draw[draw=black, postaction={decorate},
        decoration={markings,mark=at position .55 with {\arrow[draw=black, scale=2]{<}}}] (0,2) -- (0,0);
\draw[amplitude] (0,2) -- (2,2);
\draw[draw=black, postaction={decorate},
        decoration={markings,mark=at position .55 with {\arrow[draw=black, scale=2]{>}}}] (1,0) -- (1,2);
\draw[draw=black, postaction={decorate},
        decoration={markings,mark=at position .55 with {\arrow[draw=black, scale=2]{>}}}] (2,2) -- (1.5,0);
\draw[gluon] (2.25,1) -- (3.25,1);
\draw[draw=black, postaction={decorate},
        decoration={markings,mark=at position .55 with {\arrow[draw=black, scale=2]{<}}}] (2,2) -- (2.25,1);
\draw[draw=black, postaction={decorate},
        decoration={markings,mark=at position .55 with {\arrow[draw=black, scale=2]{<}}}] (2.25,1) -- (2.5,0);
\end{tikzpicture} &  $\makecell{P(ab,c) \sum_e \braket{i|L^0_0|abe} \braket{e|\overline{\Theta^{\dagger}}^K_{M_K}|c}}$   & $\makecell{P(ab,c) \sum_e (-1)^{j_i+j_c+K+J_{ab}}\hat{j_i}\hat{j_c}\\ \times 
  \left \{ \begin{array}{ll}
  j_e \;\;\;\; K \;\; j_c\\
  J_{abc} \; J_{ab} \;  j_i
  \end{array} 
\right \} \\ \times \braket{i||L^0||abe;J_{ab};J_{abe} = j_i} \\ \times\braket{e||\overline{\Theta^{\dagger}}^K||c}}$      \\
\begin{tikzpicture}
\draw[draw=black, postaction={decorate},
        decoration={markings,mark=at position .55 with {\arrow[draw=black, scale=2]{<}}}] (0,2) -- (0,0);
\draw[amplitude] (0,2) -- (2,2);
\draw[draw=black, postaction={decorate},
        decoration={markings,mark=at position .4 with {\arrow[draw=black, scale=2]{>}}}] (1,0) -- (1,1.5);
\draw[draw=black, postaction={decorate},
        decoration={markings,mark=at position .4 with {\arrow[draw=black, scale=2]{>}}}] (1,1.5) -- (1,2);
\draw[gluon] (2,1) -- (1,1);
\draw[draw=black, postaction={decorate},
        decoration={markings,mark=at position .55 with {\arrow[draw=black, scale=2]{<}}}] (2,1) .. controls (2.25, 1.5) .. (2,2);
\draw[draw=black, postaction={decorate},
        decoration={markings,mark=at position .55 with {\arrow[draw=black, scale=2]{>}}}] (2,1) .. controls (1.75, 1.5) .. (2,2);
\end{tikzpicture} & $\makecell{\frac{1}{2} P(ab) \sum_{en} \braket{n|L^0_0|aef} \braket{ef|\overline{\Theta^{\dagger}}^K_{M_K}|bn}}$ & $\makecell{\frac{1}{2} P(ab) \sum_{en} \sum_{J_{ef}, J_{an}}  (-1)^{1+j_a +  j_n +J_{an}}\\ \times \frac{\hat{J}_{an}^2\hat{j}_n}{\hat{K}}  \left \{ \begin{array}{ll}
  j_n \;\;J_{ef} \;\; j_b\\
  K \;\; j_a \;\;  J_{an}
  \end{array} 
\right \} \\ \times\braket{n||L^0||efb;J_{ef}; J_{efb}=j_n}  \\ \times \braket{ef;J_{ef}||\overline{\Theta^{\dagger}}^K||an;J_{an}}}$\\ \\
\hline
\end{xtabular*}
\end{table*}
\FloatBarrier 
\bibliography{new} 

\begin{thebibliography}{86}%
\makeatletter
\providecommand \@ifxundefined [1]{%
 \@ifx{#1\undefined}
}%
\providecommand \@ifnum [1]{%
 \ifnum #1\expandafter \@firstoftwo
 \else \expandafter \@secondoftwo
 \fi
}%
\providecommand \@ifx [1]{%
 \ifx #1\expandafter \@firstoftwo
 \else \expandafter \@secondoftwo
 \fi
}%
\providecommand \natexlab [1]{#1}%
\providecommand \enquote  [1]{``#1''}%
\providecommand \bibnamefont  [1]{#1}%
\providecommand \bibfnamefont [1]{#1}%
\providecommand \citenamefont [1]{#1}%
\providecommand \href@noop [0]{\@secondoftwo}%
\providecommand \href [0]{\begingroup \@sanitize@url \@href}%
\providecommand \@href[1]{\@@startlink{#1}\@@href}%
\providecommand \@@href[1]{\endgroup#1\@@endlink}%
\providecommand \@sanitize@url [0]{\catcode `\\12\catcode `\$12\catcode `\&12\catcode `\#12\catcode `\^12\catcode `\_12\catcode `\%12\relax}%
\providecommand \@@startlink[1]{}%
\providecommand \@@endlink[0]{}%
\providecommand \url  [0]{\begingroup\@sanitize@url \@url }%
\providecommand \@url [1]{\endgroup\@href {#1}{\urlprefix }}%
\providecommand \urlprefix  [0]{URL }%
\providecommand \Eprint [0]{\href }%
\providecommand \doibase [0]{http://dx.doi.org/}%
\providecommand \selectlanguage [0]{\@gobble}%
\providecommand \bibinfo  [0]{\@secondoftwo}%
\providecommand \bibfield  [0]{\@secondoftwo}%
\providecommand \translation [1]{[#1]}%
\providecommand \BibitemOpen [0]{}%
\providecommand \bibitemStop [0]{}%
\providecommand \bibitemNoStop [0]{.\EOS\space}%
\providecommand \EOS [0]{\spacefactor3000\relax}%
\providecommand \BibitemShut  [1]{\csname bibitem#1\endcsname}%
\let\auto@bib@innerbib\@empty
\bibitem [{\citenamefont {Goldhaber}\ and\ \citenamefont {Teller}(1948)}]{goldhaber1948}%
  \BibitemOpen
  \bibfield  {author} {\bibinfo {author} {\bibfnamefont {M.}~\bibnamefont {Goldhaber}}\ and\ \bibinfo {author} {\bibfnamefont {E.}~\bibnamefont {Teller}},\ }\bibfield  {title} {\enquote {\bibinfo {title} {On nuclear dipole vibrations},}\ }\href {\doibase 10.1103/PhysRev.74.1046} {\bibfield  {journal} {\bibinfo  {journal} {Phys. Rev.}\ }\textbf {\bibinfo {volume} {74}},\ \bibinfo {pages} {1046--1049} (\bibinfo {year} {1948})}\BibitemShut {NoStop}%
\bibitem [{\citenamefont {Steinwedel}\ and\ \citenamefont {D.Jensen}(1950)}]{steinwedel1950}%
  \BibitemOpen
  \bibfield  {author} {\bibinfo {author} {\bibfnamefont {Helmut}\ \bibnamefont {Steinwedel}}\ and\ \bibinfo {author} {\bibfnamefont {J.Hans}\ \bibnamefont {D.Jensen}},\ }\bibfield  {title} {\enquote {\bibinfo {title} {Hydrodynamik von kerndipolschwingungen},}\ }\href {\doibase doi:10.1515/zna-1950-0801} {\bibfield  {journal} {\bibinfo  {journal} {Zeitschrift für Naturforschung A}\ }\textbf {\bibinfo {volume} {5}},\ \bibinfo {pages} {413--420} (\bibinfo {year} {1950})}\BibitemShut {NoStop}%
\bibitem [{\citenamefont {Kobayashi}\ \emph {et~al.}(1989)\citenamefont {Kobayashi}, \citenamefont {Shimoura}, \citenamefont {Tanihata}, \citenamefont {Katori}, \citenamefont {Matsuta}, \citenamefont {Minamisono}, \citenamefont {Sugimoto}, \citenamefont {Müller}, \citenamefont {Olson}, \citenamefont {Symons},\ and\ \citenamefont {Wieman}}]{kobayashi1989}%
  \BibitemOpen
  \bibfield  {author} {\bibinfo {author} {\bibfnamefont {T.}~\bibnamefont {Kobayashi}}, \bibinfo {author} {\bibfnamefont {S.}~\bibnamefont {Shimoura}}, \bibinfo {author} {\bibfnamefont {I.}~\bibnamefont {Tanihata}}, \bibinfo {author} {\bibfnamefont {K.}~\bibnamefont {Katori}}, \bibinfo {author} {\bibfnamefont {K.}~\bibnamefont {Matsuta}}, \bibinfo {author} {\bibfnamefont {T.}~\bibnamefont {Minamisono}}, \bibinfo {author} {\bibfnamefont {K.}~\bibnamefont {Sugimoto}}, \bibinfo {author} {\bibfnamefont {W.}~\bibnamefont {Müller}}, \bibinfo {author} {\bibfnamefont {D.L.}\ \bibnamefont {Olson}}, \bibinfo {author} {\bibfnamefont {T.J.M.}\ \bibnamefont {Symons}}, \ and\ \bibinfo {author} {\bibfnamefont {H.}~\bibnamefont {Wieman}},\ }\bibfield  {title} {\enquote {\bibinfo {title} {Electromagnetic dissociation and soft giant dipole resonance of the neutron-dripline nucleus ${}^{11}\mathrm{Li}$},}\ }\href {\doibase https://doi.org/10.1016/0370-2693(89)90557-1} {\bibfield  {journal} {\bibinfo  {journal} {Physics Letters
  B}\ }\textbf {\bibinfo {volume} {232}},\ \bibinfo {pages} {51--55} (\bibinfo {year} {1989})}\BibitemShut {NoStop}%
\bibitem [{\citenamefont {Aumann}\ and\ \citenamefont {Nakamura}(2013)}]{aumann2013}%
  \BibitemOpen
  \bibfield  {author} {\bibinfo {author} {\bibfnamefont {T}~\bibnamefont {Aumann}}\ and\ \bibinfo {author} {\bibfnamefont {T}~\bibnamefont {Nakamura}},\ }\bibfield  {title} {\enquote {\bibinfo {title} {The electric dipole response of exotic nuclei},}\ }\href {\doibase 10.1088/0031-8949/2013/T152/014012} {\bibfield  {journal} {\bibinfo  {journal} {Physica Scripta}\ }\textbf {\bibinfo {volume} {2013}},\ \bibinfo {pages} {014012} (\bibinfo {year} {2013})}\BibitemShut {NoStop}%
\bibitem [{\citenamefont {Paar}\ \emph {et~al.}(2007)\citenamefont {Paar}, \citenamefont {Vretenar}, \citenamefont {Khan},\ and\ \citenamefont {Colò}}]{paar2007}%
  \BibitemOpen
  \bibfield  {author} {\bibinfo {author} {\bibfnamefont {Nils}\ \bibnamefont {Paar}}, \bibinfo {author} {\bibfnamefont {Dario}\ \bibnamefont {Vretenar}}, \bibinfo {author} {\bibfnamefont {Elias}\ \bibnamefont {Khan}}, \ and\ \bibinfo {author} {\bibfnamefont {Gianluca}\ \bibnamefont {Colò}},\ }\bibfield  {title} {\enquote {\bibinfo {title} {Exotic modes of excitation in atomic nuclei far from stability},}\ }\href {\doibase 10.1088/0034-4885/70/5/R02} {\bibfield  {journal} {\bibinfo  {journal} {Reports on Progress in Physics}\ }\textbf {\bibinfo {volume} {70}},\ \bibinfo {pages} {R02} (\bibinfo {year} {2007})}\BibitemShut {NoStop}%
\bibitem [{\citenamefont {Roca-Maza}\ and\ \citenamefont {Paar}(2018)}]{rocamaza2018}%
  \BibitemOpen
  \bibfield  {author} {\bibinfo {author} {\bibfnamefont {X.}~\bibnamefont {Roca-Maza}}\ and\ \bibinfo {author} {\bibfnamefont {N.}~\bibnamefont {Paar}},\ }\bibfield  {title} {\enquote {\bibinfo {title} {Nuclear equation of state from ground and collective excited state properties of nuclei},}\ }\href {\doibase https://doi.org/10.1016/j.ppnp.2018.04.001} {\bibfield  {journal} {\bibinfo  {journal} {Progress in Particle and Nuclear Physics}\ }\textbf {\bibinfo {volume} {101}},\ \bibinfo {pages} {96--176} (\bibinfo {year} {2018})}\BibitemShut {NoStop}%
\bibitem [{\citenamefont {Colo'}(2022)}]{colo2022}%
  \BibitemOpen
  \bibfield  {author} {\bibinfo {author} {\bibfnamefont {Gianluca}\ \bibnamefont {Colo'}},\ }\enquote {\bibinfo {title} {{Theoretical Methods for Giant Resonances}},}\ in\ \href {\doibase 10.1007/978-981-15-8818-1_72-1} {\emph {\bibinfo {booktitle} {{Handbook of Nuclear Physics}}}},\ \bibinfo {editor} {edited by\ \bibinfo {editor} {\bibfnamefont {Isao}\ \bibnamefont {Tanihata}}, \bibinfo {editor} {\bibfnamefont {Hiroshi}\ \bibnamefont {Toki}}, \ and\ \bibinfo {editor} {\bibfnamefont {Toshitaka}\ \bibnamefont {Kajino}}}\ (\bibinfo {year} {2022})\ pp.\ \bibinfo {pages} {1--29},\ \Eprint {http://arxiv.org/abs/2201.04578} {arXiv:2201.04578 [nucl-th]} \BibitemShut {NoStop}%
\bibitem [{\citenamefont {Ekstr\"om}\ \emph {et~al.}(2023)\citenamefont {Ekstr\"om}, \citenamefont {Forss\'en}, \citenamefont {Hagen}, \citenamefont {Jansen}, \citenamefont {Jiang},\ and\ \citenamefont {Papenbrock}}]{ekstrom2022}%
  \BibitemOpen
  \bibfield  {author} {\bibinfo {author} {\bibfnamefont {A.}~\bibnamefont {Ekstr\"om}}, \bibinfo {author} {\bibfnamefont {C.}~\bibnamefont {Forss\'en}}, \bibinfo {author} {\bibfnamefont {G.}~\bibnamefont {Hagen}}, \bibinfo {author} {\bibfnamefont {G.~R.}\ \bibnamefont {Jansen}}, \bibinfo {author} {\bibfnamefont {W.}~\bibnamefont {Jiang}}, \ and\ \bibinfo {author} {\bibfnamefont {T.}~\bibnamefont {Papenbrock}},\ }\bibfield  {title} {\enquote {\bibinfo {title} {{What is ab initio in nuclear theory?}}}\ }\href {\doibase 10.3389/fphy.2023.1129094} {\bibfield  {journal} {\bibinfo  {journal} {Front. Phys.}\ }\textbf {\bibinfo {volume} {11}},\ \bibinfo {pages} {1129094} (\bibinfo {year} {2023})},\ \Eprint {http://arxiv.org/abs/2212.11064} {arXiv:2212.11064 [nucl-th]} \BibitemShut {NoStop}%
\bibitem [{\citenamefont {Epelbaum}\ \emph {et~al.}(2009)\citenamefont {Epelbaum}, \citenamefont {Hammer},\ and\ \citenamefont {Mei\ss{}ner}}]{epelbaum2009}%
  \BibitemOpen
  \bibfield  {author} {\bibinfo {author} {\bibfnamefont {E.}~\bibnamefont {Epelbaum}}, \bibinfo {author} {\bibfnamefont {H.-W.}\ \bibnamefont {Hammer}}, \ and\ \bibinfo {author} {\bibfnamefont {Ulf-G.}\ \bibnamefont {Mei\ss{}ner}},\ }\bibfield  {title} {\enquote {\bibinfo {title} {Modern theory of nuclear forces},}\ }\href {\doibase 10.1103/RevModPhys.81.1773} {\bibfield  {journal} {\bibinfo  {journal} {Rev. Mod. Phys.}\ }\textbf {\bibinfo {volume} {81}},\ \bibinfo {pages} {1773--1825} (\bibinfo {year} {2009})}\BibitemShut {NoStop}%
\bibitem [{\citenamefont {Machleidt}\ and\ \citenamefont {Entem}(2011)}]{machleidt2011}%
  \BibitemOpen
  \bibfield  {author} {\bibinfo {author} {\bibfnamefont {R.}~\bibnamefont {Machleidt}}\ and\ \bibinfo {author} {\bibfnamefont {D.R.}\ \bibnamefont {Entem}},\ }\bibfield  {title} {\enquote {\bibinfo {title} {Chiral effective field theory and nuclear forces},}\ }\href {\doibase https://doi.org/10.1016/j.physrep.2011.02.001} {\bibfield  {journal} {\bibinfo  {journal} {Physics Reports}\ }\textbf {\bibinfo {volume} {503}},\ \bibinfo {pages} {1--75} (\bibinfo {year} {2011})}\BibitemShut {NoStop}%
\bibitem [{\citenamefont {Hammer}\ \emph {et~al.}(2020)\citenamefont {Hammer}, \citenamefont {K\"onig},\ and\ \citenamefont {van Kolck}}]{hammer2020}%
  \BibitemOpen
  \bibfield  {author} {\bibinfo {author} {\bibfnamefont {H.-W.}\ \bibnamefont {Hammer}}, \bibinfo {author} {\bibfnamefont {Sebastian}\ \bibnamefont {K\"onig}}, \ and\ \bibinfo {author} {\bibfnamefont {U.}~\bibnamefont {van Kolck}},\ }\bibfield  {title} {\enquote {\bibinfo {title} {Nuclear effective field theory: Status and perspectives},}\ }\href {\doibase 10.1103/RevModPhys.92.025004} {\bibfield  {journal} {\bibinfo  {journal} {Rev. Mod. Phys.}\ }\textbf {\bibinfo {volume} {92}},\ \bibinfo {pages} {025004} (\bibinfo {year} {2020})}\BibitemShut {NoStop}%
\bibitem [{\citenamefont {Bacca}\ \emph {et~al.}(2002)\citenamefont {Bacca}, \citenamefont {Marchisio}, \citenamefont {Barnea}, \citenamefont {Leidemann},\ and\ \citenamefont {Orlandini}}]{bacca_photon1}%
  \BibitemOpen
  \bibfield  {author} {\bibinfo {author} {\bibfnamefont {Sonia}\ \bibnamefont {Bacca}}, \bibinfo {author} {\bibfnamefont {Mario~Andrea}\ \bibnamefont {Marchisio}}, \bibinfo {author} {\bibfnamefont {Nir}\ \bibnamefont {Barnea}}, \bibinfo {author} {\bibfnamefont {Winfried}\ \bibnamefont {Leidemann}}, \ and\ \bibinfo {author} {\bibfnamefont {Giuseppina}\ \bibnamefont {Orlandini}},\ }\bibfield  {title} {\enquote {\bibinfo {title} {Microscopic calculation of six-body inelastic reactions with complete final state interaction: Photoabsorption of $^6$he and $^6$li},}\ }\href {\doibase 10.1103/PhysRevLett.89.052502} {\bibfield  {journal} {\bibinfo  {journal} {Phys. Rev. Lett.}\ }\textbf {\bibinfo {volume} {89}},\ \bibinfo {pages} {052502} (\bibinfo {year} {2002})}\BibitemShut {NoStop}%
\bibitem [{\citenamefont {Bacca}\ \emph {et~al.}(2004)\citenamefont {Bacca}, \citenamefont {Arenhövel}, \citenamefont {Barnea}, \citenamefont {Leidemann},\ and\ \citenamefont {Orlandini}}]{bacca_photon2}%
  \BibitemOpen
  \bibfield  {author} {\bibinfo {author} {\bibfnamefont {Sonia}\ \bibnamefont {Bacca}}, \bibinfo {author} {\bibfnamefont {Hartmuth}\ \bibnamefont {Arenhövel}}, \bibinfo {author} {\bibfnamefont {Nir}\ \bibnamefont {Barnea}}, \bibinfo {author} {\bibfnamefont {Winfried}\ \bibnamefont {Leidemann}}, \ and\ \bibinfo {author} {\bibfnamefont {Giuseppina}\ \bibnamefont {Orlandini}},\ }\bibfield  {title} {\enquote {\bibinfo {title} {Ab initio calculation of 7li photodisintegration},}\ }\href {\doibase https://doi.org/10.1016/j.physletb.2004.10.025} {\bibfield  {journal} {\bibinfo  {journal} {Physics Letters B}\ }\textbf {\bibinfo {volume} {603}},\ \bibinfo {pages} {159--164} (\bibinfo {year} {2004})}\BibitemShut {NoStop}%
\bibitem [{\citenamefont {Bacca}\ \emph {et~al.}(2007)\citenamefont {Bacca}, \citenamefont {Arenh\"ovel}, \citenamefont {Barnea}, \citenamefont {Leidemann},\ and\ \citenamefont {Orlandini}}]{bacca_el}%
  \BibitemOpen
  \bibfield  {author} {\bibinfo {author} {\bibfnamefont {Sonia}\ \bibnamefont {Bacca}}, \bibinfo {author} {\bibfnamefont {Hartmuth}\ \bibnamefont {Arenh\"ovel}}, \bibinfo {author} {\bibfnamefont {Nir}\ \bibnamefont {Barnea}}, \bibinfo {author} {\bibfnamefont {Winfried}\ \bibnamefont {Leidemann}}, \ and\ \bibinfo {author} {\bibfnamefont {Giuseppina}\ \bibnamefont {Orlandini}},\ }\bibfield  {title} {\enquote {\bibinfo {title} {Inclusive electron scattering off $^{4}\mathrm{He}$},}\ }\href {\doibase 10.1103/PhysRevC.76.014003} {\bibfield  {journal} {\bibinfo  {journal} {Phys. Rev. C}\ }\textbf {\bibinfo {volume} {76}},\ \bibinfo {pages} {014003} (\bibinfo {year} {2007})}\BibitemShut {NoStop}%
\bibitem [{\citenamefont {Bacca}\ \emph {et~al.}(2013{\natexlab{a}})\citenamefont {Bacca}, \citenamefont {Barnea}, \citenamefont {Leidemann},\ and\ \citenamefont {Orlandini}}]{bacca_monopole}%
  \BibitemOpen
  \bibfield  {author} {\bibinfo {author} {\bibfnamefont {Sonia}\ \bibnamefont {Bacca}}, \bibinfo {author} {\bibfnamefont {Nir}\ \bibnamefont {Barnea}}, \bibinfo {author} {\bibfnamefont {Winfried}\ \bibnamefont {Leidemann}}, \ and\ \bibinfo {author} {\bibfnamefont {Giuseppina}\ \bibnamefont {Orlandini}},\ }\bibfield  {title} {\enquote {\bibinfo {title} {Isoscalar monopole resonance of the alpha particle: A prism to nuclear hamiltonians},}\ }\href {\doibase 10.1103/PhysRevLett.110.042503} {\bibfield  {journal} {\bibinfo  {journal} {Phys. Rev. Lett.}\ }\textbf {\bibinfo {volume} {110}},\ \bibinfo {pages} {042503} (\bibinfo {year} {2013}{\natexlab{a}})}\BibitemShut {NoStop}%
\bibitem [{\citenamefont {Kegel}\ \emph {et~al.}(2023)\citenamefont {Kegel}, \citenamefont {Achenbach}, \citenamefont {Bacca}, \citenamefont {Barnea}, \citenamefont {Beri\ifmmode \check{c}\else \v{c}\fi{}i\ifmmode~\check{c}\else \v{c}\fi{}}, \citenamefont {Bosnar}, \citenamefont {Correa}, \citenamefont {Distler}, \citenamefont {Esser}, \citenamefont {Fonvieille}, \citenamefont {Fri\ifmmode \check{s}\else \v{s}\fi{}\ifmmode \check{c}\else \v{c}\fi{}i\ifmmode~\acute{c}\else \'{c}\fi{}}, \citenamefont {Heilig}, \citenamefont {Herrmann}, \citenamefont {Hoek}, \citenamefont {Klag}, \citenamefont {Kolar}, \citenamefont {Leidemann}, \citenamefont {Merkel}, \citenamefont {Mihovilovi\ifmmode~\check{c}\else \v{c}\fi{}}, \citenamefont {M\"uller}, \citenamefont {M\"uller}, \citenamefont {Orlandini}, \citenamefont {Pochodzalla}, \citenamefont {Schlimme}, \citenamefont {Schoth}, \citenamefont {Schulz}, \citenamefont {Sfienti}, \citenamefont {\ifmmode~\check{S}\else \v{S}\fi{}irca}, \citenamefont {Spreckels}, \citenamefont
  {St\"ottinger}, \citenamefont {Thiel}, \citenamefont {Tyukin}, \citenamefont {Walcher},\ and\ \citenamefont {Weber}}]{Kegel}%
  \BibitemOpen
  \bibfield  {author} {\bibinfo {author} {\bibfnamefont {S.}~\bibnamefont {Kegel}}, \bibinfo {author} {\bibfnamefont {P.}~\bibnamefont {Achenbach}}, \bibinfo {author} {\bibfnamefont {S.}~\bibnamefont {Bacca}}, \bibinfo {author} {\bibfnamefont {N.}~\bibnamefont {Barnea}}, \bibinfo {author} {\bibfnamefont {J.}~\bibnamefont {Beri\ifmmode \check{c}\else \v{c}\fi{}i\ifmmode~\check{c}\else \v{c}\fi{}}}, \bibinfo {author} {\bibfnamefont {D.}~\bibnamefont {Bosnar}}, \bibinfo {author} {\bibfnamefont {L.}~\bibnamefont {Correa}}, \bibinfo {author} {\bibfnamefont {M.~O.}\ \bibnamefont {Distler}}, \bibinfo {author} {\bibfnamefont {A.}~\bibnamefont {Esser}}, \bibinfo {author} {\bibfnamefont {H.}~\bibnamefont {Fonvieille}}, \bibinfo {author} {\bibfnamefont {I.}~\bibnamefont {Fri\ifmmode \check{s}\else \v{s}\fi{}\ifmmode \check{c}\else \v{c}\fi{}i\ifmmode~\acute{c}\else \'{c}\fi{}}}, \bibinfo {author} {\bibfnamefont {M.}~\bibnamefont {Heilig}}, \bibinfo {author} {\bibfnamefont {P.}~\bibnamefont {Herrmann}}, \bibinfo {author}
  {\bibfnamefont {M.}~\bibnamefont {Hoek}}, \bibinfo {author} {\bibfnamefont {P.}~\bibnamefont {Klag}}, \bibinfo {author} {\bibfnamefont {T.}~\bibnamefont {Kolar}}, \bibinfo {author} {\bibfnamefont {W.}~\bibnamefont {Leidemann}}, \bibinfo {author} {\bibfnamefont {H.}~\bibnamefont {Merkel}}, \bibinfo {author} {\bibfnamefont {M.}~\bibnamefont {Mihovilovi\ifmmode~\check{c}\else \v{c}\fi{}}}, \bibinfo {author} {\bibfnamefont {J.}~\bibnamefont {M\"uller}}, \bibinfo {author} {\bibfnamefont {U.}~\bibnamefont {M\"uller}}, \bibinfo {author} {\bibfnamefont {G.}~\bibnamefont {Orlandini}}, \bibinfo {author} {\bibfnamefont {J.}~\bibnamefont {Pochodzalla}}, \bibinfo {author} {\bibfnamefont {B.~S.}\ \bibnamefont {Schlimme}}, \bibinfo {author} {\bibfnamefont {M.}~\bibnamefont {Schoth}}, \bibinfo {author} {\bibfnamefont {F.}~\bibnamefont {Schulz}}, \bibinfo {author} {\bibfnamefont {C.}~\bibnamefont {Sfienti}}, \bibinfo {author} {\bibfnamefont {S.}~\bibnamefont {\ifmmode~\check{S}\else \v{S}\fi{}irca}}, \bibinfo {author}
  {\bibfnamefont {R.}~\bibnamefont {Spreckels}}, \bibinfo {author} {\bibfnamefont {Y.}~\bibnamefont {St\"ottinger}}, \bibinfo {author} {\bibfnamefont {M.}~\bibnamefont {Thiel}}, \bibinfo {author} {\bibfnamefont {A.}~\bibnamefont {Tyukin}}, \bibinfo {author} {\bibfnamefont {T.}~\bibnamefont {Walcher}}, \ and\ \bibinfo {author} {\bibfnamefont {A.}~\bibnamefont {Weber}},\ }\bibfield  {title} {\enquote {\bibinfo {title} {Measurement of the $\ensuremath{\alpha}$-particle monopole transition form factor challenges theory: A low-energy puzzle for nuclear forces?}}\ }\href {\doibase 10.1103/PhysRevLett.130.152502} {\bibfield  {journal} {\bibinfo  {journal} {Phys. Rev. Lett.}\ }\textbf {\bibinfo {volume} {130}},\ \bibinfo {pages} {152502} (\bibinfo {year} {2023})}\BibitemShut {NoStop}%
\bibitem [{\citenamefont {Acharya}\ \emph {et~al.}(2023)\citenamefont {Acharya}, \citenamefont {Bacca}, \citenamefont {Bonaiti}, \citenamefont {Li~Muli},\ and\ \citenamefont {Sobczyk}}]{acharya2023}%
  \BibitemOpen
  \bibfield  {author} {\bibinfo {author} {\bibfnamefont {Bijaya}\ \bibnamefont {Acharya}}, \bibinfo {author} {\bibfnamefont {Sonia}\ \bibnamefont {Bacca}}, \bibinfo {author} {\bibfnamefont {Francesca}\ \bibnamefont {Bonaiti}}, \bibinfo {author} {\bibfnamefont {Simone~Salvatore}\ \bibnamefont {Li~Muli}}, \ and\ \bibinfo {author} {\bibfnamefont {Joanna~E.}\ \bibnamefont {Sobczyk}},\ }\bibfield  {title} {\enquote {\bibinfo {title} {Uncertainty quantification in electromagnetic observables of nuclei},}\ }\href {\doibase 10.3389/fphy.2022.1066035} {\bibfield  {journal} {\bibinfo  {journal} {Frontiers in Physics}\ }\textbf {\bibinfo {volume} {10}} (\bibinfo {year} {2023}),\ 10.3389/fphy.2022.1066035}\BibitemShut {NoStop}%
\bibitem [{\citenamefont {Stetcu}\ \emph {et~al.}(2007)\citenamefont {Stetcu}, \citenamefont {Quaglioni}, \citenamefont {Bacca}, \citenamefont {Barrett}, \citenamefont {Johnson}, \citenamefont {Navrátil}, \citenamefont {Barnea}, \citenamefont {Leidemann},\ and\ \citenamefont {Orlandini}}]{stetcu2007}%
  \BibitemOpen
  \bibfield  {author} {\bibinfo {author} {\bibfnamefont {Ionel}\ \bibnamefont {Stetcu}}, \bibinfo {author} {\bibfnamefont {Sofia}\ \bibnamefont {Quaglioni}}, \bibinfo {author} {\bibfnamefont {Sonia}\ \bibnamefont {Bacca}}, \bibinfo {author} {\bibfnamefont {Bruce~R.}\ \bibnamefont {Barrett}}, \bibinfo {author} {\bibfnamefont {Calvin~W.}\ \bibnamefont {Johnson}}, \bibinfo {author} {\bibfnamefont {Petr}\ \bibnamefont {Navrátil}}, \bibinfo {author} {\bibfnamefont {Nir}\ \bibnamefont {Barnea}}, \bibinfo {author} {\bibfnamefont {Winfried}\ \bibnamefont {Leidemann}}, \ and\ \bibinfo {author} {\bibfnamefont {Giuseppina}\ \bibnamefont {Orlandini}},\ }\bibfield  {title} {\enquote {\bibinfo {title} {Benchmark calculation of inclusive electromagnetic responses in the four-body nuclear system},}\ }\href {\doibase https://doi.org/10.1016/j.nuclphysa.2006.12.047} {\bibfield  {journal} {\bibinfo  {journal} {Nuclear Physics A}\ }\textbf {\bibinfo {volume} {785}},\ \bibinfo {pages} {307--321} (\bibinfo {year}
  {2007})}\BibitemShut {NoStop}%
\bibitem [{\citenamefont {Quaglioni}\ and\ \citenamefont {Navrátil}(2007)}]{quaglioni2007}%
  \BibitemOpen
  \bibfield  {author} {\bibinfo {author} {\bibfnamefont {Sofia}\ \bibnamefont {Quaglioni}}\ and\ \bibinfo {author} {\bibfnamefont {Petr}\ \bibnamefont {Navrátil}},\ }\bibfield  {title} {\enquote {\bibinfo {title} {The 4he total photo-absorption cross section with two- plus three-nucleon interactions from chiral effective field theory},}\ }\href {\doibase https://doi.org/10.1016/j.physletb.2007.06.082} {\bibfield  {journal} {\bibinfo  {journal} {Physics Letters B}\ }\textbf {\bibinfo {volume} {652}},\ \bibinfo {pages} {370--375} (\bibinfo {year} {2007})}\BibitemShut {NoStop}%
\bibitem [{\citenamefont {Stumpf}\ \emph {et~al.}(2017)\citenamefont {Stumpf}, \citenamefont {Wolfgruber},\ and\ \citenamefont {Roth}}]{stumpf2017}%
  \BibitemOpen
  \bibfield  {author} {\bibinfo {author} {\bibfnamefont {Christina}\ \bibnamefont {Stumpf}}, \bibinfo {author} {\bibfnamefont {Tobias}\ \bibnamefont {Wolfgruber}}, \ and\ \bibinfo {author} {\bibfnamefont {Robert}\ \bibnamefont {Roth}},\ }\bibfield  {title} {\enquote {\bibinfo {title} {{Electromagnetic Strength Distributions from the Ab Initio No-Core Shell Model}},}\ }\href@noop {} {\  (\bibinfo {year} {2017})},\ \Eprint {http://arxiv.org/abs/1709.06840} {arXiv:1709.06840 [nucl-th]} \BibitemShut {NoStop}%
\bibitem [{\citenamefont {Baker}\ \emph {et~al.}(2020)\citenamefont {Baker}, \citenamefont {Launey}, \citenamefont {Bacca}, \citenamefont {Dinur},\ and\ \citenamefont {Dytrych}}]{baker2020}%
  \BibitemOpen
  \bibfield  {author} {\bibinfo {author} {\bibfnamefont {R.~B.}\ \bibnamefont {Baker}}, \bibinfo {author} {\bibfnamefont {K.~D.}\ \bibnamefont {Launey}}, \bibinfo {author} {\bibfnamefont {S.}~\bibnamefont {Bacca}}, \bibinfo {author} {\bibfnamefont {N.~Nevo}\ \bibnamefont {Dinur}}, \ and\ \bibinfo {author} {\bibfnamefont {T.}~\bibnamefont {Dytrych}},\ }\bibfield  {title} {\enquote {\bibinfo {title} {Benchmark calculations of electromagnetic sum rules with a symmetry-adapted basis and hyperspherical harmonics},}\ }\href {\doibase 10.1103/PhysRevC.102.014320} {\bibfield  {journal} {\bibinfo  {journal} {Phys. Rev. C}\ }\textbf {\bibinfo {volume} {102}},\ \bibinfo {pages} {014320} (\bibinfo {year} {2020})}\BibitemShut {NoStop}%
\bibitem [{\citenamefont {Coester}(1958)}]{coester1958}%
  \BibitemOpen
  \bibfield  {author} {\bibinfo {author} {\bibfnamefont {F.}~\bibnamefont {Coester}},\ }\bibfield  {title} {\enquote {\bibinfo {title} {Bound states of a many-particle system},}\ }\href {\doibase https://doi.org/10.1016/0029-5582(58)90280-3} {\bibfield  {journal} {\bibinfo  {journal} {Nuclear Physics}\ }\textbf {\bibinfo {volume} {7}},\ \bibinfo {pages} {421--424} (\bibinfo {year} {1958})}\BibitemShut {NoStop}%
\bibitem [{\citenamefont {Coester}\ and\ \citenamefont {Kümmel}(1960)}]{coester1960}%
  \BibitemOpen
  \bibfield  {author} {\bibinfo {author} {\bibfnamefont {F.}~\bibnamefont {Coester}}\ and\ \bibinfo {author} {\bibfnamefont {H.}~\bibnamefont {Kümmel}},\ }\bibfield  {title} {\enquote {\bibinfo {title} {Short-range correlations in nuclear wave functions},}\ }\href {\doibase https://doi.org/10.1016/0029-5582(60)90140-1} {\bibfield  {journal} {\bibinfo  {journal} {Nuclear Physics}\ }\textbf {\bibinfo {volume} {17}},\ \bibinfo {pages} {477--485} (\bibinfo {year} {1960})}\BibitemShut {NoStop}%
\bibitem [{\citenamefont {Dean}\ and\ \citenamefont {Hjorth-Jensen}(2004)}]{dean2004}%
  \BibitemOpen
  \bibfield  {author} {\bibinfo {author} {\bibfnamefont {D.~J.}\ \bibnamefont {Dean}}\ and\ \bibinfo {author} {\bibfnamefont {M.}~\bibnamefont {Hjorth-Jensen}},\ }\bibfield  {title} {\enquote {\bibinfo {title} {Coupled-cluster approach to nuclear physics},}\ }\href {\doibase 10.1103/PhysRevC.69.054320} {\bibfield  {journal} {\bibinfo  {journal} {Phys. Rev. C}\ }\textbf {\bibinfo {volume} {69}},\ \bibinfo {pages} {054320} (\bibinfo {year} {2004})}\BibitemShut {NoStop}%
\bibitem [{\citenamefont {W\l{}och}\ \emph {et~al.}(2005)\citenamefont {W\l{}och}, \citenamefont {Dean}, \citenamefont {Gour}, \citenamefont {Hjorth-Jensen}, \citenamefont {Kowalski}, \citenamefont {Papenbrock},\ and\ \citenamefont {Piecuch}}]{wloch2005}%
  \BibitemOpen
  \bibfield  {author} {\bibinfo {author} {\bibfnamefont {M.}~\bibnamefont {W\l{}och}}, \bibinfo {author} {\bibfnamefont {D.~J.}\ \bibnamefont {Dean}}, \bibinfo {author} {\bibfnamefont {J.~R.}\ \bibnamefont {Gour}}, \bibinfo {author} {\bibfnamefont {M.}~\bibnamefont {Hjorth-Jensen}}, \bibinfo {author} {\bibfnamefont {K.}~\bibnamefont {Kowalski}}, \bibinfo {author} {\bibfnamefont {T.}~\bibnamefont {Papenbrock}}, \ and\ \bibinfo {author} {\bibfnamefont {P.}~\bibnamefont {Piecuch}},\ }\bibfield  {title} {\enquote {\bibinfo {title} {Ab-initio coupled-cluster study of $^{16}\mathrm{O}$},}\ }\href {\doibase 10.1103/PhysRevLett.94.212501} {\bibfield  {journal} {\bibinfo  {journal} {Phys. Rev. Lett.}\ }\textbf {\bibinfo {volume} {94}},\ \bibinfo {pages} {212501} (\bibinfo {year} {2005})}\BibitemShut {NoStop}%
\bibitem [{\citenamefont {Hagen}\ \emph {et~al.}(2008)\citenamefont {Hagen}, \citenamefont {Papenbrock}, \citenamefont {Dean},\ and\ \citenamefont {Hjorth-Jensen}}]{hagen2008}%
  \BibitemOpen
  \bibfield  {author} {\bibinfo {author} {\bibfnamefont {G.}~\bibnamefont {Hagen}}, \bibinfo {author} {\bibfnamefont {T.}~\bibnamefont {Papenbrock}}, \bibinfo {author} {\bibfnamefont {D.~J.}\ \bibnamefont {Dean}}, \ and\ \bibinfo {author} {\bibfnamefont {M.}~\bibnamefont {Hjorth-Jensen}},\ }\bibfield  {title} {\enquote {\bibinfo {title} {Medium-mass nuclei from chiral nucleon-nucleon interactions},}\ }\href {\doibase 10.1103/PhysRevLett.101.092502} {\bibfield  {journal} {\bibinfo  {journal} {Phys. Rev. Lett.}\ }\textbf {\bibinfo {volume} {101}},\ \bibinfo {pages} {092502} (\bibinfo {year} {2008})}\BibitemShut {NoStop}%
\bibitem [{\citenamefont {Hagen}\ \emph {et~al.}(2010)\citenamefont {Hagen}, \citenamefont {Papenbrock}, \citenamefont {Dean},\ and\ \citenamefont {Hjorth-Jensen}}]{hagen2010}%
  \BibitemOpen
  \bibfield  {author} {\bibinfo {author} {\bibfnamefont {G.}~\bibnamefont {Hagen}}, \bibinfo {author} {\bibfnamefont {T.}~\bibnamefont {Papenbrock}}, \bibinfo {author} {\bibfnamefont {D.~J.}\ \bibnamefont {Dean}}, \ and\ \bibinfo {author} {\bibfnamefont {M.}~\bibnamefont {Hjorth-Jensen}},\ }\bibfield  {title} {\enquote {\bibinfo {title} {Ab initio coupled-cluster approach to nuclear structure with modern nucleon-nucleon interactions},}\ }\href {\doibase 10.1103/PhysRevC.82.034330} {\bibfield  {journal} {\bibinfo  {journal} {Phys. Rev. C}\ }\textbf {\bibinfo {volume} {82}},\ \bibinfo {pages} {034330} (\bibinfo {year} {2010})}\BibitemShut {NoStop}%
\bibitem [{\citenamefont {Binder}\ \emph {et~al.}(2014)\citenamefont {Binder}, \citenamefont {Langhammer}, \citenamefont {Calci},\ and\ \citenamefont {Roth}}]{binder2014}%
  \BibitemOpen
  \bibfield  {author} {\bibinfo {author} {\bibfnamefont {Sven}\ \bibnamefont {Binder}}, \bibinfo {author} {\bibfnamefont {Joachim}\ \bibnamefont {Langhammer}}, \bibinfo {author} {\bibfnamefont {Angelo}\ \bibnamefont {Calci}}, \ and\ \bibinfo {author} {\bibfnamefont {Robert}\ \bibnamefont {Roth}},\ }\bibfield  {title} {\enquote {\bibinfo {title} {Ab initio path to heavy nuclei},}\ }\href {\doibase https://doi.org/10.1016/j.physletb.2014.07.010} {\bibfield  {journal} {\bibinfo  {journal} {Physics Letters B}\ }\textbf {\bibinfo {volume} {736}},\ \bibinfo {pages} {119--123} (\bibinfo {year} {2014})}\BibitemShut {NoStop}%
\bibitem [{\citenamefont {Hagen}\ \emph {et~al.}(2014)\citenamefont {Hagen}, \citenamefont {Papenbrock}, \citenamefont {Hjorth-Jensen},\ and\ \citenamefont {Dean}}]{hagen2014}%
  \BibitemOpen
  \bibfield  {author} {\bibinfo {author} {\bibfnamefont {G}~\bibnamefont {Hagen}}, \bibinfo {author} {\bibfnamefont {T}~\bibnamefont {Papenbrock}}, \bibinfo {author} {\bibfnamefont {M}~\bibnamefont {Hjorth-Jensen}}, \ and\ \bibinfo {author} {\bibfnamefont {D~J}\ \bibnamefont {Dean}},\ }\bibfield  {title} {\enquote {\bibinfo {title} {Coupled-cluster computations of atomic nuclei},}\ }\href {\doibase 10.1088/0034-4885/77/9/096302} {\bibfield  {journal} {\bibinfo  {journal} {Reports on Progress in Physics}\ }\textbf {\bibinfo {volume} {77}},\ \bibinfo {pages} {096302} (\bibinfo {year} {2014})}\BibitemShut {NoStop}%
\bibitem [{\citenamefont {Efros}\ \emph {et~al.}(2007)\citenamefont {Efros}, \citenamefont {Leidemann}, \citenamefont {Orlandini},\ and\ \citenamefont {Barnea}}]{efros2007}%
  \BibitemOpen
  \bibfield  {author} {\bibinfo {author} {\bibfnamefont {V~D}\ \bibnamefont {Efros}}, \bibinfo {author} {\bibfnamefont {W}~\bibnamefont {Leidemann}}, \bibinfo {author} {\bibfnamefont {G}~\bibnamefont {Orlandini}}, \ and\ \bibinfo {author} {\bibfnamefont {N}~\bibnamefont {Barnea}},\ }\bibfield  {title} {\enquote {\bibinfo {title} {The lorentz integral transform (lit) method and its applications to perturbation-induced reactions},}\ }\href {\doibase 10.1088/0954-3899/34/12/R02} {\bibfield  {journal} {\bibinfo  {journal} {Journal of Physics G: Nuclear and Particle Physics}\ }\textbf {\bibinfo {volume} {34}},\ \bibinfo {pages} {R459} (\bibinfo {year} {2007})}\BibitemShut {NoStop}%
\bibitem [{\citenamefont {Bacca}\ \emph {et~al.}(2013{\natexlab{b}})\citenamefont {Bacca}, \citenamefont {Barnea}, \citenamefont {Hagen}, \citenamefont {Orlandini},\ and\ \citenamefont {Papenbrock}}]{bacca2013}%
  \BibitemOpen
  \bibfield  {author} {\bibinfo {author} {\bibfnamefont {S.}~\bibnamefont {Bacca}}, \bibinfo {author} {\bibfnamefont {N.}~\bibnamefont {Barnea}}, \bibinfo {author} {\bibfnamefont {G.}~\bibnamefont {Hagen}}, \bibinfo {author} {\bibfnamefont {G.}~\bibnamefont {Orlandini}}, \ and\ \bibinfo {author} {\bibfnamefont {T.}~\bibnamefont {Papenbrock}},\ }\bibfield  {title} {\enquote {\bibinfo {title} {First principles description of the giant dipole resonance in $^{16}\mathbf{O}$},}\ }\href {\doibase 10.1103/PhysRevLett.111.122502} {\bibfield  {journal} {\bibinfo  {journal} {Phys. Rev. Lett.}\ }\textbf {\bibinfo {volume} {111}},\ \bibinfo {pages} {122502} (\bibinfo {year} {2013}{\natexlab{b}})}\BibitemShut {NoStop}%
\bibitem [{\citenamefont {Bacca}\ \emph {et~al.}(2014)\citenamefont {Bacca}, \citenamefont {Barnea}, \citenamefont {Hagen}, \citenamefont {Miorelli}, \citenamefont {Orlandini},\ and\ \citenamefont {Papenbrock}}]{bacca2014}%
  \BibitemOpen
  \bibfield  {author} {\bibinfo {author} {\bibfnamefont {S.}~\bibnamefont {Bacca}}, \bibinfo {author} {\bibfnamefont {N.}~\bibnamefont {Barnea}}, \bibinfo {author} {\bibfnamefont {G.}~\bibnamefont {Hagen}}, \bibinfo {author} {\bibfnamefont {M.}~\bibnamefont {Miorelli}}, \bibinfo {author} {\bibfnamefont {G.}~\bibnamefont {Orlandini}}, \ and\ \bibinfo {author} {\bibfnamefont {T.}~\bibnamefont {Papenbrock}},\ }\bibfield  {title} {\enquote {\bibinfo {title} {Giant and pigmy dipole resonances in $^{4}\mathrm{He}$, $^{16,22}\mathrm{O}$, and $^{40}\mathrm{Ca}$ from chiral nucleon-nucleon interactions},}\ }\href {\doibase 10.1103/PhysRevC.90.064619} {\bibfield  {journal} {\bibinfo  {journal} {Phys. Rev. C}\ }\textbf {\bibinfo {volume} {90}},\ \bibinfo {pages} {064619} (\bibinfo {year} {2014})}\BibitemShut {NoStop}%
\bibitem [{\citenamefont {Miorelli}\ \emph {et~al.}(2016)\citenamefont {Miorelli}, \citenamefont {Bacca}, \citenamefont {Barnea}, \citenamefont {Hagen}, \citenamefont {Jansen}, \citenamefont {Orlandini},\ and\ \citenamefont {Papenbrock}}]{miorelli2016}%
  \BibitemOpen
  \bibfield  {author} {\bibinfo {author} {\bibfnamefont {M.}~\bibnamefont {Miorelli}}, \bibinfo {author} {\bibfnamefont {S.}~\bibnamefont {Bacca}}, \bibinfo {author} {\bibfnamefont {N.}~\bibnamefont {Barnea}}, \bibinfo {author} {\bibfnamefont {G.}~\bibnamefont {Hagen}}, \bibinfo {author} {\bibfnamefont {G.~R.}\ \bibnamefont {Jansen}}, \bibinfo {author} {\bibfnamefont {G.}~\bibnamefont {Orlandini}}, \ and\ \bibinfo {author} {\bibfnamefont {T.}~\bibnamefont {Papenbrock}},\ }\bibfield  {title} {\enquote {\bibinfo {title} {Electric dipole polarizability from first principles calculations},}\ }\href {\doibase 10.1103/PhysRevC.94.034317} {\bibfield  {journal} {\bibinfo  {journal} {Phys. Rev. C}\ }\textbf {\bibinfo {volume} {94}},\ \bibinfo {pages} {034317} (\bibinfo {year} {2016})}\BibitemShut {NoStop}%
\bibitem [{\citenamefont {Miorelli}\ \emph {et~al.}(2018)\citenamefont {Miorelli}, \citenamefont {Bacca}, \citenamefont {Hagen},\ and\ \citenamefont {Papenbrock}}]{miorelli2018}%
  \BibitemOpen
  \bibfield  {author} {\bibinfo {author} {\bibfnamefont {M.}~\bibnamefont {Miorelli}}, \bibinfo {author} {\bibfnamefont {S.}~\bibnamefont {Bacca}}, \bibinfo {author} {\bibfnamefont {G.}~\bibnamefont {Hagen}}, \ and\ \bibinfo {author} {\bibfnamefont {T.}~\bibnamefont {Papenbrock}},\ }\bibfield  {title} {\enquote {\bibinfo {title} {Computing the dipole polarizability of $^{48}\mathrm{Ca}$ with increased precision},}\ }\href {\doibase 10.1103/PhysRevC.98.014324} {\bibfield  {journal} {\bibinfo  {journal} {Phys. Rev. C}\ }\textbf {\bibinfo {volume} {98}},\ \bibinfo {pages} {014324} (\bibinfo {year} {2018})}\BibitemShut {NoStop}%
\bibitem [{\citenamefont {Simonis}\ \emph {et~al.}(2019)\citenamefont {Simonis}, \citenamefont {Bacca},\ and\ \citenamefont {Hagen}}]{simonis2019}%
  \BibitemOpen
  \bibfield  {author} {\bibinfo {author} {\bibfnamefont {J.}~\bibnamefont {Simonis}}, \bibinfo {author} {\bibfnamefont {S.}~\bibnamefont {Bacca}}, \ and\ \bibinfo {author} {\bibfnamefont {G.}~\bibnamefont {Hagen}},\ }\bibfield  {title} {\enquote {\bibinfo {title} {First principles electromagnetic responses in medium-mass nuclei - recent progress from coupled-cluster theory},}\ }\href {\doibase 10.1140/epja/i2019-12825-0} {\bibfield  {journal} {\bibinfo  {journal} {Eur. Phys. J. A}\ }\textbf {\bibinfo {volume} {55}},\ \bibinfo {pages} {241} (\bibinfo {year} {2019})}\BibitemShut {NoStop}%
\bibitem [{\citenamefont {Sobczyk}\ \emph {et~al.}(2021)\citenamefont {Sobczyk}, \citenamefont {Acharya}, \citenamefont {Bacca},\ and\ \citenamefont {Hagen}}]{sobczyk2021}%
  \BibitemOpen
  \bibfield  {author} {\bibinfo {author} {\bibfnamefont {J.~E.}\ \bibnamefont {Sobczyk}}, \bibinfo {author} {\bibfnamefont {B.}~\bibnamefont {Acharya}}, \bibinfo {author} {\bibfnamefont {S.}~\bibnamefont {Bacca}}, \ and\ \bibinfo {author} {\bibfnamefont {G.}~\bibnamefont {Hagen}},\ }\bibfield  {title} {\enquote {\bibinfo {title} {{Ab initio computation of the longitudinal response function in $^{40}$Ca}},}\ }\href {\doibase 10.1103/PhysRevLett.127.072501} {\bibfield  {journal} {\bibinfo  {journal} {Phys. Rev. Lett.}\ }\textbf {\bibinfo {volume} {127}},\ \bibinfo {pages} {072501} (\bibinfo {year} {2021})},\ \Eprint {http://arxiv.org/abs/2103.06786} {arXiv:2103.06786 [nucl-th]} \BibitemShut {NoStop}%
\bibitem [{\citenamefont {Sobczyk}\ \emph {et~al.}(2024)\citenamefont {Sobczyk}, \citenamefont {Acharya}, \citenamefont {Bacca},\ and\ \citenamefont {Hagen}}]{sobczyk2023}%
  \BibitemOpen
  \bibfield  {author} {\bibinfo {author} {\bibfnamefont {J.~E.}\ \bibnamefont {Sobczyk}}, \bibinfo {author} {\bibfnamefont {B.}~\bibnamefont {Acharya}}, \bibinfo {author} {\bibfnamefont {S.}~\bibnamefont {Bacca}}, \ and\ \bibinfo {author} {\bibfnamefont {G.}~\bibnamefont {Hagen}},\ }\bibfield  {title} {\enquote {\bibinfo {title} {$^{40}\mathrm{Ca}$ transverse response function from coupled-cluster theory},}\ }\href {\doibase 10.1103/PhysRevC.109.025502} {\bibfield  {journal} {\bibinfo  {journal} {Phys. Rev. C}\ }\textbf {\bibinfo {volume} {109}},\ \bibinfo {pages} {025502} (\bibinfo {year} {2024})}\BibitemShut {NoStop}%
\bibitem [{\citenamefont {Raimondi}\ and\ \citenamefont {Barbieri}(2019)}]{raimondi2019}%
  \BibitemOpen
  \bibfield  {author} {\bibinfo {author} {\bibfnamefont {Francesco}\ \bibnamefont {Raimondi}}\ and\ \bibinfo {author} {\bibfnamefont {Carlo}\ \bibnamefont {Barbieri}},\ }\bibfield  {title} {\enquote {\bibinfo {title} {Nuclear electromagnetic dipole response with the self-consistent green's function formalism},}\ }\href {\doibase 10.1103/PhysRevC.99.054327} {\bibfield  {journal} {\bibinfo  {journal} {Phys. Rev. C}\ }\textbf {\bibinfo {volume} {99}},\ \bibinfo {pages} {054327} (\bibinfo {year} {2019})}\BibitemShut {NoStop}%
\bibitem [{\citenamefont {Burrows}\ \emph {et~al.}(2023)\citenamefont {Burrows}, \citenamefont {Baker}, \citenamefont {Bacca}, \citenamefont {Launey}, \citenamefont {Dytrych},\ and\ \citenamefont {Langr}}]{burrows2023}%
  \BibitemOpen
  \bibfield  {author} {\bibinfo {author} {\bibfnamefont {M.}~\bibnamefont {Burrows}}, \bibinfo {author} {\bibfnamefont {R.~B.}\ \bibnamefont {Baker}}, \bibinfo {author} {\bibfnamefont {S.}~\bibnamefont {Bacca}}, \bibinfo {author} {\bibfnamefont {K.~D.}\ \bibnamefont {Launey}}, \bibinfo {author} {\bibfnamefont {T.}~\bibnamefont {Dytrych}}, \ and\ \bibinfo {author} {\bibfnamefont {D.}~\bibnamefont {Langr}},\ }\bibfield  {title} {\enquote {\bibinfo {title} {{Response functions and giant monopole resonances for light to medium-mass nuclei from the \textbackslash{}textit{ab initio} symmetry-adapted no-core shell model}},}\ }\href@noop {} {\  (\bibinfo {year} {2023})},\ \Eprint {http://arxiv.org/abs/2312.09782} {arXiv:2312.09782 [nucl-th]} \BibitemShut {NoStop}%
\bibitem [{\citenamefont {Beaujeault-Taudi\`ere}\ \emph {et~al.}(2023)\citenamefont {Beaujeault-Taudi\`ere}, \citenamefont {Frosini}, \citenamefont {Ebran}, \citenamefont {Duguet}, \citenamefont {Roth},\ and\ \citenamefont {Som\`a}}]{taudiere2023}%
  \BibitemOpen
  \bibfield  {author} {\bibinfo {author} {\bibfnamefont {Y.}~\bibnamefont {Beaujeault-Taudi\`ere}}, \bibinfo {author} {\bibfnamefont {M.}~\bibnamefont {Frosini}}, \bibinfo {author} {\bibfnamefont {J.-P.}\ \bibnamefont {Ebran}}, \bibinfo {author} {\bibfnamefont {T.}~\bibnamefont {Duguet}}, \bibinfo {author} {\bibfnamefont {R.}~\bibnamefont {Roth}}, \ and\ \bibinfo {author} {\bibfnamefont {V.}~\bibnamefont {Som\`a}},\ }\bibfield  {title} {\enquote {\bibinfo {title} {Zero- and finite-temperature electromagnetic strength distributions in closed- and open-shell nuclei from first principles},}\ }\href {\doibase 10.1103/PhysRevC.107.L021302} {\bibfield  {journal} {\bibinfo  {journal} {Phys. Rev. C}\ }\textbf {\bibinfo {volume} {107}},\ \bibinfo {pages} {L021302} (\bibinfo {year} {2023})}\BibitemShut {NoStop}%
\bibitem [{\citenamefont {Porro}\ \emph {et~al.}(2024{\natexlab{a}})\citenamefont {Porro}, \citenamefont {Duguet}, \citenamefont {Ebran}, \citenamefont {Frosini}, \citenamefont {Roth},\ and\ \citenamefont {Som{\`a}}}]{porro2024a}%
  \BibitemOpen
  \bibfield  {author} {\bibinfo {author} {\bibfnamefont {A}~\bibnamefont {Porro}}, \bibinfo {author} {\bibfnamefont {T}~\bibnamefont {Duguet}}, \bibinfo {author} {\bibfnamefont {J-P}\ \bibnamefont {Ebran}}, \bibinfo {author} {\bibfnamefont {M}~\bibnamefont {Frosini}}, \bibinfo {author} {\bibfnamefont {R}~\bibnamefont {Roth}}, \ and\ \bibinfo {author} {\bibfnamefont {V}~\bibnamefont {Som{\`a}}},\ }\bibfield  {title} {\enquote {\bibinfo {title} {Ab initio description of monopole resonances in light- and medium-mass nuclei},}\ }\href@noop {} {\bibfield  {journal} {\bibinfo  {journal} {The European Physical Journal A}\ }\textbf {\bibinfo {volume} {60}},\ \bibinfo {pages} {133} (\bibinfo {year} {2024}{\natexlab{a}})}\BibitemShut {NoStop}%
\bibitem [{\citenamefont {Porro}\ \emph {et~al.}(2024{\natexlab{b}})\citenamefont {Porro}, \citenamefont {Duguet}, \citenamefont {Ebran}, \citenamefont {Frosini}, \citenamefont {Roth},\ and\ \citenamefont {Som{\`a}}}]{porro2024b}%
  \BibitemOpen
  \bibfield  {author} {\bibinfo {author} {\bibfnamefont {A}~\bibnamefont {Porro}}, \bibinfo {author} {\bibfnamefont {T}~\bibnamefont {Duguet}}, \bibinfo {author} {\bibfnamefont {J-P}\ \bibnamefont {Ebran}}, \bibinfo {author} {\bibfnamefont {M}~\bibnamefont {Frosini}}, \bibinfo {author} {\bibfnamefont {R}~\bibnamefont {Roth}}, \ and\ \bibinfo {author} {\bibfnamefont {V}~\bibnamefont {Som{\`a}}},\ }\bibfield  {title} {\enquote {\bibinfo {title} {Ab initio description of monopole resonances in light- and medium-mass nuclei},}\ }\href@noop {} {\bibfield  {journal} {\bibinfo  {journal} {The European Physical Journal A}\ }\textbf {\bibinfo {volume} {60}},\ \bibinfo {pages} {134} (\bibinfo {year} {2024}{\natexlab{b}})}\BibitemShut {NoStop}%
\bibitem [{\citenamefont {Roca-Maza}\ \emph {et~al.}(2013)\citenamefont {Roca-Maza}, \citenamefont {Centelles}, \citenamefont {Vi\~nas}, \citenamefont {Brenna}, \citenamefont {Col\`o}, \citenamefont {Agrawal}, \citenamefont {Paar}, \citenamefont {Piekarewicz},\ and\ \citenamefont {Vretenar}}]{rocamaza2013}%
  \BibitemOpen
  \bibfield  {author} {\bibinfo {author} {\bibfnamefont {X.}~\bibnamefont {Roca-Maza}}, \bibinfo {author} {\bibfnamefont {M.}~\bibnamefont {Centelles}}, \bibinfo {author} {\bibfnamefont {X.}~\bibnamefont {Vi\~nas}}, \bibinfo {author} {\bibfnamefont {M.}~\bibnamefont {Brenna}}, \bibinfo {author} {\bibfnamefont {G.}~\bibnamefont {Col\`o}}, \bibinfo {author} {\bibfnamefont {B.~K.}\ \bibnamefont {Agrawal}}, \bibinfo {author} {\bibfnamefont {N.}~\bibnamefont {Paar}}, \bibinfo {author} {\bibfnamefont {J.}~\bibnamefont {Piekarewicz}}, \ and\ \bibinfo {author} {\bibfnamefont {D.}~\bibnamefont {Vretenar}},\ }\bibfield  {title} {\enquote {\bibinfo {title} {{Electric dipole polarizability in $^{208}Pb$: Insights from the droplet model}},}\ }\href {\doibase 10.1103/PhysRevC.88.024316} {\bibfield  {journal} {\bibinfo  {journal} {Phys. Rev. C}\ }\textbf {\bibinfo {volume} {88}},\ \bibinfo {pages} {024316} (\bibinfo {year} {2013})},\ \Eprint {http://arxiv.org/abs/1307.4806} {arXiv:1307.4806 [nucl-th]} \BibitemShut {NoStop}%
\bibitem [{\citenamefont {Roca-Maza}\ \emph {et~al.}(2015)\citenamefont {Roca-Maza}, \citenamefont {Vi\~nas}, \citenamefont {Centelles}, \citenamefont {Agrawal}, \citenamefont {Col\`o}, \citenamefont {Paar}, \citenamefont {Piekarewicz},\ and\ \citenamefont {Vretenar}}]{rocamaza2015}%
  \BibitemOpen
  \bibfield  {author} {\bibinfo {author} {\bibfnamefont {X.}~\bibnamefont {Roca-Maza}}, \bibinfo {author} {\bibfnamefont {X.}~\bibnamefont {Vi\~nas}}, \bibinfo {author} {\bibfnamefont {M.}~\bibnamefont {Centelles}}, \bibinfo {author} {\bibfnamefont {B.~K.}\ \bibnamefont {Agrawal}}, \bibinfo {author} {\bibfnamefont {G.}~\bibnamefont {Col\`o}}, \bibinfo {author} {\bibfnamefont {N.}~\bibnamefont {Paar}}, \bibinfo {author} {\bibfnamefont {J.}~\bibnamefont {Piekarewicz}}, \ and\ \bibinfo {author} {\bibfnamefont {D.}~\bibnamefont {Vretenar}},\ }\bibfield  {title} {\enquote {\bibinfo {title} {Neutron skin thickness from the measured electric dipole polarizability in $^{68}\text{Ni}$, $^{120}\text{Sn}$, and $^{208}\text{Pb}$},}\ }\href {\doibase 10.1103/PhysRevC.92.064304} {\bibfield  {journal} {\bibinfo  {journal} {Phys. Rev. C}\ }\textbf {\bibinfo {volume} {92}},\ \bibinfo {pages} {064304} (\bibinfo {year} {2015})}\BibitemShut {NoStop}%
\bibitem [{\citenamefont {Piekarewicz}(2021)}]{piekarewicz2021}%
  \BibitemOpen
  \bibfield  {author} {\bibinfo {author} {\bibfnamefont {J.}~\bibnamefont {Piekarewicz}},\ }\bibfield  {title} {\enquote {\bibinfo {title} {Implications of prex-2 on the electric dipole polarizability of neutron-rich nuclei},}\ }\href {\doibase 10.1103/PhysRevC.104.024329} {\bibfield  {journal} {\bibinfo  {journal} {Phys. Rev. C}\ }\textbf {\bibinfo {volume} {104}},\ \bibinfo {pages} {024329} (\bibinfo {year} {2021})}\BibitemShut {NoStop}%
\bibitem [{\citenamefont {Birkhan}\ \emph {et~al.}(2017)\citenamefont {Birkhan}, \citenamefont {Miorelli}, \citenamefont {Bacca}, \citenamefont {Bassauer}, \citenamefont {Bertulani}, \citenamefont {Hagen}, \citenamefont {Matsubara}, \citenamefont {von Neumann-Cosel}, \citenamefont {Papenbrock}, \citenamefont {Pietralla}, \citenamefont {Ponomarev}, \citenamefont {Richter}, \citenamefont {Schwenk},\ and\ \citenamefont {Tamii}}]{birkhan2017}%
  \BibitemOpen
  \bibfield  {author} {\bibinfo {author} {\bibfnamefont {J.}~\bibnamefont {Birkhan}}, \bibinfo {author} {\bibfnamefont {M.}~\bibnamefont {Miorelli}}, \bibinfo {author} {\bibfnamefont {S.}~\bibnamefont {Bacca}}, \bibinfo {author} {\bibfnamefont {S.}~\bibnamefont {Bassauer}}, \bibinfo {author} {\bibfnamefont {C.~A.}\ \bibnamefont {Bertulani}}, \bibinfo {author} {\bibfnamefont {G.}~\bibnamefont {Hagen}}, \bibinfo {author} {\bibfnamefont {H.}~\bibnamefont {Matsubara}}, \bibinfo {author} {\bibfnamefont {P.}~\bibnamefont {von Neumann-Cosel}}, \bibinfo {author} {\bibfnamefont {T.}~\bibnamefont {Papenbrock}}, \bibinfo {author} {\bibfnamefont {N.}~\bibnamefont {Pietralla}}, \bibinfo {author} {\bibfnamefont {V.~Yu.}\ \bibnamefont {Ponomarev}}, \bibinfo {author} {\bibfnamefont {A.}~\bibnamefont {Richter}}, \bibinfo {author} {\bibfnamefont {A.}~\bibnamefont {Schwenk}}, \ and\ \bibinfo {author} {\bibfnamefont {A.}~\bibnamefont {Tamii}},\ }\bibfield  {title} {\enquote {\bibinfo {title} {Electric dipole polarizability of
  $^{48}\mathrm{Ca}$ and implications for the neutron skin},}\ }\href {\doibase 10.1103/PhysRevLett.118.252501} {\bibfield  {journal} {\bibinfo  {journal} {Phys. Rev. Lett.}\ }\textbf {\bibinfo {volume} {118}},\ \bibinfo {pages} {252501} (\bibinfo {year} {2017})}\BibitemShut {NoStop}%
\bibitem [{\citenamefont {Fearick}\ \emph {et~al.}(2023)\citenamefont {Fearick}, \citenamefont {von Neumann-Cosel}, \citenamefont {Bacca}, \citenamefont {Birkhan}, \citenamefont {Bonaiti}, \citenamefont {Brandherm}, \citenamefont {Hagen}, \citenamefont {Matsubara}, \citenamefont {Nazarewicz}, \citenamefont {Pietralla}, \citenamefont {Ponomarev}, \citenamefont {Reinhard}, \citenamefont {Roca-Maza}, \citenamefont {Richter}, \citenamefont {Schwenk}, \citenamefont {Simonis},\ and\ \citenamefont {Tamii}}]{fearick2023}%
  \BibitemOpen
  \bibfield  {author} {\bibinfo {author} {\bibfnamefont {R.~W.}\ \bibnamefont {Fearick}}, \bibinfo {author} {\bibfnamefont {P.}~\bibnamefont {von Neumann-Cosel}}, \bibinfo {author} {\bibfnamefont {S.}~\bibnamefont {Bacca}}, \bibinfo {author} {\bibfnamefont {J.}~\bibnamefont {Birkhan}}, \bibinfo {author} {\bibfnamefont {F.}~\bibnamefont {Bonaiti}}, \bibinfo {author} {\bibfnamefont {I.}~\bibnamefont {Brandherm}}, \bibinfo {author} {\bibfnamefont {G.}~\bibnamefont {Hagen}}, \bibinfo {author} {\bibfnamefont {H.}~\bibnamefont {Matsubara}}, \bibinfo {author} {\bibfnamefont {W.}~\bibnamefont {Nazarewicz}}, \bibinfo {author} {\bibfnamefont {N.}~\bibnamefont {Pietralla}}, \bibinfo {author} {\bibfnamefont {V.~Yu.}\ \bibnamefont {Ponomarev}}, \bibinfo {author} {\bibfnamefont {P.-G.}\ \bibnamefont {Reinhard}}, \bibinfo {author} {\bibfnamefont {X.}~\bibnamefont {Roca-Maza}}, \bibinfo {author} {\bibfnamefont {A.}~\bibnamefont {Richter}}, \bibinfo {author} {\bibfnamefont {A.}~\bibnamefont {Schwenk}}, \bibinfo {author}
  {\bibfnamefont {J.}~\bibnamefont {Simonis}}, \ and\ \bibinfo {author} {\bibfnamefont {A.}~\bibnamefont {Tamii}},\ }\bibfield  {title} {\enquote {\bibinfo {title} {Electric dipole polarizability of $^{40}\mathrm{Ca}$},}\ }\href {\doibase 10.1103/PhysRevResearch.5.L022044} {\bibfield  {journal} {\bibinfo  {journal} {Phys. Rev. Res.}\ }\textbf {\bibinfo {volume} {5}},\ \bibinfo {pages} {L022044} (\bibinfo {year} {2023})}\BibitemShut {NoStop}%
\bibitem [{\citenamefont {Kaufmann}\ \emph {et~al.}(2020)\citenamefont {Kaufmann} \emph {et~al.}}]{kaufmann2020}%
  \BibitemOpen
  \bibfield  {author} {\bibinfo {author} {\bibfnamefont {S.}~\bibnamefont {Kaufmann}} \emph {et~al.},\ }\bibfield  {title} {\enquote {\bibinfo {title} {{Charge Radius of the Short-Lived $^{68}$Ni and Correlation with the Dipole Polarizability}},}\ }\href {\doibase 10.1103/PhysRevLett.124.132502} {\bibfield  {journal} {\bibinfo  {journal} {Phys. Rev. Lett.}\ }\textbf {\bibinfo {volume} {124}},\ \bibinfo {pages} {132502} (\bibinfo {year} {2020})},\ \Eprint {http://arxiv.org/abs/2003.06353} {arXiv:2003.06353 [nucl-ex]} \BibitemShut {NoStop}%
\bibitem [{\citenamefont {Bonaiti}\ \emph {et~al.}(2022)\citenamefont {Bonaiti}, \citenamefont {Bacca},\ and\ \citenamefont {Hagen}}]{bonaiti2022}%
  \BibitemOpen
  \bibfield  {author} {\bibinfo {author} {\bibfnamefont {F.}~\bibnamefont {Bonaiti}}, \bibinfo {author} {\bibfnamefont {S.}~\bibnamefont {Bacca}}, \ and\ \bibinfo {author} {\bibfnamefont {G.}~\bibnamefont {Hagen}},\ }\bibfield  {title} {\enquote {\bibinfo {title} {Ab initio coupled-cluster calculations of ground and dipole excited states in $^{8}\mathrm{He}$},}\ }\href {\doibase 10.1103/PhysRevC.105.034313} {\bibfield  {journal} {\bibinfo  {journal} {Phys. Rev. C}\ }\textbf {\bibinfo {volume} {105}},\ \bibinfo {pages} {034313} (\bibinfo {year} {2022})}\BibitemShut {NoStop}%
\bibitem [{\citenamefont {Bonaiti}\ and\ \citenamefont {Bacca}(2024)}]{bonaiti2024}%
  \BibitemOpen
  \bibfield  {author} {\bibinfo {author} {\bibfnamefont {Francesca}\ \bibnamefont {Bonaiti}}\ and\ \bibinfo {author} {\bibfnamefont {Sonia}\ \bibnamefont {Bacca}},\ }\bibfield  {title} {\enquote {\bibinfo {title} {{Low-Energy} {Dipole Strength} in {$^8$He}},}\ }\href@noop {} {\bibfield  {journal} {\bibinfo  {journal} {Few-Body Systems}\ }\textbf {\bibinfo {volume} {65}},\ \bibinfo {pages} {54} (\bibinfo {year} {2024})}\BibitemShut {NoStop}%
\bibitem [{\citenamefont {Von Neumann-Cosel}()}]{vnc_private}%
  \BibitemOpen
  \bibfield  {author} {\bibinfo {author} {\bibfnamefont {P.}~\bibnamefont {Von Neumann-Cosel}},\ }\href@noop {} {\bibinfo  {journal} {private communication}\ }\BibitemShut {NoStop}%
\bibitem [{\citenamefont {Community}()}]{frib400}%
  \BibitemOpen
\bibfield  {journal} {  }\bibfield  {author} {\bibinfo {author} {\bibfnamefont {FRIB~Science}\ \bibnamefont {Community}},\ }\href@noop {} {\enquote {\bibinfo {title} {{FRIB400} - {The Scientific Case for the $400$ $\mathrm{MeV/u}$ Energy Upgrade} of {FRIB}},}\ }\bibinfo {howpublished} {\url{https://frib.msu.edu/_files/pdfs/frib400_final.pdf}},\ \bibinfo {note} {accessed: 2023-09-26}\BibitemShut {NoStop}%
\bibitem [{\citenamefont {Stanton}\ and\ \citenamefont {Bartlett}(1993)}]{stanton1993}%
  \BibitemOpen
  \bibfield  {author} {\bibinfo {author} {\bibfnamefont {John~F.}\ \bibnamefont {Stanton}}\ and\ \bibinfo {author} {\bibfnamefont {Rodney~J.}\ \bibnamefont {Bartlett}},\ }\bibfield  {title} {\enquote {\bibinfo {title} {The equation of motion coupled‐cluster method. a systematic biorthogonal approach to molecular excitation energies, transition probabilities, and excited state properties},}\ }\href {\doibase 10.1063/1.464746} {\bibfield  {journal} {\bibinfo  {journal} {The Journal of Chemical Physics}\ }\textbf {\bibinfo {volume} {98}},\ \bibinfo {pages} {7029--7039} (\bibinfo {year} {1993})},\ \Eprint {http://arxiv.org/abs/https://doi.org/10.1063/1.464746} {https://doi.org/10.1063/1.464746} \BibitemShut {NoStop}%
\bibitem [{\citenamefont {Henderson}\ \emph {et~al.}(2003)\citenamefont {Henderson}, \citenamefont {Runge},\ and\ \citenamefont {Bartlett}}]{bartlett2003}%
  \BibitemOpen
  \bibfield  {author} {\bibinfo {author} {\bibfnamefont {Thomas~M.}\ \bibnamefont {Henderson}}, \bibinfo {author} {\bibfnamefont {Keith}\ \bibnamefont {Runge}}, \ and\ \bibinfo {author} {\bibfnamefont {Rodney~J.}\ \bibnamefont {Bartlett}},\ }\bibfield  {title} {\enquote {\bibinfo {title} {Excited states in artificial atoms via equation-of-motion coupled cluster theory},}\ }\href {\doibase 10.1103/PhysRevB.67.045320} {\bibfield  {journal} {\bibinfo  {journal} {Phys. Rev. B}\ }\textbf {\bibinfo {volume} {67}},\ \bibinfo {pages} {045320} (\bibinfo {year} {2003})}\BibitemShut {NoStop}%
\bibitem [{\citenamefont {Bartlett}\ and\ \citenamefont {Musia\l{}}(2007)}]{bartlett2007}%
  \BibitemOpen
  \bibfield  {author} {\bibinfo {author} {\bibfnamefont {Rodney~J.}\ \bibnamefont {Bartlett}}\ and\ \bibinfo {author} {\bibfnamefont {Monika}\ \bibnamefont {Musia\l{}}},\ }\bibfield  {title} {\enquote {\bibinfo {title} {Coupled-cluster theory in quantum chemistry},}\ }\href {\doibase 10.1103/RevModPhys.79.291} {\bibfield  {journal} {\bibinfo  {journal} {Rev. Mod. Phys.}\ }\textbf {\bibinfo {volume} {79}},\ \bibinfo {pages} {291--352} (\bibinfo {year} {2007})}\BibitemShut {NoStop}%
\bibitem [{\citenamefont {Gour}\ \emph {et~al.}(2005)\citenamefont {Gour}, \citenamefont {Piecuch},\ and\ \citenamefont {W{\l}och}}]{gour2005}%
  \BibitemOpen
  \bibfield  {author} {\bibinfo {author} {\bibfnamefont {J.~R.}\ \bibnamefont {Gour}}, \bibinfo {author} {\bibfnamefont {P.}~\bibnamefont {Piecuch}}, \ and\ \bibinfo {author} {\bibfnamefont {M.}~\bibnamefont {W{\l}och}},\ }\bibfield  {title} {\enquote {\bibinfo {title} {Active-space equation-of-motion coupled-cluster methods for excited states of radicals and other open-shell systems: Ea-eomccsdt and ip-eomccsdt},}\ }\href {\doibase 10.1063/1.2042452} {\bibfield  {journal} {\bibinfo  {journal} {J. Chem. Phys.}\ }\textbf {\bibinfo {volume} {123}},\ \bibinfo {eid} {134113} (\bibinfo {year} {2005})}\BibitemShut {NoStop}%
\bibitem [{\citenamefont {Gour}\ \emph {et~al.}(2006)\citenamefont {Gour}, \citenamefont {Piecuch}, \citenamefont {Hjorth-Jensen}, \citenamefont {W\l{}och},\ and\ \citenamefont {Dean}}]{gour2006}%
  \BibitemOpen
  \bibfield  {author} {\bibinfo {author} {\bibfnamefont {J.~R.}\ \bibnamefont {Gour}}, \bibinfo {author} {\bibfnamefont {P.}~\bibnamefont {Piecuch}}, \bibinfo {author} {\bibfnamefont {M.}~\bibnamefont {Hjorth-Jensen}}, \bibinfo {author} {\bibfnamefont {M.}~\bibnamefont {W\l{}och}}, \ and\ \bibinfo {author} {\bibfnamefont {D.~J.}\ \bibnamefont {Dean}},\ }\bibfield  {title} {\enquote {\bibinfo {title} {Coupled-cluster calculations for valence systems around $^{16}\mathrm{O}$},}\ }\href {\doibase 10.1103/PhysRevC.74.024310} {\bibfield  {journal} {\bibinfo  {journal} {Phys. Rev. C}\ }\textbf {\bibinfo {volume} {74}},\ \bibinfo {pages} {024310} (\bibinfo {year} {2006})}\BibitemShut {NoStop}%
\bibitem [{\citenamefont {Jansen}\ \emph {et~al.}(2011)\citenamefont {Jansen}, \citenamefont {Hjorth-Jensen}, \citenamefont {Hagen},\ and\ \citenamefont {Papenbrock}}]{jansen2011}%
  \BibitemOpen
  \bibfield  {author} {\bibinfo {author} {\bibfnamefont {G.~R.}\ \bibnamefont {Jansen}}, \bibinfo {author} {\bibfnamefont {M.}~\bibnamefont {Hjorth-Jensen}}, \bibinfo {author} {\bibfnamefont {G.}~\bibnamefont {Hagen}}, \ and\ \bibinfo {author} {\bibfnamefont {T.}~\bibnamefont {Papenbrock}},\ }\bibfield  {title} {\enquote {\bibinfo {title} {Toward open-shell nuclei with coupled-cluster theory},}\ }\href {\doibase 10.1103/PhysRevC.83.054306} {\bibfield  {journal} {\bibinfo  {journal} {Phys. Rev. C}\ }\textbf {\bibinfo {volume} {83}},\ \bibinfo {pages} {054306} (\bibinfo {year} {2011})}\BibitemShut {NoStop}%
\bibitem [{\citenamefont {Jansen}(2013)}]{jansen2013}%
  \BibitemOpen
  \bibfield  {author} {\bibinfo {author} {\bibfnamefont {G.~R.}\ \bibnamefont {Jansen}},\ }\bibfield  {title} {\enquote {\bibinfo {title} {Spherical coupled-cluster theory for open-shell nuclei},}\ }\href {\doibase 10.1103/PhysRevC.88.024305} {\bibfield  {journal} {\bibinfo  {journal} {Phys. Rev. C}\ }\textbf {\bibinfo {volume} {88}},\ \bibinfo {pages} {024305} (\bibinfo {year} {2013})}\BibitemShut {NoStop}%
\bibitem [{\citenamefont {Shen}\ and\ \citenamefont {Piecuch}(2013)}]{shen2013}%
  \BibitemOpen
  \bibfield  {author} {\bibinfo {author} {\bibfnamefont {Jun}\ \bibnamefont {Shen}}\ and\ \bibinfo {author} {\bibfnamefont {Piotr}\ \bibnamefont {Piecuch}},\ }\bibfield  {title} {\enquote {\bibinfo {title} {Doubly electron-attached and doubly ionized equation-of-motion coupled-cluster methods with 4-particle–2-hole and 4-hole–2-particle excitations and their active-space extensions},}\ }\href {\doibase 10.1063/1.4803883} {\bibfield  {journal} {\bibinfo  {journal} {J. Chem. Phys.}\ }\textbf {\bibinfo {volume} {138}},\ \bibinfo {eid} {194102} (\bibinfo {year} {2013})}\BibitemShut {NoStop}%
\bibitem [{\citenamefont {Shen}\ and\ \citenamefont {Piecuch}(2014)}]{shen2014}%
  \BibitemOpen
  \bibfield  {author} {\bibinfo {author} {\bibfnamefont {J.}~\bibnamefont {Shen}}\ and\ \bibinfo {author} {\bibfnamefont {P.}~\bibnamefont {Piecuch}},\ }\bibfield  {title} {\enquote {\bibinfo {title} {Doubly electron-attached and doubly ionised equation-of-motion coupled-cluster methods with full and active-space treatments of 4-particle–2-hole and 4-hole–2-particle excitations: the role of orbital choices},}\ }\href {\doibase 10.1080/00268976.2014.886397} {\bibfield  {journal} {\bibinfo  {journal} {Molecular Physics}\ }\textbf {\bibinfo {volume} {112}},\ \bibinfo {pages} {868--885} (\bibinfo {year} {2014})}\BibitemShut {NoStop}%
\bibitem [{\citenamefont {Ajala}\ \emph {et~al.}(2017)\citenamefont {Ajala}, \citenamefont {Shen},\ and\ \citenamefont {Piecuch}}]{ajala2017}%
  \BibitemOpen
  \bibfield  {author} {\bibinfo {author} {\bibfnamefont {Adeayo~O.}\ \bibnamefont {Ajala}}, \bibinfo {author} {\bibfnamefont {Jun}\ \bibnamefont {Shen}}, \ and\ \bibinfo {author} {\bibfnamefont {Piotr}\ \bibnamefont {Piecuch}},\ }\bibfield  {title} {\enquote {\bibinfo {title} {Economical doubly electron-attached equation-of-motion coupled-cluster methods with an active-space treatment of three-particle--one-hole and four-particle--two-hole excitations},}\ }\href {\doibase 10.1021/acs.jpca.6b11393} {\bibfield  {journal} {\bibinfo  {journal} {The Journal of Physical Chemistry A}\ }\textbf {\bibinfo {volume} {121}},\ \bibinfo {pages} {3469--3485} (\bibinfo {year} {2017})}\BibitemShut {NoStop}%
\bibitem [{\citenamefont {Jiang}\ \emph {et~al.}(2020)\citenamefont {Jiang}, \citenamefont {Ekstr\"om}, \citenamefont {Forss\'en}, \citenamefont {Hagen}, \citenamefont {Jansen},\ and\ \citenamefont {Papenbrock}}]{jiang2020}%
  \BibitemOpen
  \bibfield  {author} {\bibinfo {author} {\bibfnamefont {W.~G.}\ \bibnamefont {Jiang}}, \bibinfo {author} {\bibfnamefont {A.}~\bibnamefont {Ekstr\"om}}, \bibinfo {author} {\bibfnamefont {C.}~\bibnamefont {Forss\'en}}, \bibinfo {author} {\bibfnamefont {G.}~\bibnamefont {Hagen}}, \bibinfo {author} {\bibfnamefont {G.~R.}\ \bibnamefont {Jansen}}, \ and\ \bibinfo {author} {\bibfnamefont {T.}~\bibnamefont {Papenbrock}},\ }\bibfield  {title} {\enquote {\bibinfo {title} {Accurate bulk properties of nuclei from $a=2$ to $\ensuremath{\infty}$ from potentials with $\mathrm{\ensuremath{\Delta}}$ isobars},}\ }\href {\doibase 10.1103/PhysRevC.102.054301} {\bibfield  {journal} {\bibinfo  {journal} {Phys. Rev. C}\ }\textbf {\bibinfo {volume} {102}},\ \bibinfo {pages} {054301} (\bibinfo {year} {2020})}\BibitemShut {NoStop}%
\bibitem [{\citenamefont {Hu}\ \emph {et~al.}(2022)\citenamefont {Hu}, \citenamefont {Jiang}, \citenamefont {Miyagi}, \citenamefont {Sun}, \citenamefont {Ekstr{\"o}m}, \citenamefont {Forss{\'e}n}, \citenamefont {Hagen}, \citenamefont {Holt}, \citenamefont {Papenbrock}, \citenamefont {Stroberg},\ and\ \citenamefont {Vernon}}]{hu2022}%
  \BibitemOpen
  \bibfield  {author} {\bibinfo {author} {\bibfnamefont {Baishan}\ \bibnamefont {Hu}}, \bibinfo {author} {\bibfnamefont {Weiguang}\ \bibnamefont {Jiang}}, \bibinfo {author} {\bibfnamefont {Takayuki}\ \bibnamefont {Miyagi}}, \bibinfo {author} {\bibfnamefont {Zhonghao}\ \bibnamefont {Sun}}, \bibinfo {author} {\bibfnamefont {Andreas}\ \bibnamefont {Ekstr{\"o}m}}, \bibinfo {author} {\bibfnamefont {Christian}\ \bibnamefont {Forss{\'e}n}}, \bibinfo {author} {\bibfnamefont {Gaute}\ \bibnamefont {Hagen}}, \bibinfo {author} {\bibfnamefont {Jason~D}\ \bibnamefont {Holt}}, \bibinfo {author} {\bibfnamefont {Thomas}\ \bibnamefont {Papenbrock}}, \bibinfo {author} {\bibfnamefont {S~Ragnar}\ \bibnamefont {Stroberg}}, \ and\ \bibinfo {author} {\bibfnamefont {Ian}\ \bibnamefont {Vernon}},\ }\bibfield  {title} {\enquote {\bibinfo {title} {Ab initio predictions link the neutron skin of 208pb to nuclear forces},}\ }\href@noop {} {\bibfield  {journal} {\bibinfo  {journal} {Nature Physics}\ }\textbf {\bibinfo {volume} {18}},\
  \bibinfo {pages} {1196--1200} (\bibinfo {year} {2022})}\BibitemShut {NoStop}%
\bibitem [{\citenamefont {Kondo}\ \emph {et~al.}(2023)\citenamefont {Kondo}, \citenamefont {Achouri}, \citenamefont {Falou}, \citenamefont {Atar}, \citenamefont {Aumann}, \citenamefont {Baba}, \citenamefont {Boretzky}, \citenamefont {Caesar}, \citenamefont {Calvet}, \citenamefont {Chae}, \citenamefont {Chiga}, \citenamefont {Corsi}, \citenamefont {Delaunay}, \citenamefont {Delbart}, \citenamefont {Deshayes}, \citenamefont {Dombr{\'a}di}, \citenamefont {Douma}, \citenamefont {Ekstr{\"o}m}, \citenamefont {Elekes}, \citenamefont {Forss{\'e}n}, \citenamefont {Ga{\v s}pari{\'c}}, \citenamefont {Gheller}, \citenamefont {Gibelin}, \citenamefont {Gillibert}, \citenamefont {Hagen}, \citenamefont {Harakeh}, \citenamefont {Hirayama}, \citenamefont {Hoffman}, \citenamefont {Holl}, \citenamefont {Horvat}, \citenamefont {Horv{\'a}th}, \citenamefont {Hwang}, \citenamefont {Isobe}, \citenamefont {Jiang}, \citenamefont {Kahlbow}, \citenamefont {Kalantar-Nayestanaki}, \citenamefont {Kawase}, \citenamefont {Kim}, \citenamefont
  {Kisamori}, \citenamefont {Kobayashi}, \citenamefont {K{\"o}rper}, \citenamefont {Koyama}, \citenamefont {Kuti}, \citenamefont {Lapoux}, \citenamefont {Lindberg}, \citenamefont {Marqu{\'e}s}, \citenamefont {Masuoka}, \citenamefont {Mayer}, \citenamefont {Miki}, \citenamefont {Murakami}, \citenamefont {Najafi}, \citenamefont {Nakamura}, \citenamefont {Nakano}, \citenamefont {Nakatsuka}, \citenamefont {Nilsson}, \citenamefont {Obertelli}, \citenamefont {Ogata}, \citenamefont {de~Oliveira~Santos}, \citenamefont {Orr}, \citenamefont {Otsu}, \citenamefont {Otsuka}, \citenamefont {Ozaki}, \citenamefont {Panin}, \citenamefont {Papenbrock}, \citenamefont {Paschalis}, \citenamefont {Revel}, \citenamefont {Rossi}, \citenamefont {Saito}, \citenamefont {Saito}, \citenamefont {Sasano}, \citenamefont {Sato}, \citenamefont {Satou}, \citenamefont {Scheit}, \citenamefont {Schindler}, \citenamefont {Schrock}, \citenamefont {Shikata}, \citenamefont {Shimizu}, \citenamefont {Shimizu}, \citenamefont {Simon}, \citenamefont
  {Sohler}, \citenamefont {Sorlin}, \citenamefont {Stuhl}, \citenamefont {Sun}, \citenamefont {Takeuchi}, \citenamefont {Tanaka}, \citenamefont {Thoennessen}, \citenamefont {T{\"o}rnqvist}, \citenamefont {Togano}, \citenamefont {Tomai}, \citenamefont {Tscheuschner}, \citenamefont {Tsubota}, \citenamefont {Tsunoda}, \citenamefont {Uesaka}, \citenamefont {Utsuno}, \citenamefont {Vernon}, \citenamefont {Wang}, \citenamefont {Yang}, \citenamefont {Yasuda}, \citenamefont {Yoneda},\ and\ \citenamefont {Yoshida}}]{kondo2023}%
  \BibitemOpen
  \bibfield  {author} {\bibinfo {author} {\bibfnamefont {Y}~\bibnamefont {Kondo}}, \bibinfo {author} {\bibfnamefont {N~L}\ \bibnamefont {Achouri}}, \bibinfo {author} {\bibfnamefont {H~Al}\ \bibnamefont {Falou}}, \bibinfo {author} {\bibfnamefont {L}~\bibnamefont {Atar}}, \bibinfo {author} {\bibfnamefont {T}~\bibnamefont {Aumann}}, \bibinfo {author} {\bibfnamefont {H}~\bibnamefont {Baba}}, \bibinfo {author} {\bibfnamefont {K}~\bibnamefont {Boretzky}}, \bibinfo {author} {\bibfnamefont {C}~\bibnamefont {Caesar}}, \bibinfo {author} {\bibfnamefont {D}~\bibnamefont {Calvet}}, \bibinfo {author} {\bibfnamefont {H}~\bibnamefont {Chae}}, \bibinfo {author} {\bibfnamefont {N}~\bibnamefont {Chiga}}, \bibinfo {author} {\bibfnamefont {A}~\bibnamefont {Corsi}}, \bibinfo {author} {\bibfnamefont {F}~\bibnamefont {Delaunay}}, \bibinfo {author} {\bibfnamefont {A}~\bibnamefont {Delbart}}, \bibinfo {author} {\bibfnamefont {Q}~\bibnamefont {Deshayes}}, \bibinfo {author} {\bibfnamefont {Zs}~\bibnamefont {Dombr{\'a}di}}, \bibinfo
  {author} {\bibfnamefont {C~A}\ \bibnamefont {Douma}}, \bibinfo {author} {\bibfnamefont {A}~\bibnamefont {Ekstr{\"o}m}}, \bibinfo {author} {\bibfnamefont {Z}~\bibnamefont {Elekes}}, \bibinfo {author} {\bibfnamefont {C}~\bibnamefont {Forss{\'e}n}}, \bibinfo {author} {\bibfnamefont {I}~\bibnamefont {Ga{\v s}pari{\'c}}}, \bibinfo {author} {\bibfnamefont {J-M}\ \bibnamefont {Gheller}}, \bibinfo {author} {\bibfnamefont {J}~\bibnamefont {Gibelin}}, \bibinfo {author} {\bibfnamefont {A}~\bibnamefont {Gillibert}}, \bibinfo {author} {\bibfnamefont {G}~\bibnamefont {Hagen}}, \bibinfo {author} {\bibfnamefont {M~N}\ \bibnamefont {Harakeh}}, \bibinfo {author} {\bibfnamefont {A}~\bibnamefont {Hirayama}}, \bibinfo {author} {\bibfnamefont {C~R}\ \bibnamefont {Hoffman}}, \bibinfo {author} {\bibfnamefont {M}~\bibnamefont {Holl}}, \bibinfo {author} {\bibfnamefont {A}~\bibnamefont {Horvat}}, \bibinfo {author} {\bibfnamefont {{\'A}}~\bibnamefont {Horv{\'a}th}}, \bibinfo {author} {\bibfnamefont {J~W}\ \bibnamefont {Hwang}},
  \bibinfo {author} {\bibfnamefont {T}~\bibnamefont {Isobe}}, \bibinfo {author} {\bibfnamefont {W~G}\ \bibnamefont {Jiang}}, \bibinfo {author} {\bibfnamefont {J}~\bibnamefont {Kahlbow}}, \bibinfo {author} {\bibfnamefont {N}~\bibnamefont {Kalantar-Nayestanaki}}, \bibinfo {author} {\bibfnamefont {S}~\bibnamefont {Kawase}}, \bibinfo {author} {\bibfnamefont {S}~\bibnamefont {Kim}}, \bibinfo {author} {\bibfnamefont {K}~\bibnamefont {Kisamori}}, \bibinfo {author} {\bibfnamefont {T}~\bibnamefont {Kobayashi}}, \bibinfo {author} {\bibfnamefont {D}~\bibnamefont {K{\"o}rper}}, \bibinfo {author} {\bibfnamefont {S}~\bibnamefont {Koyama}}, \bibinfo {author} {\bibfnamefont {I}~\bibnamefont {Kuti}}, \bibinfo {author} {\bibfnamefont {V}~\bibnamefont {Lapoux}}, \bibinfo {author} {\bibfnamefont {S}~\bibnamefont {Lindberg}}, \bibinfo {author} {\bibfnamefont {F~M}\ \bibnamefont {Marqu{\'e}s}}, \bibinfo {author} {\bibfnamefont {S}~\bibnamefont {Masuoka}}, \bibinfo {author} {\bibfnamefont {J}~\bibnamefont {Mayer}}, \bibinfo
  {author} {\bibfnamefont {K}~\bibnamefont {Miki}}, \bibinfo {author} {\bibfnamefont {T}~\bibnamefont {Murakami}}, \bibinfo {author} {\bibfnamefont {M}~\bibnamefont {Najafi}}, \bibinfo {author} {\bibfnamefont {T}~\bibnamefont {Nakamura}}, \bibinfo {author} {\bibfnamefont {K}~\bibnamefont {Nakano}}, \bibinfo {author} {\bibfnamefont {N}~\bibnamefont {Nakatsuka}}, \bibinfo {author} {\bibfnamefont {T}~\bibnamefont {Nilsson}}, \bibinfo {author} {\bibfnamefont {A}~\bibnamefont {Obertelli}}, \bibinfo {author} {\bibfnamefont {K}~\bibnamefont {Ogata}}, \bibinfo {author} {\bibfnamefont {F}~\bibnamefont {de~Oliveira~Santos}}, \bibinfo {author} {\bibfnamefont {N~A}\ \bibnamefont {Orr}}, \bibinfo {author} {\bibfnamefont {H}~\bibnamefont {Otsu}}, \bibinfo {author} {\bibfnamefont {T}~\bibnamefont {Otsuka}}, \bibinfo {author} {\bibfnamefont {T}~\bibnamefont {Ozaki}}, \bibinfo {author} {\bibfnamefont {V}~\bibnamefont {Panin}}, \bibinfo {author} {\bibfnamefont {T}~\bibnamefont {Papenbrock}}, \bibinfo {author} {\bibfnamefont
  {S}~\bibnamefont {Paschalis}}, \bibinfo {author} {\bibfnamefont {A}~\bibnamefont {Revel}}, \bibinfo {author} {\bibfnamefont {D}~\bibnamefont {Rossi}}, \bibinfo {author} {\bibfnamefont {A~T}\ \bibnamefont {Saito}}, \bibinfo {author} {\bibfnamefont {T~Y}\ \bibnamefont {Saito}}, \bibinfo {author} {\bibfnamefont {M}~\bibnamefont {Sasano}}, \bibinfo {author} {\bibfnamefont {H}~\bibnamefont {Sato}}, \bibinfo {author} {\bibfnamefont {Y}~\bibnamefont {Satou}}, \bibinfo {author} {\bibfnamefont {H}~\bibnamefont {Scheit}}, \bibinfo {author} {\bibfnamefont {F}~\bibnamefont {Schindler}}, \bibinfo {author} {\bibfnamefont {P}~\bibnamefont {Schrock}}, \bibinfo {author} {\bibfnamefont {M}~\bibnamefont {Shikata}}, \bibinfo {author} {\bibfnamefont {N}~\bibnamefont {Shimizu}}, \bibinfo {author} {\bibfnamefont {Y}~\bibnamefont {Shimizu}}, \bibinfo {author} {\bibfnamefont {H}~\bibnamefont {Simon}}, \bibinfo {author} {\bibfnamefont {D}~\bibnamefont {Sohler}}, \bibinfo {author} {\bibfnamefont {O}~\bibnamefont {Sorlin}}, \bibinfo
  {author} {\bibfnamefont {L}~\bibnamefont {Stuhl}}, \bibinfo {author} {\bibfnamefont {Z~H}\ \bibnamefont {Sun}}, \bibinfo {author} {\bibfnamefont {S}~\bibnamefont {Takeuchi}}, \bibinfo {author} {\bibfnamefont {M}~\bibnamefont {Tanaka}}, \bibinfo {author} {\bibfnamefont {M}~\bibnamefont {Thoennessen}}, \bibinfo {author} {\bibfnamefont {H}~\bibnamefont {T{\"o}rnqvist}}, \bibinfo {author} {\bibfnamefont {Y}~\bibnamefont {Togano}}, \bibinfo {author} {\bibfnamefont {T}~\bibnamefont {Tomai}}, \bibinfo {author} {\bibfnamefont {J}~\bibnamefont {Tscheuschner}}, \bibinfo {author} {\bibfnamefont {J}~\bibnamefont {Tsubota}}, \bibinfo {author} {\bibfnamefont {N}~\bibnamefont {Tsunoda}}, \bibinfo {author} {\bibfnamefont {T}~\bibnamefont {Uesaka}}, \bibinfo {author} {\bibfnamefont {Y}~\bibnamefont {Utsuno}}, \bibinfo {author} {\bibfnamefont {I}~\bibnamefont {Vernon}}, \bibinfo {author} {\bibfnamefont {H}~\bibnamefont {Wang}}, \bibinfo {author} {\bibfnamefont {Z}~\bibnamefont {Yang}}, \bibinfo {author} {\bibfnamefont
  {M}~\bibnamefont {Yasuda}}, \bibinfo {author} {\bibfnamefont {K}~\bibnamefont {Yoneda}}, \ and\ \bibinfo {author} {\bibfnamefont {S}~\bibnamefont {Yoshida}},\ }\bibfield  {title} {\enquote {\bibinfo {title} {First observation of {28O}},}\ }\href@noop {} {\bibfield  {journal} {\bibinfo  {journal} {Nature}\ }\textbf {\bibinfo {volume} {620}},\ \bibinfo {pages} {965--970} (\bibinfo {year} {2023})}\BibitemShut {NoStop}%
\bibitem [{\citenamefont {Sun}\ \emph {et~al.}(2024)\citenamefont {Sun}, \citenamefont {Ekstr\"om}, \citenamefont {Forss\'en}, \citenamefont {Hagen}, \citenamefont {Jansen},\ and\ \citenamefont {Papenbrock}}]{sun2024}%
  \BibitemOpen
  \bibfield  {author} {\bibinfo {author} {\bibfnamefont {Z.~H.}\ \bibnamefont {Sun}}, \bibinfo {author} {\bibfnamefont {A.}~\bibnamefont {Ekstr\"om}}, \bibinfo {author} {\bibfnamefont {C.}~\bibnamefont {Forss\'en}}, \bibinfo {author} {\bibfnamefont {G.}~\bibnamefont {Hagen}}, \bibinfo {author} {\bibfnamefont {G.~R.}\ \bibnamefont {Jansen}}, \ and\ \bibinfo {author} {\bibfnamefont {T.}~\bibnamefont {Papenbrock}},\ }\bibfield  {title} {\enquote {\bibinfo {title} {{Multiscale physics of atomic nuclei from first principles}},}\ \ }(\bibinfo {year} {2024})\ \Eprint {http://arxiv.org/abs/2404.00058} {arXiv:2404.00058 [nucl-th]} \BibitemShut {NoStop}%
\bibitem [{\citenamefont {Efros}\ \emph {et~al.}(1994)\citenamefont {Efros}, \citenamefont {Leidemann},\ and\ \citenamefont {Orlandini}}]{efros1994}%
  \BibitemOpen
  \bibfield  {author} {\bibinfo {author} {\bibfnamefont {Victor~D.}\ \bibnamefont {Efros}}, \bibinfo {author} {\bibfnamefont {Winfred}\ \bibnamefont {Leidemann}}, \ and\ \bibinfo {author} {\bibfnamefont {Giuseppina}\ \bibnamefont {Orlandini}},\ }\bibfield  {title} {\enquote {\bibinfo {title} {Response functions from integral transforms with a lorentz kernel},}\ }\href {\doibase 10.1016/0370-2693(94)91355-2} {\bibfield  {journal} {\bibinfo  {journal} {Phys. Lett. B}\ }\textbf {\bibinfo {volume} {338}},\ \bibinfo {pages} {130 -- 133} (\bibinfo {year} {1994})}\BibitemShut {NoStop}%
\bibitem [{\citenamefont {{\v {C}}{\'\i}{\v z}ek}(1966)}]{cizek1966}%
  \BibitemOpen
  \bibfield  {author} {\bibinfo {author} {\bibfnamefont {Jiri}\ \bibnamefont {{\v {C}}{\'\i}{\v z}ek}},\ }\bibfield  {title} {\enquote {\bibinfo {title} {{On the Correlation Problem in Atomic and Molecular Systems. Calculation of Wavefunction Components in Ursell-Type Expansion Using Quantum-Field Theoretical Methods}},}\ }\href {\doibase 10.1063/1.1727484} {\bibfield  {journal} {\bibinfo  {journal} {J. Chem. Phys.}\ }\textbf {\bibinfo {volume} {45}},\ \bibinfo {pages} {4256--4266} (\bibinfo {year} {1966})}\BibitemShut {NoStop}%
\bibitem [{\citenamefont {K{\"u}mmel}\ \emph {et~al.}(1978)\citenamefont {K{\"u}mmel}, \citenamefont {L{\"u}hrmann},\ and\ \citenamefont {Zabolitzky}}]{kuemmel1978}%
  \BibitemOpen
  \bibfield  {author} {\bibinfo {author} {\bibfnamefont {H.}~\bibnamefont {K{\"u}mmel}}, \bibinfo {author} {\bibfnamefont {K.~H.}\ \bibnamefont {L{\"u}hrmann}}, \ and\ \bibinfo {author} {\bibfnamefont {J.~G.}\ \bibnamefont {Zabolitzky}},\ }\bibfield  {title} {\enquote {\bibinfo {title} {{Many-fermion theory in expS- (or coupled cluster) form}},}\ }\href {\doibase 10.1016/0370-1573(78)90081-9} {\bibfield  {journal} {\bibinfo  {journal} {Physics Reports}\ }\textbf {\bibinfo {volume} {36}},\ \bibinfo {pages} {1 -- 63} (\bibinfo {year} {1978})}\BibitemShut {NoStop}%
\bibitem [{\citenamefont {Hagen}\ \emph {et~al.}(2007)\citenamefont {Hagen}, \citenamefont {Papenbrock}, \citenamefont {Dean}, \citenamefont {Schwenk}, \citenamefont {Nogga}, \citenamefont {W\l{}och},\ and\ \citenamefont {Piecuch}}]{hagen2007}%
  \BibitemOpen
  \bibfield  {author} {\bibinfo {author} {\bibfnamefont {G.}~\bibnamefont {Hagen}}, \bibinfo {author} {\bibfnamefont {T.}~\bibnamefont {Papenbrock}}, \bibinfo {author} {\bibfnamefont {D.~J.}\ \bibnamefont {Dean}}, \bibinfo {author} {\bibfnamefont {A.}~\bibnamefont {Schwenk}}, \bibinfo {author} {\bibfnamefont {A.}~\bibnamefont {Nogga}}, \bibinfo {author} {\bibfnamefont {M.}~\bibnamefont {W\l{}och}}, \ and\ \bibinfo {author} {\bibfnamefont {P.}~\bibnamefont {Piecuch}},\ }\bibfield  {title} {\enquote {\bibinfo {title} {Coupled-cluster theory for three-body hamiltonians},}\ }\href {\doibase 10.1103/PhysRevC.76.034302} {\bibfield  {journal} {\bibinfo  {journal} {Phys. Rev. C}\ }\textbf {\bibinfo {volume} {76}},\ \bibinfo {pages} {034302} (\bibinfo {year} {2007})}\BibitemShut {NoStop}%
\bibitem [{\citenamefont {Roth}\ \emph {et~al.}(2012)\citenamefont {Roth}, \citenamefont {Binder}, \citenamefont {Vobig}, \citenamefont {Calci}, \citenamefont {Langhammer},\ and\ \citenamefont {Navr\'atil}}]{roth2012}%
  \BibitemOpen
  \bibfield  {author} {\bibinfo {author} {\bibfnamefont {Robert}\ \bibnamefont {Roth}}, \bibinfo {author} {\bibfnamefont {Sven}\ \bibnamefont {Binder}}, \bibinfo {author} {\bibfnamefont {Klaus}\ \bibnamefont {Vobig}}, \bibinfo {author} {\bibfnamefont {Angelo}\ \bibnamefont {Calci}}, \bibinfo {author} {\bibfnamefont {Joachim}\ \bibnamefont {Langhammer}}, \ and\ \bibinfo {author} {\bibfnamefont {Petr}\ \bibnamefont {Navr\'atil}},\ }\bibfield  {title} {\enquote {\bibinfo {title} {Medium-mass nuclei with normal-ordered chiral $nn\mathbf{+}3n$ interactions},}\ }\href {\doibase 10.1103/PhysRevLett.109.052501} {\bibfield  {journal} {\bibinfo  {journal} {Phys. Rev. Lett.}\ }\textbf {\bibinfo {volume} {109}},\ \bibinfo {pages} {052501} (\bibinfo {year} {2012})}\BibitemShut {NoStop}%
\bibitem [{\citenamefont {Arponen}(1982)}]{arponen1982}%
  \BibitemOpen
  \bibfield  {author} {\bibinfo {author} {\bibfnamefont {J.}~\bibnamefont {Arponen}},\ }\bibfield  {title} {\enquote {\bibinfo {title} {The method of stationary cluster amplitudes and the phase transition in the lipkin pseudospin model},}\ }\href {\doibase 10.1088/0305-4616/8/8/004} {\bibfield  {journal} {\bibinfo  {journal} {Journal of Physics G: Nuclear Physics}\ }\textbf {\bibinfo {volume} {8}},\ \bibinfo {pages} {L129} (\bibinfo {year} {1982})}\BibitemShut {NoStop}%
\bibitem [{\citenamefont {Arponen}(1983)}]{arponen1983}%
  \BibitemOpen
  \bibfield  {author} {\bibinfo {author} {\bibfnamefont {Jouko}\ \bibnamefont {Arponen}},\ }\bibfield  {title} {\enquote {\bibinfo {title} {{Variational principles and linked-cluster exp S expansions for static and dynamic many-body problems}},}\ }\href {\doibase 10.1016/0003-4916(83)90284-1} {\bibfield  {journal} {\bibinfo  {journal} {Ann. Phys.}\ }\textbf {\bibinfo {volume} {151}},\ \bibinfo {pages} {311 -- 382} (\bibinfo {year} {1983})}\BibitemShut {NoStop}%
\bibitem [{\citenamefont {Watts}\ \emph {et~al.}(1993)\citenamefont {Watts}, \citenamefont {Gauss},\ and\ \citenamefont {Bartlett}}]{watts1993}%
  \BibitemOpen
  \bibfield  {author} {\bibinfo {author} {\bibfnamefont {John~D.}\ \bibnamefont {Watts}}, \bibinfo {author} {\bibfnamefont {Jürgen}\ \bibnamefont {Gauss}}, \ and\ \bibinfo {author} {\bibfnamefont {Rodney~J.}\ \bibnamefont {Bartlett}},\ }\bibfield  {title} {\enquote {\bibinfo {title} {Coupled‐cluster methods with noniterative triple excitations for restricted open‐shell hartree–fock and other general single determinant reference functions. energies and analytical gradients},}\ }\href {\doibase 10.1063/1.464480} {\bibfield  {journal} {\bibinfo  {journal} {The Journal of Chemical Physics}\ }\textbf {\bibinfo {volume} {98}},\ \bibinfo {pages} {8718--8733} (\bibinfo {year} {1993})},\ \Eprint {http://arxiv.org/abs/https://doi.org/10.1063/1.464480} {https://doi.org/10.1063/1.464480} \BibitemShut {NoStop}%
\bibitem [{Note1()}]{Note1}%
  \BibitemOpen
  \bibinfo {note} {Note that in the case of electric dipole transitions, if the nuclear ground state has $J^{\pi } = 0$, $r_0$ and $l_0$ are identically zero.}\BibitemShut {Stop}%
\bibitem [{\citenamefont {Cullum}(1995)}]{cullum1995}%
  \BibitemOpen
  \bibfield  {author} {\bibinfo {author} {\bibfnamefont {J.~K.}\ \bibnamefont {Cullum}},\ }\bibfield  {title} {\enquote {\bibinfo {title} {Arnoldi versus nonsymmetric lanczos algorithms for solving nonsymmetric matrix eigenvalue problems},}\ }\href@noop {} {\ \textbf {\bibinfo {volume} {Technical Report TR-3576, University of Maryland}} (\bibinfo {year} {1995})}\BibitemShut {NoStop}%
\bibitem [{ntc()}]{ntcl}%
  \BibitemOpen
  \bibfield  {title} {\enquote {\bibinfo {title} {{NTCL: The Nuclear Tensor Contraction Library}},}\ }\href {\doibase https://gitlab.com/ntcl/ntcl} {\ https://gitlab.com/ntcl/ntcl}\BibitemShut {NoStop}%
\bibitem [{\citenamefont {Shen}\ and\ \citenamefont {Piecuch}(2021)}]{shen2021}%
  \BibitemOpen
  \bibfield  {author} {\bibinfo {author} {\bibfnamefont {Jun}\ \bibnamefont {Shen}}\ and\ \bibinfo {author} {\bibfnamefont {Piotr}\ \bibnamefont {Piecuch}},\ }\bibfield  {title} {\enquote {\bibinfo {title} {Double electron-attachment equation-of-motion coupled-cluster methods with up to 4-particle–2-hole excitations: improved implementation and application to singlet–triplet gaps in ortho-, meta-, and para-benzyne isomers},}\ }\href {\doibase 10.1080/00268976.2021.1966534} {\bibfield  {journal} {\bibinfo  {journal} {Molecular Physics}\ }\textbf {\bibinfo {volume} {119}},\ \bibinfo {pages} {e1966534} (\bibinfo {year} {2021})},\ \Eprint {http://arxiv.org/abs/https://doi.org/10.1080/00268976.2021.1966534} {https://doi.org/10.1080/00268976.2021.1966534} \BibitemShut {NoStop}%
\bibitem [{\citenamefont {Stumpf}(2018)}]{stumpf2018}%
  \BibitemOpen
  \bibfield  {author} {\bibinfo {author} {\bibfnamefont {Christina}\ \bibnamefont {Stumpf}},\ }\emph {\bibinfo {title} {{Nuclear Spectra and Strength Distributions from Importance-Truncated Configuration-Interaction Methods}}},\ \href@noop {} {Ph.D. thesis},\ \bibinfo  {school} {Technische Universit\"at Darmstadt} (\bibinfo {year} {2018})\BibitemShut {NoStop}%
\bibitem [{\citenamefont {Ahrens}\ \emph {et~al.}(1975)\citenamefont {Ahrens}, \citenamefont {Borchert}, \citenamefont {Czock}, \citenamefont {Eppler}, \citenamefont {Gimm}, \citenamefont {Gundrum}, \citenamefont {Kröning}, \citenamefont {Riehn}, \citenamefont {{Sita Ram}}, \citenamefont {Zieger},\ and\ \citenamefont {Ziegler}}]{ahrens1975}%
  \BibitemOpen
  \bibfield  {author} {\bibinfo {author} {\bibfnamefont {J.}~\bibnamefont {Ahrens}}, \bibinfo {author} {\bibfnamefont {H.}~\bibnamefont {Borchert}}, \bibinfo {author} {\bibfnamefont {K.H.}\ \bibnamefont {Czock}}, \bibinfo {author} {\bibfnamefont {H.B.}\ \bibnamefont {Eppler}}, \bibinfo {author} {\bibfnamefont {H.}~\bibnamefont {Gimm}}, \bibinfo {author} {\bibfnamefont {H.}~\bibnamefont {Gundrum}}, \bibinfo {author} {\bibfnamefont {M.}~\bibnamefont {Kröning}}, \bibinfo {author} {\bibfnamefont {P.}~\bibnamefont {Riehn}}, \bibinfo {author} {\bibfnamefont {G.}~\bibnamefont {{Sita Ram}}}, \bibinfo {author} {\bibfnamefont {A.}~\bibnamefont {Zieger}}, \ and\ \bibinfo {author} {\bibfnamefont {B.}~\bibnamefont {Ziegler}},\ }\bibfield  {title} {\enquote {\bibinfo {title} {Total nuclear photon absorption cross sections for some light elements},}\ }\href {\doibase https://doi.org/10.1016/0375-9474(75)90543-6} {\bibfield  {journal} {\bibinfo  {journal} {Nuclear Physics A}\ }\textbf {\bibinfo {volume} {251}},\ \bibinfo
  {pages} {479--492} (\bibinfo {year} {1975})}\BibitemShut {NoStop}%
\bibitem [{\citenamefont {Ishkhanov}\ \emph {et~al.}(2002)\citenamefont {Ishkhanov}, \citenamefont {Kapitonov}, \citenamefont {Lileeva}, \citenamefont {Shirokov}, \citenamefont {Erokhova}, \citenamefont {Elkin},\ and\ \citenamefont {Izotova}}]{ishkanov2002}%
  \BibitemOpen
  \bibfield  {author} {\bibinfo {author} {\bibfnamefont {B.~S.}\ \bibnamefont {Ishkhanov}}, \bibinfo {author} {\bibfnamefont {I.~M.}\ \bibnamefont {Kapitonov}}, \bibinfo {author} {\bibfnamefont {E.~I.}\ \bibnamefont {Lileeva}}, \bibinfo {author} {\bibfnamefont {E.~V.}\ \bibnamefont {Shirokov}}, \bibinfo {author} {\bibfnamefont {V.~A.}\ \bibnamefont {Erokhova}}, \bibinfo {author} {\bibfnamefont {M.~A.}\ \bibnamefont {Elkin}}, \ and\ \bibinfo {author} {\bibfnamefont {A.~V.}\ \bibnamefont {Izotova}},\ }\bibfield  {title} {\enquote {\bibinfo {title} {Cross sections of photon absorption by nuclei with nucleon numbers 12 - 65},}\ }\href@noop {} {\ \textbf {\bibinfo {volume} {Tech. Rep. MSU-INP-2002-27/711 (Institute of Nuclear Physics, Moscow State University}} (\bibinfo {year} {2002})}\BibitemShut {NoStop}%
\bibitem [{\citenamefont {Leistenschneider}\ \emph {et~al.}(2001)\citenamefont {Leistenschneider}, \citenamefont {Aumann}, \citenamefont {Boretzky}, \citenamefont {Cortina}, \citenamefont {Cub}, \citenamefont {Pramanik}, \citenamefont {Dostal}, \citenamefont {Elze}, \citenamefont {Emling}, \citenamefont {Geissel}, \citenamefont {Gr\"unschlo\ss{}}, \citenamefont {Hellstr}, \citenamefont {Holzmann}, \citenamefont {Ilievski}, \citenamefont {Iwasa}, \citenamefont {Kaspar}, \citenamefont {Kleinb\"ohl}, \citenamefont {Kratz}, \citenamefont {Kulessa}, \citenamefont {Leifels}, \citenamefont {Lubkiewicz}, \citenamefont {M\"unzenberg}, \citenamefont {Reiter}, \citenamefont {Rejmund}, \citenamefont {Scheidenberger}, \citenamefont {Schlegel}, \citenamefont {Simon}, \citenamefont {Stroth}, \citenamefont {S\"ummerer}, \citenamefont {Wajda}, \citenamefont {Wal\'us},\ and\ \citenamefont {Wan}}]{leistenschneider2001}%
  \BibitemOpen
  \bibfield  {author} {\bibinfo {author} {\bibfnamefont {A.}~\bibnamefont {Leistenschneider}}, \bibinfo {author} {\bibfnamefont {T.}~\bibnamefont {Aumann}}, \bibinfo {author} {\bibfnamefont {K.}~\bibnamefont {Boretzky}}, \bibinfo {author} {\bibfnamefont {D.}~\bibnamefont {Cortina}}, \bibinfo {author} {\bibfnamefont {J.}~\bibnamefont {Cub}}, \bibinfo {author} {\bibfnamefont {U.~Datta}\ \bibnamefont {Pramanik}}, \bibinfo {author} {\bibfnamefont {W.}~\bibnamefont {Dostal}}, \bibinfo {author} {\bibfnamefont {Th.~W.}\ \bibnamefont {Elze}}, \bibinfo {author} {\bibfnamefont {H.}~\bibnamefont {Emling}}, \bibinfo {author} {\bibfnamefont {H.}~\bibnamefont {Geissel}}, \bibinfo {author} {\bibfnamefont {A.}~\bibnamefont {Gr\"unschlo\ss{}}}, \bibinfo {author} {\bibfnamefont {M.}~\bibnamefont {Hellstr}}, \bibinfo {author} {\bibfnamefont {R.}~\bibnamefont {Holzmann}}, \bibinfo {author} {\bibfnamefont {S.}~\bibnamefont {Ilievski}}, \bibinfo {author} {\bibfnamefont {N.}~\bibnamefont {Iwasa}}, \bibinfo {author} {\bibfnamefont
  {M.}~\bibnamefont {Kaspar}}, \bibinfo {author} {\bibfnamefont {A.}~\bibnamefont {Kleinb\"ohl}}, \bibinfo {author} {\bibfnamefont {J.~V.}\ \bibnamefont {Kratz}}, \bibinfo {author} {\bibfnamefont {R.}~\bibnamefont {Kulessa}}, \bibinfo {author} {\bibfnamefont {Y.}~\bibnamefont {Leifels}}, \bibinfo {author} {\bibfnamefont {E.}~\bibnamefont {Lubkiewicz}}, \bibinfo {author} {\bibfnamefont {G.}~\bibnamefont {M\"unzenberg}}, \bibinfo {author} {\bibfnamefont {P.}~\bibnamefont {Reiter}}, \bibinfo {author} {\bibfnamefont {M.}~\bibnamefont {Rejmund}}, \bibinfo {author} {\bibfnamefont {C.}~\bibnamefont {Scheidenberger}}, \bibinfo {author} {\bibfnamefont {C.}~\bibnamefont {Schlegel}}, \bibinfo {author} {\bibfnamefont {H.}~\bibnamefont {Simon}}, \bibinfo {author} {\bibfnamefont {J.}~\bibnamefont {Stroth}}, \bibinfo {author} {\bibfnamefont {K.}~\bibnamefont {S\"ummerer}}, \bibinfo {author} {\bibfnamefont {E.}~\bibnamefont {Wajda}}, \bibinfo {author} {\bibfnamefont {W.}~\bibnamefont {Wal\'us}}, \ and\ \bibinfo {author}
  {\bibfnamefont {S.}~\bibnamefont {Wan}},\ }\bibfield  {title} {\enquote {\bibinfo {title} {Photoneutron cross sections for unstable neutron-rich oxygen isotopes},}\ }\href {\doibase 10.1103/PhysRevLett.86.5442} {\bibfield  {journal} {\bibinfo  {journal} {Phys. Rev. Lett.}\ }\textbf {\bibinfo {volume} {86}},\ \bibinfo {pages} {5442--5445} (\bibinfo {year} {2001})}\BibitemShut {NoStop}%
\bibitem [{\citenamefont {Hebeler}\ \emph {et~al.}(2011)\citenamefont {Hebeler}, \citenamefont {Bogner}, \citenamefont {Furnstahl}, \citenamefont {Nogga},\ and\ \citenamefont {Schwenk}}]{hebeler2011}%
  \BibitemOpen
  \bibfield  {author} {\bibinfo {author} {\bibfnamefont {K.}~\bibnamefont {Hebeler}}, \bibinfo {author} {\bibfnamefont {S.~K.}\ \bibnamefont {Bogner}}, \bibinfo {author} {\bibfnamefont {R.~J.}\ \bibnamefont {Furnstahl}}, \bibinfo {author} {\bibfnamefont {A.}~\bibnamefont {Nogga}}, \ and\ \bibinfo {author} {\bibfnamefont {A.}~\bibnamefont {Schwenk}},\ }\bibfield  {title} {\enquote {\bibinfo {title} {Improved nuclear matter calculations from chiral low-momentum interactions},}\ }\href {\doibase 10.1103/PhysRevC.83.031301} {\bibfield  {journal} {\bibinfo  {journal} {Phys. Rev. C}\ }\textbf {\bibinfo {volume} {83}},\ \bibinfo {pages} {031301} (\bibinfo {year} {2011})}\BibitemShut {NoStop}%
\bibitem [{\citenamefont {Novario}\ \emph {et~al.}(2020)\citenamefont {Novario}, \citenamefont {Hagen}, \citenamefont {Jansen},\ and\ \citenamefont {Papenbrock}}]{novario2020}%
  \BibitemOpen
  \bibfield  {author} {\bibinfo {author} {\bibfnamefont {S.~J.}\ \bibnamefont {Novario}}, \bibinfo {author} {\bibfnamefont {G.}~\bibnamefont {Hagen}}, \bibinfo {author} {\bibfnamefont {G.~R.}\ \bibnamefont {Jansen}}, \ and\ \bibinfo {author} {\bibfnamefont {T.}~\bibnamefont {Papenbrock}},\ }\bibfield  {title} {\enquote {\bibinfo {title} {Charge radii of exotic neon and magnesium isotopes},}\ }\href {\doibase 10.1103/PhysRevC.102.051303} {\bibfield  {journal} {\bibinfo  {journal} {Phys. Rev. C}\ }\textbf {\bibinfo {volume} {102}},\ \bibinfo {pages} {051303} (\bibinfo {year} {2020})}\BibitemShut {NoStop}%
\bibitem [{\citenamefont {Hagen}\ \emph {et~al.}(2022)\citenamefont {Hagen}, \citenamefont {Novario}, \citenamefont {Sun}, \citenamefont {Papenbrock}, \citenamefont {Jansen}, \citenamefont {Lietz}, \citenamefont {Duguet},\ and\ \citenamefont {Tichai}}]{hagen2022}%
  \BibitemOpen
  \bibfield  {author} {\bibinfo {author} {\bibfnamefont {G.}~\bibnamefont {Hagen}}, \bibinfo {author} {\bibfnamefont {S.~J.}\ \bibnamefont {Novario}}, \bibinfo {author} {\bibfnamefont {Z.~H.}\ \bibnamefont {Sun}}, \bibinfo {author} {\bibfnamefont {T.}~\bibnamefont {Papenbrock}}, \bibinfo {author} {\bibfnamefont {G.~R.}\ \bibnamefont {Jansen}}, \bibinfo {author} {\bibfnamefont {J.~G.}\ \bibnamefont {Lietz}}, \bibinfo {author} {\bibfnamefont {T.}~\bibnamefont {Duguet}}, \ and\ \bibinfo {author} {\bibfnamefont {A.}~\bibnamefont {Tichai}},\ }\bibfield  {title} {\enquote {\bibinfo {title} {Angular-momentum projection in coupled-cluster theory: Structure of $^{34}\mathrm{Mg}$},}\ }\href {\doibase 10.1103/PhysRevC.105.064311} {\bibfield  {journal} {\bibinfo  {journal} {Phys. Rev. C}\ }\textbf {\bibinfo {volume} {105}},\ \bibinfo {pages} {064311} (\bibinfo {year} {2022})}\BibitemShut {NoStop}%
\bibitem [{\citenamefont {Porro}(2023)}]{porro}%
  \BibitemOpen
  \bibfield  {author} {\bibinfo {author} {\bibfnamefont {Andrea}\ \bibnamefont {Porro}},\ }\emph {\bibinfo {title} {{Ab-initio description of monopole resonances in light- and medium-mass nuclei}}},\ \href@noop {} {Ph.D. thesis},\ \bibinfo  {school} {Institut de Recherches sur les lois Fondamentales de l'Univers, France} (\bibinfo {year} {2023})\BibitemShut {NoStop}%
\end{thebibliography}%

\end{document}